\documentclass[a4j]{article}

\usepackage{amsmath}
\usepackage{amssymb}
\begin{document}
\begin{flushright}
January 1999\\
Revised in June 1999
\end{flushright}
\vspace{44pt}
\begin{center}
\begin{Large}
\bf
Spontaneous Polarization of \\
the Kondo problem associated with \\
the higher-spin six-vertex model
\end{Large}\\[6pt]
\vspace{25pt}
Nobuhisa Fukushima 
\vspace{10pt}

{\it Department of Mathematics\\
School of Science and Engineering,
Waseda University,\\
 3-4-1,Okubo Shinjuku-ku,
Tokyo 169, JAPAN}

\vspace{20pt}
Takeo Kojima\\
\vspace{10pt}

{\it Department of Mathematics\\
College of Science and Technology,
Nihon University,\\
 1-8,Kanda-Surugadai,
Chiyoda-ku, Tokyo 101, JAPAN}

\vspace{60pt}
\underline{Abstract}
\end{center}
We study the multi-channel Kondo model associated 
with an integrable higher-spin analog
of the anti-ferroelectric 6-vertex model,
which is constructed by inserting a spin $\frac{1}{2}$
to spin $1$ lines :
$\cdots 
{\mathbb C}^3
\otimes 
{\mathbb C}^3 \otimes {\mathbb C}^3 \otimes
{\mathbb C}^2 \otimes {\mathbb C}^3
\otimes {\mathbb C}^3 \otimes 
{\mathbb C}^3
\cdots $.
We formulate the problem in terms of representation theory
of quantum affine algebra $U_q(\widehat{sl_2})$ \cite{JM}.
We derive an exact formula of
the spontaneous staggered polarization
for our model, which corresponds to
Baxter's formula \cite{Bax1} for the 6-vertex model.
\vspace{15pt}

\newpage
\section{Introduction}
R. Baxter \cite{Bax1} studied spontaneous 
staggered polarization
of the 6-vertex model in 1973.
He derived an exact formula of this quantity 
by the Transfer Matrix Method :
\begin{eqnarray}
\frac{(q^2;q^2)_\infty^{~2}}
{(-q^2;q^2)_\infty^{~2}}.
\label{eq:Baxter}
\end{eqnarray}
Here we have used the standard notation
\begin{eqnarray}
(z;p)_\infty=\prod_{n=0}^\infty (1-p^n z).
\nonumber
\end{eqnarray}
In 1976 R. Baxter \cite{Bax2}
invented the Corner Transfer Matrix Method.
The calculation of the spontaneous staggered polarization
is reduced to counting the multiplicities of
the eigenvalues of the Corner Transfer Matrix.
It was recognized that in many interesting cases
the eigenvalue of the Corner Transfer Matrix
can be described in terms of the characters of affine
Lie algebras.
Kyoto-school \cite{DFJMN}, \cite{JM}
gave the mathematical explanations of the Corner 
Transfer Matrix Method,
and at the same time they invented
the representation theoretical approach to
solvable lattice models.
Kyoto-school's approach reproduces Baxter's formula
(\ref{eq:Baxter}) and makes it possible to calculate
the quantities which cannot be calculated by
the Corner Transfer Matrix Method.
Kyoto-school's methods have been applied to
various problems \cite{LP}, \cite{Iz}, \cite{JKKKM}, 
\cite{LaP}.
A. Nakayashiki \cite{N} 
introduced new-type vertex operators
and gave the mathematical formulation of 
the usual Kondo model.

In this article we consider the multi-channel Kondo 
problem \cite{FG} associated with an integrable higher-spin 
analog of the anti-ferroelectric 6-vertex model,
which is constructed by inserting a spin $\frac{1}{2}$
to spin $1$ lines :
\begin{eqnarray}
\cdots 
{\mathbb C}^3
\otimes 
{\mathbb C}^3\otimes {\mathbb C}^3\otimes
{\mathbb C}^2\otimes {\mathbb C}^3
\otimes {\mathbb C}^3\otimes 
{\mathbb C}^3
\cdots.\nonumber
\end{eqnarray}
This problem has quantum affine
symmetry $U_q(\widehat{sl_2})$.
Our main result is an exact formula
of spontaneous staggered polarization :
\begin{eqnarray}
\displaystyle
-\frac{1}{1-q^4}
\frac{(q^{16};q^{16})_\infty}
{(q^{4};q^{4})_\infty}
\left\{(1+q^{4})\frac{(-q^4;q^8)_\infty}
{(-q^4;q^4)_\infty^2}
-2q^{2}\frac{(-q^8;q^{16})_\infty^2}
{(-q^2;q^4)_\infty^2}
-4q^4\frac{(-q^{16};q^{16})_\infty^2}
{(-q^2;q^4)_\infty^2}\right\}.\nonumber
\end{eqnarray}
Now a few words about the organization of the paper.
In section 2 we define the problem
and state the main result.
In section 3 we derive an exact formula
of spontaneous staggered polarization.

\section{Problem and Result}
The purpose of this section is to set
the problem and summarize the main result.

\subsection{Quantum affine algebra $U_q(\widehat{sl_2})$}
We follow the notation of \cite{JM}.
We give definitions of quantum affine Lie algebras
$U_q(\widehat{sl_2}),$ highest weight modules,
and principal evaluation modules.

Consider a free abelian group on the letters
$\Lambda_0,\Lambda_1,\delta$ :
\begin{eqnarray}
P={\mathbb Z}\Lambda_0
\oplus
{\mathbb Z}\Lambda_1\oplus
{\mathbb Z}\delta.\nonumber
\end{eqnarray}
Define the simple roots $\alpha_0, \alpha_1$ and an element
$\rho$ by
\begin{eqnarray}
\alpha_0+\alpha_1=\delta,~~
\Lambda_1=\Lambda_0+\frac{\alpha_1}{2},~~
\rho=\Lambda_0+\Lambda_1.\nonumber
\end{eqnarray}
Let $(h_0,h_1,d)$ be an basis of $P^*
=\hbox{Hom}(P,{\mathbb Z})$ dual to 
$(\Lambda_0,\Lambda_1,\delta)$.
Define a symmetric bilinear form by
\begin{eqnarray}
(\Lambda_0,\Lambda_0)=0,&(\Lambda_0,\alpha_1)=0,&
(\Lambda_0,\delta)=1,\nonumber \\
(\alpha_1,\alpha_1)=2,&(\alpha_1,\delta)=0,&
(\delta,\delta)=0.\nonumber
\end{eqnarray}
Regarding $P^* \subset P$ via this bilinear form
we have the identification
\begin{eqnarray}
h_0=\alpha_0,~~h_1=\alpha_1,~~d=\Lambda_0.\nonumber
\end{eqnarray}
We use the symbol
$$[n]=\frac{q^n-q^{-n}}{q-q^{-1}}.$$
The quantum affine algebra
$U_q(\widehat{sl_2})$
is an algebra with $1$ over ${\mathbb C}$,
defined on the generators $e_0,e_1,f_0,f_1$ and $q^h~
(h \in P^*)$ through the defining relations :
$$q^hq^{h'}=q^{h+h'}, \quad q^0=1,$$
$$q^he_iq^{-h}=q^{(\alpha_i,h )}e_i, \quad
q^hf_iq^{-h}=q^{-( \alpha_i,h )}f_i,$$
$$[e_i,f_j]=\delta_{ij}\frac{t_i-t_i^{-1}}{q-q^{-1}},$$
$$e_i^3e_j-[3]e_i^2e_je_i+[3]e_ie_je_i^2-e_je_i^3=0 \ 
~~(i \neq j),$$
$$f_i^3f_j-[3]f_i^2f_jf_i+[3]f_if_jf_i^2-f_jf_i^3=0 \ 
~~(i \neq j).$$
Here $t_i=q^{h_i}$.
We write $U_q'(\widehat{sl_2})$ 
for subalgebra of $U_q(\widehat{sl_2})$
generated by $e_0,e_1,f_0,f_1,$ \newline 
$q^{h_0},q^{h_1},$
and $U_q({sl_2})$ by $e_1,f_1,q^{h_1}$.
We define the coproduct $\Delta$ by
$$\Delta(q^h)=q^h \otimes q^h,~
\Delta(e_i)=e_i \otimes 1+t_i \otimes e_i,~
\Delta(f_i)=f_i \otimes t_i^{-1}+1 \otimes f_i.$$

We define the irreducible highest weight module.
Set $P_+={\mathbb Z}_{\geq 0} \Lambda_0 \oplus
{\mathbb Z}_{\geq 0} \Lambda_1$.
For $\lambda \in P_+$, a $U_q(\widehat{sl_2})$ module
$V(\lambda)$ is called an irreducible highest weight 
module with highest weight $\lambda$
if the following conditions are satisfied :
there exists a nonzero vector $|\lambda\rangle
\in V(\lambda)$, called the highest weight vector,
such that $q^h|\lambda\rangle=q^{(\lambda,h)}
|\lambda\rangle (h \in P^*)$,
$e_i|\lambda\rangle=f_i^{(\lambda,h_i)+1}
|\lambda\rangle=0 (i=0,1)$,
and $V(\lambda)=U_q(\widehat{sl_2})|\lambda\rangle$.
We say that $V(\lambda)$ has level $k$ if
$t_0 t_1|\lambda\rangle=q^k |\lambda\rangle$.
When  $V(\lambda)$ has level $k$,
the weight $\lambda=(k-m)\Lambda_0+m \Lambda_1
(m=0,\cdots,k)$.
In this article we use level $2$ modules :
$V(2\Lambda_0),V(\Lambda_0+\Lambda_1),
V(2\Lambda_1)$.

We define the principal evaluation modules
$V_\zeta$ of the subalgebra $U_q'(\widehat{sl_2})$.
Let $V$ be a module of $U_q({sl_2})$.
We equip $V_\zeta$ with a $U_q'(\widehat{sl_2})$-module
structure by setting
\begin{eqnarray}
e_0(v_\epsilon \otimes \zeta^m)=(f_1 v_\epsilon) 
\otimes \zeta^{m+1},
& 
e_1(v_\epsilon \otimes \zeta^m)=(e_1 v_\epsilon) 
\otimes \zeta^{m+1},
\nonumber \\
f_0(v_\epsilon \otimes \zeta^m)=(e_1 v_\epsilon) 
\otimes \zeta^{m-1},
& 
f_1(v_\epsilon \otimes \zeta^m)=(f_1 v_\epsilon) 
\otimes \zeta^{m-1},
\nonumber \\
t_0=t_1^{-1}, &t_1(v_\epsilon \otimes \zeta^m)=
(t_1 v_\epsilon)\otimes \zeta^m.\nonumber
\end{eqnarray}

\subsection{R-matrix and Lattice Model}
In this subsection we will define
our two-dimensional lattice model,
and summarize the main result.
Let $V^{(1)}_{\zeta} \simeq {\mathbb C}^3$
and
$V^{\left(\frac{1}{2}\right)}_{\zeta}
\simeq {\mathbb C}^2$ be the 
$U_q(\widehat{sl_2})$ principal modules.
We fix real numbers $q$ and $\zeta$ as 
$$-1 < q < 0,~~~1<\zeta<(-q)^{-1},$$ 
in the following.
The Boltzmann weights of our model
are specified
by the spin $(1,1)$ R-matrix intertwiner
$R^{(1,1)}\left(\zeta\right)$
and the spin
$\left(\frac{1}{2},1\right)$ R-matrix intertwiner
$R^{\left(\frac{1}{2},1\right)}\left(\zeta\right)$.
The spin $(1,1)$ R-matrix intertwiner
$R^{(1,1)}\left(\zeta_1/\zeta_2\right)
:V^{(1)}_{\zeta_1}\otimes V^{(1)}_{\zeta_2}
\to V^{(1)}_{\zeta_2}\otimes V^{(1)}_{\zeta_1}$ 
is given by
\begin{eqnarray}
R^{(1,1)}\left(\zeta\right)&=&
\displaystyle
\frac{1}{
{\kappa^{(1,1)}(\zeta)}}\left(
\begin{array}{ccccccccc}
a_1&~&~&~&~&~&~&~&~\\
~&a_2&~&a_3&~&~&~&~&~\\
~&~&a_4&~&a_5&~&a_6&~&~\\
~&a_3&~&a_2&~&~&~&~&~\\
~&~&a_5&~&a_7&~&a_5&~&~\\
~&~&~&~&~&a_2&~&a_3&~\\
~&~&a_6&~&a_5&~&a_4&~&~\\
~&~&~&~&~&a_3&~&a_2&~\\
~&~&~&~&~&~&~&~&a_1
\end{array}\right).
\end{eqnarray}
Here we set
\begin{eqnarray}
\kappa^{(1,1)}(\zeta)
=\zeta^2\frac{1-q^2\zeta^{-2}}{1-q^2\zeta^2},
\nonumber
\end{eqnarray}
and
\begin{eqnarray}
&&a_1=1,~~
a_2=(1-\zeta^2)q^2/d_4,~~
a_3=(1-q^4)\zeta/d_4,\nonumber\\
&&a_4=(1-\zeta^2)(q^2-\zeta^2)q^2/d_2 d_4,~~
a_5=(1-\zeta^2)(1-q^4)q\zeta/d_2 d_4,\nonumber\\
&&a_6=(1-q^2)(1-q^4)\zeta^2/d_2 d_4,\nonumber\\
&&a_7=a_2+a_6,~~~
d_2=1-q^2\zeta^2,~~~
d_4=1-q^4\zeta^2.\nonumber
\end{eqnarray}
It is the Boltzmann weight $a_6$ that dominates 
at low temperature, i.e., when $q$ is nearly 
equal to $0$.
The R-matrix $R^{(1,1)}(\zeta)$ satisfies 
unitarity and crossing-symmetry :
\begin{eqnarray}
R^{(1,1)}(\zeta)R^{(1,1)}(\zeta^{-1})=I,~~~
R^{(1,1)}(-q^{-1}\zeta)_{k,l}^{k',l'}=
R^{(1,1)}(\zeta^{-1})_{2-k',l}^{2-k,l'}.\nonumber
\end{eqnarray}
Let us define
the spin $\left(\frac{1}{2},1\right)$ R-matrix intertwiner
$R^{\left(\frac{1}{2},1\right)}\left(\zeta_1/\zeta_2\right)
:V^{\left(\frac{1}{2}\right)}_{\zeta_1}\otimes 
V^{(1)}_{\zeta_2}
\to V^{(1)}_{\zeta_2}\otimes 
V^{\left(\frac{1}{2}\right)}_{\zeta_1}$ by
\begin{eqnarray}
R^{\left(\frac{1}{2},1\right)}\left(\zeta\right)
&=&\displaystyle
\frac{1}{\kappa^{\left(\frac{1}{2},1\right)}(\zeta)}
\left(
\begin{array}{cccccc}
b_1&~&~&~&~&~\\
~&b_2&~&b_4&~&~\\
~&~&b_3&~&b_4&~\\
~&b_4&~&b_3&~&~\\
~&~&b_4&~&b_2&~\\
~&~&~&~&~&b_1
\end{array}\right).
\end{eqnarray}
Here we set
\begin{eqnarray}
\kappa^{\left(\frac{1}{2},1\right)}(\zeta)
=\zeta
\frac{(q^5\zeta^2;q^4)_\infty (q^3\zeta^{-2};q^4)_\infty}
{(q^5\zeta^{-2};q^4)_\infty (q^3\zeta^2
;q^4)_\infty},\nonumber
\end{eqnarray}
and
\begin{eqnarray}
b_1=1,~~~
b_2=\frac{(1-q\zeta^2)q}{1-q^3\zeta^2},~~
b_3=\frac{(q-\zeta^2)q}{1-q^3\zeta^2},~~
b_4=\sqrt{1+q^2}~\frac{(1-q^2)\zeta}{1-q^3\zeta^2}.\nonumber
\end{eqnarray}
It is the Boltzmann weight $b_4$ that dominates 
at low temperature, i.e., when $q$ is nearly 
equal to $0$.
The R-matrix $R^{\left(\frac{1}{2},1\right)}
\left(\zeta\right)$ satisfies unitarity and
crossing-symmetry :
\begin{eqnarray}
R^{\left(\frac{1}{2},1\right)}\left(\zeta\right)
R^{\left(\frac{1}{2},1\right)}\left(\zeta^{-1}\right)
=I,~~~
R^{\left(\frac{1}{2},1\right)}\left(-q^{-1}\zeta\right)
_{k,l}^{k',l'}=
R^{\left(\frac{1}{2},1\right)}\left(\zeta^{-1}\right)
_{1-k',l}^{1-k,l}.\nonumber
\end{eqnarray}

A lattice vertex associated with the interaction of
a spin $1$ and spin $1$ line has spin variables
$i,i'=(0,1,2)$
and $j,j'=(0,1,2)$, and spectral parameters
$\zeta_1, \zeta_2 \in {\mathbb C}$.
A Boltzmann weight
$R^{\left(1,1\right)}\left(\zeta_1/\zeta_2\right)
_{i',j'}^{i,j}$
is attached to the configuration of these variables
shown in figure 1.
A lattice vertex associated with the interaction of
a spin $\frac{1}{2}$ and spin $1$ line has spin variables
$i,i'=(0,1)$
and $j,j'=(0,1,2)$, and spectral parameters
$\zeta_1, \zeta_2 \in {\mathbb C}$.
A Boltzmann weight
$R^{\left(\frac{1}{2},1\right)}\left(\zeta_1/\zeta_2\right)
_{i',j'}^{i,j}$
is attached to the configuration of these variables
shown in figure 2.

~\\

%WinTpicVersion2.13
\unitlength 0.1in
\begin{picture}(38.65,17.00)(5.30,-18.85)
% VECTOR 2 0 3 0
% 4 1400 810 1400 2010 2000 1410 800 1410
% 
\special{pn 8}%
\special{pa 1400 410}%
\special{pa 1400 1610}%
\special{fp}%
\special{sh 1}%
\special{pa 1400 1610}%
\special{pa 1420 1543}%
\special{pa 1400 1557}%
\special{pa 1380 1543}%
\special{pa 1400 1610}%
\special{fp}%
\special{pa 2000 1010}%
\special{pa 800 1010}%
\special{fp}%
\special{sh 1}%
\special{pa 800 1010}%
\special{pa 867 1030}%
\special{pa 853 1010}%
\special{pa 867 990}%
\special{pa 800 1010}%
\special{fp}%
% STR 2 0 3 0
% 3 2110 1310 2110 1410 5 0
% $j$
\put(21.1000,-10.1000){\makebox(0,0){$j$}}%
% STR 2 0 3 0
% 3 710 1310 710 1410 5 0
% $j'$
\put(7.1000,-10.1000){\makebox(0,0){$j'$}}%
% STR 2 0 3 0
% 3 1400 550 1400 650 5 0
% 
\put(14.0000,-2.5000){\makebox(0,0){}}%
% STR 2 0 3 0
% 3 1410 580 1410 680 5 0
% $i$
\put(14.1000,-2.8000){\makebox(0,0){$i$}}%
% STR 2 0 3 0
% 3 1400 1970 1400 2070 5 0
% $i'$
\put(14.0000,-16.7000){\makebox(0,0){$i'$}}%
% STR 2 0 3 0
% 3 1290 1020 1290 1120 5 0
% $\zeta_1$
\put(12.9000,-7.2000){\makebox(0,0){$\zeta_1$}}%
% STR 2 0 3 0
% 3 1670 1450 1670 1550 5 0
% $\zeta_2$
\put(16.7000,-11.5000){\makebox(0,0){$\zeta_2$}}%
% VECTOR 2 0 3 0
% 2 4390 1410 3190 1410
% 
\special{pn 8}%
\special{pa 4390 1010}%
\special{pa 3190 1010}%
\special{fp}%
\special{sh 1}%
\special{pa 3190 1010}%
\special{pa 3257 1030}%
\special{pa 3243 1010}%
\special{pa 3257 990}%
\special{pa 3190 1010}%
\special{fp}%
% VECTOR 0 0 3 0
% 2 3810 810 3810 2010
% 
\special{pn 20}%
\special{pa 3810 410}%
\special{pa 3810 1610}%
\special{fp}%
\special{sh 1}%
\special{pa 3810 1610}%
\special{pa 3830 1543}%
\special{pa 3810 1557}%
\special{pa 3790 1543}%
\special{pa 3810 1610}%
\special{fp}%
% STR 2 0 3 0
% 3 3810 570 3810 670 5 0
% $i$
\put(38.1000,-2.7000){\makebox(0,0){$i$}}%
% STR 2 0 3 0
% 3 3810 2010 3810 2110 5 0
% $i'$
\put(38.1000,-17.1000){\makebox(0,0){$i'$}}%
% STR 2 0 3 0
% 3 4530 1310 4530 1410 5 0
% $j$
\put(45.3000,-10.1000){\makebox(0,0){$j$}}%
% STR 2 0 3 0
% 3 3130 1310 3130 1410 5 0
% $j'$
\put(31.3000,-10.1000){\makebox(0,0){$j'$}}%
% STR 2 0 3 0
% 3 3650 1020 3650 1120 5 0
% $\zeta_1$
\put(36.5000,-7.2000){\makebox(0,0){$\zeta_1$}}%
% STR 2 0 3 0
% 3 4080 1460 4080 1560 5 0
% $\zeta_2$
\put(40.8000,-11.6000){\makebox(0,0){$\zeta_2$}}%
% STR 2 0 3 0
% 3 1390 2270 1390 2370 5 0
% \bf Figure 1
\put(13.9000,-19.7000){\makebox(0,0){\bf Figure 1}}%
% STR 2 0 3 0
% 3 3790 2270 3790 2370 5 0
% \bf Figure 2
\put(37.9000,-19.7000){\makebox(0,0){\bf Figure 2}}%
\end{picture}%

\vspace{24pt}

Now we consider the finite lattice in figure 3
under special boundary conditions.
\vspace{24pt}

%WinTpicVersion2.13
\unitlength 0.1in
\begin{picture}(49.30,38.85)(10.60,-40.00)
% VECTOR 1 0 3 0
% 2 3800 600 3800 4400
% 
\special{pn 13}%
\special{pa 3800 200}%
\special{pa 3800 4000}%
\special{fp}%
\special{sh 1}%
\special{pa 3800 4000}%
\special{pa 3820 3933}%
\special{pa 3800 3947}%
\special{pa 3780 3933}%
\special{pa 3800 4000}%
\special{fp}%
% VECTOR 2 0 3 0
% 2 4200 600 4200 4400
% 
\special{pn 8}%
\special{pa 4200 200}%
\special{pa 4200 4000}%
\special{fp}%
\special{sh 1}%
\special{pa 4200 4000}%
\special{pa 4220 3933}%
\special{pa 4200 3947}%
\special{pa 4180 3933}%
\special{pa 4200 4000}%
\special{fp}%
% VECTOR 2 0 3 0
% 2 3400 600 3400 4400
% 
\special{pn 8}%
\special{pa 3400 200}%
\special{pa 3400 4000}%
\special{fp}%
\special{sh 1}%
\special{pa 3400 4000}%
\special{pa 3420 3933}%
\special{pa 3400 3947}%
\special{pa 3380 3933}%
\special{pa 3400 4000}%
\special{fp}%
% VECTOR 2 0 3 0
% 8 4600 1000 4600 4000 5000 1400 5000 3600 5400 1800 5400 3200 5800 2200 5800 2800
% 
\special{pn 8}%
\special{pa 4600 600}%
\special{pa 4600 3600}%
\special{fp}%
\special{sh 1}%
\special{pa 4600 3600}%
\special{pa 4620 3533}%
\special{pa 4600 3547}%
\special{pa 4580 3533}%
\special{pa 4600 3600}%
\special{fp}%
\special{pa 5000 1000}%
\special{pa 5000 3200}%
\special{fp}%
\special{sh 1}%
\special{pa 5000 3200}%
\special{pa 5020 3133}%
\special{pa 5000 3147}%
\special{pa 4980 3133}%
\special{pa 5000 3200}%
\special{fp}%
\special{pa 5400 1400}%
\special{pa 5400 2800}%
\special{fp}%
\special{sh 1}%
\special{pa 5400 2800}%
\special{pa 5420 2733}%
\special{pa 5400 2747}%
\special{pa 5380 2733}%
\special{pa 5400 2800}%
\special{fp}%
\special{pa 5800 1800}%
\special{pa 5800 2400}%
\special{fp}%
\special{sh 1}%
\special{pa 5800 2400}%
\special{pa 5820 2333}%
\special{pa 5800 2347}%
\special{pa 5780 2333}%
\special{pa 5800 2400}%
\special{fp}%
% VECTOR 2 0 3 0
% 8 2990 1000 2990 4000 2590 1400 2590 3600 2190 1800 2190 3200 1790 2200 1790 2800
% 
\special{pn 8}%
\special{pa 2990 600}%
\special{pa 2990 3600}%
\special{fp}%
\special{sh 1}%
\special{pa 2990 3600}%
\special{pa 3010 3533}%
\special{pa 2990 3547}%
\special{pa 2970 3533}%
\special{pa 2990 3600}%
\special{fp}%
\special{pa 2590 1000}%
\special{pa 2590 3200}%
\special{fp}%
\special{sh 1}%
\special{pa 2590 3200}%
\special{pa 2610 3133}%
\special{pa 2590 3147}%
\special{pa 2570 3133}%
\special{pa 2590 3200}%
\special{fp}%
\special{pa 2190 1400}%
\special{pa 2190 2800}%
\special{fp}%
\special{sh 1}%
\special{pa 2190 2800}%
\special{pa 2210 2733}%
\special{pa 2190 2747}%
\special{pa 2170 2733}%
\special{pa 2190 2800}%
\special{fp}%
\special{pa 1790 1800}%
\special{pa 1790 2400}%
\special{fp}%
\special{sh 1}%
\special{pa 1790 2400}%
\special{pa 1810 2333}%
\special{pa 1790 2347}%
\special{pa 1770 2333}%
\special{pa 1790 2400}%
\special{fp}%
% VECTOR 2 0 3 0
% 4 5990 2400 1590 2400 5990 2600 1590 2600
% 
\special{pn 8}%
\special{pa 5990 2000}%
\special{pa 1590 2000}%
\special{fp}%
\special{sh 1}%
\special{pa 1590 2000}%
\special{pa 1657 2020}%
\special{pa 1643 2000}%
\special{pa 1657 1980}%
\special{pa 1590 2000}%
\special{fp}%
\special{pa 5990 2200}%
\special{pa 1590 2200}%
\special{fp}%
\special{sh 1}%
\special{pa 1590 2200}%
\special{pa 1657 2220}%
\special{pa 1643 2200}%
\special{pa 1657 2180}%
\special{pa 1590 2200}%
\special{fp}%
% VECTOR 2 0 3 0
% 16 5600 2010 2000 2010 5200 1610 2400 1610 4800 1210 2800 1210 4400 810 3200 810 5600 3010 2000 3010 5200 3410 2400 3410 4800 3810 2800 3810 4400 4210 3200 4210
% 
\special{pn 8}%
\special{pa 5600 1610}%
\special{pa 2000 1610}%
\special{fp}%
\special{sh 1}%
\special{pa 2000 1610}%
\special{pa 2067 1630}%
\special{pa 2053 1610}%
\special{pa 2067 1590}%
\special{pa 2000 1610}%
\special{fp}%
\special{pa 5200 1210}%
\special{pa 2400 1210}%
\special{fp}%
\special{sh 1}%
\special{pa 2400 1210}%
\special{pa 2467 1230}%
\special{pa 2453 1210}%
\special{pa 2467 1190}%
\special{pa 2400 1210}%
\special{fp}%
\special{pa 4800 810}%
\special{pa 2800 810}%
\special{fp}%
\special{sh 1}%
\special{pa 2800 810}%
\special{pa 2867 830}%
\special{pa 2853 810}%
\special{pa 2867 790}%
\special{pa 2800 810}%
\special{fp}%
\special{pa 4400 410}%
\special{pa 3200 410}%
\special{fp}%
\special{sh 1}%
\special{pa 3200 410}%
\special{pa 3267 430}%
\special{pa 3253 410}%
\special{pa 3267 390}%
\special{pa 3200 410}%
\special{fp}%
\special{pa 5600 2610}%
\special{pa 2000 2610}%
\special{fp}%
\special{sh 1}%
\special{pa 2000 2610}%
\special{pa 2067 2630}%
\special{pa 2053 2610}%
\special{pa 2067 2590}%
\special{pa 2000 2610}%
\special{fp}%
\special{pa 5200 3010}%
\special{pa 2400 3010}%
\special{fp}%
\special{sh 1}%
\special{pa 2400 3010}%
\special{pa 2467 3030}%
\special{pa 2453 3010}%
\special{pa 2467 2990}%
\special{pa 2400 3010}%
\special{fp}%
\special{pa 4800 3410}%
\special{pa 2800 3410}%
\special{fp}%
\special{sh 1}%
\special{pa 2800 3410}%
\special{pa 2867 3430}%
\special{pa 2853 3410}%
\special{pa 2867 3390}%
\special{pa 2800 3410}%
\special{fp}%
\special{pa 4400 3810}%
\special{pa 3200 3810}%
\special{fp}%
\special{sh 1}%
\special{pa 3200 3810}%
\special{pa 3267 3830}%
\special{pa 3253 3810}%
\special{pa 3267 3790}%
\special{pa 3200 3810}%
\special{fp}%
% VECTOR 0 0 3 0
% 2 3800 600 3800 4400
% 
\special{pn 20}%
\special{pa 3800 200}%
\special{pa 3800 4000}%
\special{fp}%
\special{sh 1}%
\special{pa 3800 4000}%
\special{pa 3820 3933}%
\special{pa 3800 3947}%
\special{pa 3780 3933}%
\special{pa 3800 4000}%
\special{fp}%
% STR 2 0 3 0
% 3 1480 2300 1480 2400 5 0
% $b_{2N}$
\put(14.8000,-20.0000){\makebox(0,0){$b_{2N}$}}%
% STR 2 0 3 0
% 3 1480 2500 1480 2600 5 0
% $c_{2N}$
\put(14.8000,-22.0000){\makebox(0,0){$c_{2N}$}}%
% STR 2 0 3 0
% 3 6120 2500 6120 2600 5 0
% $d_{2N}$
\put(61.2000,-22.0000){\makebox(0,0){$d_{2N}$}}%
% STR 2 0 3 0
% 3 6120 2300 6120 2400 5 0
% $a_{2N}$
\put(61.2000,-20.0000){\makebox(0,0){$a_{2N}$}}%
% STR 2 0 3 0
% 3 6010 2090 6010 2190 5 0
% $a_{2N-1}$
\put(60.1000,-17.9000){\makebox(0,0){$a_{2N-1}$}}%
% STR 2 0 3 0
% 3 6000 2690 6000 2790 5 0
% $d_{2N-1}$
\put(60.0000,-23.9000){\makebox(0,0){$d_{2N-1}$}}%
% STR 2 0 3 0
% 3 1600 2090 1600 2190 5 0
% $b_{2N-1}$
\put(16.0000,-17.9000){\makebox(0,0){$b_{2N-1}$}}%
% STR 2 0 3 0
% 3 1600 2690 1600 2790 5 0
% $c_{2N-1}$
\put(16.0000,-23.9000){\makebox(0,0){$c_{2N-1}$}}%
% STR 2 0 3 0
% 3 5800 1890 5800 1990 5 0
% $a_{2N-2}$
\put(58.0000,-15.9000){\makebox(0,0){$a_{2N-2}$}}%
% STR 2 0 3 0
% 3 5800 2890 5800 2990 5 0
% $d_{2N-2}$
\put(58.0000,-25.9000){\makebox(0,0){$d_{2N-2}$}}%
% STR 2 0 3 0
% 3 1800 2890 1800 2990 5 0
% $c_{2N-2}$
\put(18.0000,-25.9000){\makebox(0,0){$c_{2N-2}$}}%
% STR 2 0 3 0
% 3 1770 1900 1770 2000 5 0
% $b_{2N-2}$
\put(17.7000,-16.0000){\makebox(0,0){$b_{2N-2}$}}%
% STR 2 0 3 0
% 3 1970 1700 1970 1800 5 0
% $b_{2N-3}$
\put(19.7000,-14.0000){\makebox(0,0){$b_{2N-3}$}}%
% STR 2 0 3 0
% 3 1970 3100 1970 3200 5 0
% $c_{2N-3}$
\put(19.7000,-28.0000){\makebox(0,0){$c_{2N-3}$}}%
% STR 2 0 3 0
% 3 5600 3100 5600 3200 5 0
% $d_{2N-3}$
\put(56.0000,-28.0000){\makebox(0,0){$d_{2N-3}$}}%
% STR 2 0 3 0
% 3 5600 1700 5600 1800 5 0
% $a_{2N-3}$
\put(56.0000,-14.0000){\makebox(0,0){$a_{2N-3}$}}%
% STR 2 0 3 0
% 3 4400 500 4400 600 5 0
% $a_1$
\put(44.0000,-2.0000){\makebox(0,0){$a_1$}}%
% STR 2 0 3 0
% 3 3210 510 3210 610 5 0
% $b_1$
\put(32.1000,-2.1000){\makebox(0,0){$b_1$}}%
% STR 2 0 3 0
% 3 3210 4310 3210 4410 5 0
% $c_1$
\put(32.1000,-40.1000){\makebox(0,0){$c_1$}}%
% STR 2 0 3 0
% 3 4410 4310 4410 4410 5 0
% $d_1$
\put(44.1000,-40.1000){\makebox(0,0){$d_1$}}%
% STR 2 0 3 0
% 3 4810 910 4810 1010 5 0
% $a_3$
\put(48.1000,-6.1000){\makebox(0,0){$a_3$}}%
% STR 2 0 3 0
% 3 5010 1110 5010 1210 5 0
% $a_4$
\put(50.1000,-8.1000){\makebox(0,0){$a_4$}}%
% STR 2 0 3 0
% 3 4610 710 4610 810 5 0
% $a_2$
\put(46.1000,-4.1000){\makebox(0,0){$a_2$}}%
% STR 2 0 3 0
% 3 2990 710 2990 810 5 0
% $b_2$
\put(29.9000,-4.1000){\makebox(0,0){$b_2$}}%
% STR 2 0 3 0
% 3 2790 910 2790 1010 5 0
% $b_3$
\put(27.9000,-6.1000){\makebox(0,0){$b_3$}}%
% STR 2 0 3 0
% 3 2590 1110 2590 1210 5 0
% $b_4$
\put(25.9000,-8.1000){\makebox(0,0){$b_4$}}%
% STR 2 0 3 0
% 3 2990 4110 2990 4210 5 0
% $c_2$
\put(29.9000,-38.1000){\makebox(0,0){$c_2$}}%
% STR 2 0 3 0
% 3 2790 3910 2790 4010 5 0
% $c_3$
\put(27.9000,-36.1000){\makebox(0,0){$c_3$}}%
% STR 2 0 3 0
% 3 2590 3710 2590 3810 5 0
% $c_4$
\put(25.9000,-34.1000){\makebox(0,0){$c_4$}}%
% STR 2 0 3 0
% 3 4600 4110 4600 4210 5 0
% $d_2$
\put(46.0000,-38.1000){\makebox(0,0){$d_2$}}%
% STR 2 0 3 0
% 3 4800 3910 4800 4010 5 0
% $d_3$
\put(48.0000,-36.1000){\makebox(0,0){$d_3$}}%
% STR 2 0 3 0
% 3 5000 3710 5000 3810 5 0
% $d_4$
\put(50.0000,-34.1000){\makebox(0,0){$d_4$}}%
% STR 2 0 3 0
% 3 5400 1510 5400 1610 5 0
% $\cdots$
\put(54.0000,-12.1000){\makebox(0,0){$\cdots$}}%
% STR 2 0 3 0
% 3 5200 1310 5200 1410 5 0
% $\cdots$
\put(52.0000,-10.1000){\makebox(0,0){$\cdots$}}%
% STR 2 0 3 0
% 3 2400 1310 2400 1410 5 0
% $\cdots$
\put(24.0000,-10.1000){\makebox(0,0){$\cdots$}}%
% STR 2 0 3 0
% 3 2200 1510 2200 1610 5 0
% $\cdots$
\put(22.0000,-12.1000){\makebox(0,0){$\cdots$}}%
% STR 2 0 3 0
% 3 2200 3310 2200 3410 5 0
% $\cdots$
\put(22.0000,-30.1000){\makebox(0,0){$\cdots$}}%
% STR 2 0 3 0
% 3 2400 3510 2400 3610 5 0
% $\cdots$
\put(24.0000,-32.1000){\makebox(0,0){$\cdots$}}%
% STR 2 0 3 0
% 3 5200 3510 5200 3610 5 0
% $\cdots$
\put(52.0000,-32.1000){\makebox(0,0){$\cdots$}}%
% STR 2 0 3 0
% 3 5420 3320 5420 3420 5 0
% $\cdots$
\put(54.2000,-30.2000){\makebox(0,0){$\cdots$}}%
% VECTOR 0 0 3 0
% 2 3830 4380 3830 4380
% 
\special{pn 20}%
\special{pa 3830 3980}%
\special{pa 3830 3980}%
\special{fp}%
% STR 2 0 3 0
% 3 4030 2390 4030 2490 5 0
% $\epsilon(C)$
\put(40.3000,-20.9000){\makebox(0,0){$\epsilon(C)$}}%
% STR 2 0 3 0
% 3 1600 4080 1600 4180 5 0
% \bf Figure 3
\put(16.0000,-37.8000){\makebox(0,0){\bf Figure 3}}%
\end{picture}%

\vspace{24pt}

Our model has $2N+1$ vertical lines with spectral
parameter $\zeta$ and $2N$ horizontal lines
with spectral parameter $1$, where $N \in {\mathbb N}$.
The boundary conditions $a_j,b_j,c_j,d_j~(j=1,2,\cdots,2N)$
are fixed as following $4$ cases, and their ground 
states are shown in figure 4.

\begin{enumerate}
\item 
$(\Lambda_0+\Lambda_1,2\Lambda_0)$-Case :
\begin{eqnarray}
a_j=1+(-1)^{N+j+1},~~b_j=1,~~c_j=1,~~d_j=1+(-1)^N,
\nonumber
\end{eqnarray}
\item
$(\Lambda_0+\Lambda_1,2\Lambda_1)$-Case :
\begin{eqnarray}
a_j=1+(-1)^{N+j},~~b_j=1,~~c_j=1,~~d_j=1+(-1)^{N+1},
\nonumber
\end{eqnarray}
\item
$(2\Lambda_0,\Lambda_0+\Lambda_1)$-Case :
\begin{eqnarray}
a_j=1,~~b_j=1+(-1)^{N+1},~~
c_j=1+(-1)^{N+j},~~d_j=1,
\nonumber
\end{eqnarray}
\item
$(2\Lambda_1,\Lambda_0+\Lambda_1)$-Case :
\begin{eqnarray}a_j=1,~~
b_j=1+(-1)^N,~~
c_j=1+(-1)^{N+j+1},~~d_j=1
.\nonumber
\end{eqnarray}
\end{enumerate}

%WinTpicVersion2.16
\unitlength 0.1in
\begin{picture}(27.70,20.17)(-4.82,-25.46)
% LINE 0 0 3 0
% 2 1273 1141 1273 2814
% 
\special{pn 20}%
\special{pa 1273 741}%
\special{pa 1273 2414}%
\special{fp}%
% VECTOR 1 0 3 0
% 2 1273 2793 1273 2821
% 
\special{pn 13}%
\special{pa 1273 2393}%
\special{pa 1273 2421}%
\special{fp}%
\special{sh 1}%
\special{pa 1273 2421}%
\special{pa 1293 2354}%
\special{pa 1273 2368}%
\special{pa 1253 2354}%
\special{pa 1273 2421}%
\special{fp}%
% VECTOR 2 0 3 0
% 6 1553 1134 1553 2807 1840 1414 1840 2534 2113 1694 2113 2254
% 
\special{pn 8}%
\special{pa 1553 734}%
\special{pa 1553 2407}%
\special{fp}%
\special{sh 1}%
\special{pa 1553 2407}%
\special{pa 1573 2340}%
\special{pa 1553 2354}%
\special{pa 1533 2340}%
\special{pa 1553 2407}%
\special{fp}%
\special{pa 1840 1014}%
\special{pa 1840 2134}%
\special{fp}%
\special{sh 1}%
\special{pa 1840 2134}%
\special{pa 1860 2067}%
\special{pa 1840 2081}%
\special{pa 1820 2067}%
\special{pa 1840 2134}%
\special{fp}%
\special{pa 2113 1294}%
\special{pa 2113 1854}%
\special{fp}%
\special{sh 1}%
\special{pa 2113 1854}%
\special{pa 2133 1787}%
\special{pa 2113 1801}%
\special{pa 2093 1787}%
\special{pa 2113 1854}%
\special{fp}%
% VECTOR 2 0 3 0
% 6 993 1134 993 2814 713 1414 713 2527 433 1694 433 2247
% 
\special{pn 8}%
\special{pa 993 734}%
\special{pa 993 2414}%
\special{fp}%
\special{sh 1}%
\special{pa 993 2414}%
\special{pa 1013 2347}%
\special{pa 993 2361}%
\special{pa 973 2347}%
\special{pa 993 2414}%
\special{fp}%
\special{pa 713 1014}%
\special{pa 713 2127}%
\special{fp}%
\special{sh 1}%
\special{pa 713 2127}%
\special{pa 733 2060}%
\special{pa 713 2074}%
\special{pa 693 2060}%
\special{pa 713 2127}%
\special{fp}%
\special{pa 433 1294}%
\special{pa 433 1847}%
\special{fp}%
\special{sh 1}%
\special{pa 433 1847}%
\special{pa 453 1780}%
\special{pa 433 1794}%
\special{pa 413 1780}%
\special{pa 433 1847}%
\special{fp}%
% VECTOR 2 0 3 0
% 6 2260 1834 293 1834 1973 1547 580 1547 1693 1274 853 1274
% 
\special{pn 8}%
\special{pa 2260 1434}%
\special{pa 293 1434}%
\special{fp}%
\special{sh 1}%
\special{pa 293 1434}%
\special{pa 360 1454}%
\special{pa 346 1434}%
\special{pa 360 1414}%
\special{pa 293 1434}%
\special{fp}%
\special{pa 1973 1147}%
\special{pa 580 1147}%
\special{fp}%
\special{sh 1}%
\special{pa 580 1147}%
\special{pa 647 1167}%
\special{pa 633 1147}%
\special{pa 647 1127}%
\special{pa 580 1147}%
\special{fp}%
\special{pa 1693 874}%
\special{pa 853 874}%
\special{fp}%
\special{sh 1}%
\special{pa 853 874}%
\special{pa 920 894}%
\special{pa 906 874}%
\special{pa 920 854}%
\special{pa 853 874}%
\special{fp}%
% VECTOR 2 0 3 0
% 6 2253 2114 293 2114 1973 2394 580 2394 1693 2674 853 2674
% 
\special{pn 8}%
\special{pa 2253 1714}%
\special{pa 293 1714}%
\special{fp}%
\special{sh 1}%
\special{pa 293 1714}%
\special{pa 360 1734}%
\special{pa 346 1714}%
\special{pa 360 1694}%
\special{pa 293 1714}%
\special{fp}%
\special{pa 1973 1994}%
\special{pa 580 1994}%
\special{fp}%
\special{sh 1}%
\special{pa 580 1994}%
\special{pa 647 2014}%
\special{pa 633 1994}%
\special{pa 647 1974}%
\special{pa 580 1994}%
\special{fp}%
\special{pa 1693 2274}%
\special{pa 853 2274}%
\special{fp}%
\special{sh 1}%
\special{pa 853 2274}%
\special{pa 920 2294}%
\special{pa 906 2274}%
\special{pa 920 2254}%
\special{pa 853 2274}%
\special{fp}%
% STR 2 0 3 0
% 3 1273 2961 1273 3031 5 0
% $(\Lambda_0+\Lambda_1,2\Lambda_0)$-Case
\put(12.7300,-26.3100){\makebox(0,0){$(\Lambda_0+\Lambda_1,2\Lambda_0)$-Case}}%
% STR 2 0 3 0
% 3 1574 1043 1574 1113 2 0
% $0$
\put(15.7400,-7.1300){\makebox(0,0)[lb]{$0$}}%
% STR 2 0 3 0
% 3 1721 1176 1721 1246 2 0
% $2$
\put(17.2100,-8.4600){\makebox(0,0)[lb]{$2$}}%
% STR 2 0 3 0
% 3 1868 1323 1868 1393 2 0
% $0$
\put(18.6800,-9.9300){\makebox(0,0)[lb]{$0$}}%
% STR 2 0 3 0
% 3 1994 1456 1994 1526 2 0
% $2$
\put(19.9400,-11.2600){\makebox(0,0)[lb]{$2$}}%
% STR 2 0 3 0
% 3 2134 1603 2134 1673 2 0
% $0$
\put(21.3400,-12.7300){\makebox(0,0)[lb]{$0$}}%
% STR 2 0 3 0
% 3 2288 1736 2288 1806 2 0
% $2$
\put(22.8800,-14.0600){\makebox(0,0)[lb]{$2$}}%
% STR 2 0 3 0
% 3 1413 1169 1413 1239 5 0
% $0$
\put(14.1300,-8.3900){\makebox(0,0){$0$}}%
% STR 2 0 3 0
% 3 1693 1442 1693 1512 5 0
% $0$
\put(16.9300,-11.1200){\makebox(0,0){$0$}}%
% STR 2 0 3 0
% 3 1980 1722 1980 1792 5 0
% $0$
\put(19.8000,-13.9200){\makebox(0,0){$0$}}%
% STR 2 0 3 0
% 3 1308 1344 1308 1414 5 0
% $0$
\put(13.0800,-10.1400){\makebox(0,0){$0$}}%
% STR 2 0 3 0
% 3 1595 1344 1595 1414 5 0
% $2$
\put(15.9500,-10.1400){\makebox(0,0){$2$}}%
% STR 2 0 3 0
% 3 1413 1435 1413 1505 5 0
% $2$
\put(14.1300,-11.0500){\makebox(0,0){$2$}}%
% STR 2 0 3 0
% 3 1308 1624 1308 1694 5 0
% $1$
\put(13.0800,-12.9400){\makebox(0,0){$1$}}%
% STR 2 0 3 0
% 3 1308 1904 1308 1974 5 0
% $0$
\put(13.0800,-15.7400){\makebox(0,0){$0$}}%
% STR 2 0 3 0
% 3 1308 2184 1308 2254 5 0
% $1$
\put(13.0800,-18.5400){\makebox(0,0){$1$}}%
% STR 2 0 3 0
% 3 1308 2464 1308 2534 5 0
% $0$
\put(13.0800,-21.3400){\makebox(0,0){$0$}}%
% STR 2 0 3 0
% 3 1413 1715 1413 1785 5 0
% $0$
\put(14.1300,-13.8500){\makebox(0,0){$0$}}%
% STR 2 0 3 0
% 3 1595 1624 1595 1694 5 0
% $0$
\put(15.9500,-12.9400){\makebox(0,0){$0$}}%
% STR 2 0 3 0
% 3 1882 1624 1882 1694 5 0
% $2$
\put(18.8200,-12.9400){\makebox(0,0){$2$}}%
% STR 2 0 3 0
% 3 1693 1715 1693 1785 5 0
% $2$
\put(16.9300,-13.8500){\makebox(0,0){$2$}}%
% STR 2 0 3 0
% 3 1595 1904 1595 1974 5 0
% $2$
\put(15.9500,-15.7400){\makebox(0,0){$2$}}%
% STR 2 0 3 0
% 3 1413 2009 1413 2079 5 0
% $2$
\put(14.1300,-16.7900){\makebox(0,0){$2$}}%
% STR 2 0 3 0
% 3 1875 1904 1875 1974 5 0
% $0$
\put(18.7500,-15.7400){\makebox(0,0){$0$}}%
% STR 2 0 3 0
% 3 1588 2184 1588 2254 5 0
% $0$
\put(15.8800,-18.5400){\makebox(0,0){$0$}}%
% STR 2 0 3 0
% 3 1413 2282 1413 2352 5 0
% $0$
\put(14.1300,-19.5200){\makebox(0,0){$0$}}%
% STR 2 0 3 0
% 3 2148 1904 2148 1974 5 0
% $2$
\put(21.4800,-15.7400){\makebox(0,0){$2$}}%
% STR 2 0 3 0
% 3 1973 2002 1973 2072 5 0
% $2$
\put(19.7300,-16.7200){\makebox(0,0){$2$}}%
% STR 2 0 3 0
% 3 1875 2184 1875 2254 5 0
% $2$
\put(18.7500,-18.5400){\makebox(0,0){$2$}}%
% STR 2 0 3 0
% 3 1693 2275 1693 2345 5 0
% $2$
\put(16.9300,-19.4500){\makebox(0,0){$2$}}%
% STR 2 0 3 0
% 3 1693 2009 1693 2079 5 0
% $0$
\put(16.9300,-16.7900){\makebox(0,0){$0$}}%
% STR 2 0 3 0
% 3 1588 2464 1588 2534 5 0
% $2$
\put(15.8800,-21.3400){\makebox(0,0){$2$}}%
% STR 2 0 3 0
% 3 1413 2562 1413 2632 5 0
% $2$
\put(14.1300,-22.3200){\makebox(0,0){$2$}}%
% STR 2 0 3 0
% 3 2288 2072 2288 2142 2 0
% $0$
\put(22.8800,-17.4200){\makebox(0,0)[lb]{$0$}}%
% STR 2 0 3 0
% 3 2141 2240 2141 2310 2 0
% $0$
\put(21.4100,-19.1000){\makebox(0,0)[lb]{$0$}}%
% STR 2 0 3 0
% 3 2015 2373 2015 2443 2 0
% $0$
\put(20.1500,-20.4300){\makebox(0,0)[lb]{$0$}}%
% STR 2 0 3 0
% 3 1868 2548 1868 2618 2 0
% $0$
\put(18.6800,-22.1800){\makebox(0,0)[lb]{$0$}}%
% STR 2 0 3 0
% 3 1721 2646 1721 2716 2 0
% $0$
\put(17.2100,-23.1600){\makebox(0,0)[lb]{$0$}}%
% STR 2 0 3 0
% 3 1588 2807 1588 2877 2 0
% $0$
\put(15.8800,-24.7700){\makebox(0,0)[lb]{$0$}}%
% STR 2 0 3 0
% 3 1133 1162 1133 1232 5 0
% $1$
\put(11.3300,-8.3200){\makebox(0,0){$1$}}%
% STR 2 0 3 0
% 3 1028 1344 1028 1414 5 0
% $1$
\put(10.2800,-10.1400){\makebox(0,0){$1$}}%
% STR 2 0 3 0
% 3 755 1624 755 1694 5 0
% $1$
\put(7.5500,-12.9400){\makebox(0,0){$1$}}%
% STR 2 0 3 0
% 3 853 1442 853 1512 5 0
% $1$
\put(8.5300,-11.1200){\makebox(0,0){$1$}}%
% STR 2 0 3 0
% 3 580 1722 580 1792 5 0
% $1$
\put(5.8000,-13.9200){\makebox(0,0){$1$}}%
% STR 2 0 3 0
% 3 475 1904 475 1974 5 0
% $1$
\put(4.7500,-15.7400){\makebox(0,0){$1$}}%
% STR 2 0 3 0
% 3 1133 1435 1133 1505 5 0
% $1$
\put(11.3300,-11.0500){\makebox(0,0){$1$}}%
% STR 2 0 3 0
% 3 1028 1624 1028 1694 5 0
% $1$
\put(10.2800,-12.9400){\makebox(0,0){$1$}}%
% STR 2 0 3 0
% 3 853 1715 853 1785 5 0
% $1$
\put(8.5300,-13.8500){\makebox(0,0){$1$}}%
% STR 2 0 3 0
% 3 755 1904 755 1974 5 0
% $1$
\put(7.5500,-15.7400){\makebox(0,0){$1$}}%
% STR 2 0 3 0
% 3 580 2009 580 2079 5 0
% $1$
\put(5.8000,-16.7900){\makebox(0,0){$1$}}%
% STR 2 0 3 0
% 3 1133 1715 1133 1785 5 0
% $1$
\put(11.3300,-13.8500){\makebox(0,0){$1$}}%
% STR 2 0 3 0
% 3 1028 1904 1028 1974 5 0
% $1$
\put(10.2800,-15.7400){\makebox(0,0){$1$}}%
% STR 2 0 3 0
% 3 853 2002 853 2072 5 0
% $1$
\put(8.5300,-16.7200){\makebox(0,0){$1$}}%
% STR 2 0 3 0
% 3 755 2184 755 2254 5 0
% $1$
\put(7.5500,-18.5400){\makebox(0,0){$1$}}%
% STR 2 0 3 0
% 3 1133 2002 1133 2072 5 0
% $1$
\put(11.3300,-16.7200){\makebox(0,0){$1$}}%
% STR 2 0 3 0
% 3 1028 2184 1028 2254 5 0
% $1$
\put(10.2800,-18.5400){\makebox(0,0){$1$}}%
% STR 2 0 3 0
% 3 853 2282 853 2352 5 0
% $1$
\put(8.5300,-19.5200){\makebox(0,0){$1$}}%
% STR 2 0 3 0
% 3 1133 2275 1133 2345 5 0
% $1$
\put(11.3300,-19.4500){\makebox(0,0){$1$}}%
% STR 2 0 3 0
% 3 1028 2464 1028 2534 5 0
% $1$
\put(10.2800,-21.3400){\makebox(0,0){$1$}}%
% STR 2 0 3 0
% 3 1133 2562 1133 2632 5 0
% $1$
\put(11.3300,-22.3200){\makebox(0,0){$1$}}%
% STR 2 0 3 0
% 3 951 1029 951 1099 2 0
% $1$
\put(9.5100,-6.9900){\makebox(0,0)[lb]{$1$}}%
% STR 2 0 3 0
% 3 811 1169 811 1239 2 0
% $1$
\put(8.1100,-8.3900){\makebox(0,0)[lb]{$1$}}%
% STR 2 0 3 0
% 3 664 1323 664 1393 2 0
% $1$
\put(6.6400,-9.9300){\makebox(0,0)[lb]{$1$}}%
% STR 2 0 3 0
% 3 538 1456 538 1526 2 0
% $1$
\put(5.3800,-11.2600){\makebox(0,0)[lb]{$1$}}%
% STR 2 0 3 0
% 3 391 1610 391 1680 2 0
% $1$
\put(3.9100,-12.8000){\makebox(0,0)[lb]{$1$}}%
% STR 2 0 3 0
% 3 244 1736 244 1806 2 0
% $1$
\put(2.4400,-14.0600){\makebox(0,0)[lb]{$1$}}%
% STR 2 0 3 0
% 3 230 2072 230 2142 2 0
% $1$
\put(2.3000,-17.4200){\makebox(0,0)[lb]{$1$}}%
% STR 2 0 3 0
% 3 384 2268 384 2338 2 0
% $1$
\put(3.8400,-19.3800){\makebox(0,0)[lb]{$1$}}%
% STR 2 0 3 0
% 3 517 2380 517 2450 2 0
% $1$
\put(5.1700,-20.5000){\makebox(0,0)[lb]{$1$}}%
% STR 2 0 3 0
% 3 657 2548 657 2618 2 0
% $1$
\put(6.5700,-22.1800){\makebox(0,0)[lb]{$1$}}%
% STR 2 0 3 0
% 3 783 2681 783 2751 2 0
% $1$
\put(7.8300,-23.5100){\makebox(0,0)[lb]{$1$}}%
% STR 2 0 3 0
% 3 937 2821 937 2891 2 0
% $1$
\put(9.3700,-24.9100){\makebox(0,0)[lb]{$1$}}%
% STR 2 0 3 0
% 3 1273 994 1273 1064 5 0
% $1$
\put(12.7300,-6.6400){\makebox(0,0){$1$}}%
% STR 2 0 3 0
% 3 1273 2828 1273 2898 5 0
% $1$
\put(12.7300,-24.9800){\makebox(0,0){$1$}}%
\end{picture}%
%WinTpicVersion2.16
\unitlength 0.1in
\begin{picture}(27.70,20.17)(-4.82,-25.46)
% LINE 0 0 3 0
% 2 1273 1141 1273 2814
% 
\special{pn 20}%
\special{pa 1273 741}%
\special{pa 1273 2414}%
\special{fp}%
% VECTOR 1 0 3 0
% 2 1273 2793 1273 2821
% 
\special{pn 13}%
\special{pa 1273 2393}%
\special{pa 1273 2421}%
\special{fp}%
\special{sh 1}%
\special{pa 1273 2421}%
\special{pa 1293 2354}%
\special{pa 1273 2368}%
\special{pa 1253 2354}%
\special{pa 1273 2421}%
\special{fp}%
% VECTOR 2 0 3 0
% 6 1553 1134 1553 2807 1840 1414 1840 2534 2113 1694 2113 2254
% 
\special{pn 8}%
\special{pa 1553 734}%
\special{pa 1553 2407}%
\special{fp}%
\special{sh 1}%
\special{pa 1553 2407}%
\special{pa 1573 2340}%
\special{pa 1553 2354}%
\special{pa 1533 2340}%
\special{pa 1553 2407}%
\special{fp}%
\special{pa 1840 1014}%
\special{pa 1840 2134}%
\special{fp}%
\special{sh 1}%
\special{pa 1840 2134}%
\special{pa 1860 2067}%
\special{pa 1840 2081}%
\special{pa 1820 2067}%
\special{pa 1840 2134}%
\special{fp}%
\special{pa 2113 1294}%
\special{pa 2113 1854}%
\special{fp}%
\special{sh 1}%
\special{pa 2113 1854}%
\special{pa 2133 1787}%
\special{pa 2113 1801}%
\special{pa 2093 1787}%
\special{pa 2113 1854}%
\special{fp}%
% VECTOR 2 0 3 0
% 6 993 1134 993 2814 713 1414 713 2527 433 1694 433 2247
% 
\special{pn 8}%
\special{pa 993 734}%
\special{pa 993 2414}%
\special{fp}%
\special{sh 1}%
\special{pa 993 2414}%
\special{pa 1013 2347}%
\special{pa 993 2361}%
\special{pa 973 2347}%
\special{pa 993 2414}%
\special{fp}%
\special{pa 713 1014}%
\special{pa 713 2127}%
\special{fp}%
\special{sh 1}%
\special{pa 713 2127}%
\special{pa 733 2060}%
\special{pa 713 2074}%
\special{pa 693 2060}%
\special{pa 713 2127}%
\special{fp}%
\special{pa 433 1294}%
\special{pa 433 1847}%
\special{fp}%
\special{sh 1}%
\special{pa 433 1847}%
\special{pa 453 1780}%
\special{pa 433 1794}%
\special{pa 413 1780}%
\special{pa 433 1847}%
\special{fp}%
% VECTOR 2 0 3 0
% 6 2260 1834 293 1834 1973 1547 580 1547 1693 1274 853 1274
% 
\special{pn 8}%
\special{pa 2260 1434}%
\special{pa 293 1434}%
\special{fp}%
\special{sh 1}%
\special{pa 293 1434}%
\special{pa 360 1454}%
\special{pa 346 1434}%
\special{pa 360 1414}%
\special{pa 293 1434}%
\special{fp}%
\special{pa 1973 1147}%
\special{pa 580 1147}%
\special{fp}%
\special{sh 1}%
\special{pa 580 1147}%
\special{pa 647 1167}%
\special{pa 633 1147}%
\special{pa 647 1127}%
\special{pa 580 1147}%
\special{fp}%
\special{pa 1693 874}%
\special{pa 853 874}%
\special{fp}%
\special{sh 1}%
\special{pa 853 874}%
\special{pa 920 894}%
\special{pa 906 874}%
\special{pa 920 854}%
\special{pa 853 874}%
\special{fp}%
% VECTOR 2 0 3 0
% 6 2253 2114 293 2114 1973 2394 580 2394 1693 2674 853 2674
% 
\special{pn 8}%
\special{pa 2253 1714}%
\special{pa 293 1714}%
\special{fp}%
\special{sh 1}%
\special{pa 293 1714}%
\special{pa 360 1734}%
\special{pa 346 1714}%
\special{pa 360 1694}%
\special{pa 293 1714}%
\special{fp}%
\special{pa 1973 1994}%
\special{pa 580 1994}%
\special{fp}%
\special{sh 1}%
\special{pa 580 1994}%
\special{pa 647 2014}%
\special{pa 633 1994}%
\special{pa 647 1974}%
\special{pa 580 1994}%
\special{fp}%
\special{pa 1693 2274}%
\special{pa 853 2274}%
\special{fp}%
\special{sh 1}%
\special{pa 853 2274}%
\special{pa 920 2294}%
\special{pa 906 2274}%
\special{pa 920 2254}%
\special{pa 853 2274}%
\special{fp}%
% STR 2 0 3 0
% 3 1273 2961 1273 3031 5 0
% $(\Lambda_0+\Lambda_1,2\Lambda_1)$-Case
\put(12.7300,-26.3100){\makebox(0,0){$(\Lambda_0+\Lambda_1,2\Lambda_1)$-Case}}%
% STR 2 0 3 0
% 3 1574 1043 1574 1113 2 0
% $2$
\put(15.7400,-7.1300){\makebox(0,0)[lb]{$2$}}%
% STR 2 0 3 0
% 3 1721 1176 1721 1246 2 0
% $0$
\put(17.2100,-8.4600){\makebox(0,0)[lb]{$0$}}%
% STR 2 0 3 0
% 3 1868 1323 1868 1393 2 0
% $2$
\put(18.6800,-9.9300){\makebox(0,0)[lb]{$2$}}%
% STR 2 0 3 0
% 3 1994 1456 1994 1526 2 0
% $0$
\put(19.9400,-11.2600){\makebox(0,0)[lb]{$0$}}%
% STR 2 0 3 0
% 3 2134 1603 2134 1673 2 0
% $2$
\put(21.3400,-12.7300){\makebox(0,0)[lb]{$2$}}%
% STR 2 0 3 0
% 3 2288 1736 2288 1806 2 0
% $0$
\put(22.8800,-14.0600){\makebox(0,0)[lb]{$0$}}%
% STR 2 0 3 0
% 3 1413 1169 1413 1239 5 0
% $2$
\put(14.1300,-8.3900){\makebox(0,0){$2$}}%
% STR 2 0 3 0
% 3 1693 1442 1693 1512 5 0
% $2$
\put(16.9300,-11.1200){\makebox(0,0){$2$}}%
% STR 2 0 3 0
% 3 1980 1722 1980 1792 5 0
% $2$
\put(19.8000,-13.9200){\makebox(0,0){$2$}}%
% STR 2 0 3 0
% 3 1308 1344 1308 1414 5 0
% $1$
\put(13.0800,-10.1400){\makebox(0,0){$1$}}%
% STR 2 0 3 0
% 3 1595 1344 1595 1414 5 0
% $0$
\put(15.9500,-10.1400){\makebox(0,0){$0$}}%
% STR 2 0 3 0
% 3 1413 1435 1413 1505 5 0
% $0$
\put(14.1300,-11.0500){\makebox(0,0){$0$}}%
% STR 2 0 3 0
% 3 1308 1624 1308 1694 5 0
% $0$
\put(13.0800,-12.9400){\makebox(0,0){$0$}}%
% STR 2 0 3 0
% 3 1308 1904 1308 1974 5 0
% $1$
\put(13.0800,-15.7400){\makebox(0,0){$1$}}%
% STR 2 0 3 0
% 3 1308 2184 1308 2254 5 0
% $0$
\put(13.0800,-18.5400){\makebox(0,0){$0$}}%
% STR 2 0 3 0
% 3 1308 2464 1308 2534 5 0
% $1$
\put(13.0800,-21.3400){\makebox(0,0){$1$}}%
% STR 2 0 3 0
% 3 1413 1715 1413 1785 5 0
% $2$
\put(14.1300,-13.8500){\makebox(0,0){$2$}}%
% STR 2 0 3 0
% 3 1595 1624 1595 1694 5 0
% $2$
\put(15.9500,-12.9400){\makebox(0,0){$2$}}%
% STR 2 0 3 0
% 3 1882 1624 1882 1694 5 0
% $0$
\put(18.8200,-12.9400){\makebox(0,0){$0$}}%
% STR 2 0 3 0
% 3 1693 1715 1693 1785 5 0
% $0$
\put(16.9300,-13.8500){\makebox(0,0){$0$}}%
% STR 2 0 3 0
% 3 1595 1904 1595 1974 5 0
% $0$
\put(15.9500,-15.7400){\makebox(0,0){$0$}}%
% STR 2 0 3 0
% 3 1413 2009 1413 2079 5 0
% $0$
\put(14.1300,-16.7900){\makebox(0,0){$0$}}%
% STR 2 0 3 0
% 3 1875 1904 1875 1974 5 0
% $2$
\put(18.7500,-15.7400){\makebox(0,0){$2$}}%
% STR 2 0 3 0
% 3 1588 2184 1588 2254 5 0
% $2$
\put(15.8800,-18.5400){\makebox(0,0){$2$}}%
% STR 2 0 3 0
% 3 1413 2282 1413 2352 5 0
% $2$
\put(14.1300,-19.5200){\makebox(0,0){$2$}}%
% STR 2 0 3 0
% 3 2148 1904 2148 1974 5 0
% $0$
\put(21.4800,-15.7400){\makebox(0,0){$0$}}%
% STR 2 0 3 0
% 3 1973 2002 1973 2072 5 0
% $0$
\put(19.7300,-16.7200){\makebox(0,0){$0$}}%
% STR 2 0 3 0
% 3 1875 2184 1875 2254 5 0
% $0$
\put(18.7500,-18.5400){\makebox(0,0){$0$}}%
% STR 2 0 3 0
% 3 1693 2275 1693 2345 5 0
% $0$
\put(16.9300,-19.4500){\makebox(0,0){$0$}}%
% STR 2 0 3 0
% 3 1693 2009 1693 2079 5 0
% $2$
\put(16.9300,-16.7900){\makebox(0,0){$2$}}%
% STR 2 0 3 0
% 3 1588 2464 1588 2534 5 0
% $0$
\put(15.8800,-21.3400){\makebox(0,0){$0$}}%
% STR 2 0 3 0
% 3 1413 2562 1413 2632 5 0
% $0$
\put(14.1300,-22.3200){\makebox(0,0){$0$}}%
% STR 2 0 3 0
% 3 2288 2072 2288 2142 2 0
% $2$
\put(22.8800,-17.4200){\makebox(0,0)[lb]{$2$}}%
% STR 2 0 3 0
% 3 2141 2240 2141 2310 2 0
% $2$
\put(21.4100,-19.1000){\makebox(0,0)[lb]{$2$}}%
% STR 2 0 3 0
% 3 2015 2373 2015 2443 2 0
% $2$
\put(20.1500,-20.4300){\makebox(0,0)[lb]{$2$}}%
% STR 2 0 3 0
% 3 1868 2548 1868 2618 2 0
% $2$
\put(18.6800,-22.1800){\makebox(0,0)[lb]{$2$}}%
% STR 2 0 3 0
% 3 1721 2646 1721 2716 2 0
% $2$
\put(17.2100,-23.1600){\makebox(0,0)[lb]{$2$}}%
% STR 2 0 3 0
% 3 1588 2807 1588 2877 2 0
% $2$
\put(15.8800,-24.7700){\makebox(0,0)[lb]{$2$}}%
% STR 2 0 3 0
% 3 1133 1162 1133 1232 5 0
% $1$
\put(11.3300,-8.3200){\makebox(0,0){$1$}}%
% STR 2 0 3 0
% 3 1028 1344 1028 1414 5 0
% $1$
\put(10.2800,-10.1400){\makebox(0,0){$1$}}%
% STR 2 0 3 0
% 3 755 1624 755 1694 5 0
% $1$
\put(7.5500,-12.9400){\makebox(0,0){$1$}}%
% STR 2 0 3 0
% 3 853 1442 853 1512 5 0
% $1$
\put(8.5300,-11.1200){\makebox(0,0){$1$}}%
% STR 2 0 3 0
% 3 580 1722 580 1792 5 0
% $1$
\put(5.8000,-13.9200){\makebox(0,0){$1$}}%
% STR 2 0 3 0
% 3 475 1904 475 1974 5 0
% $1$
\put(4.7500,-15.7400){\makebox(0,0){$1$}}%
% STR 2 0 3 0
% 3 1133 1435 1133 1505 5 0
% $1$
\put(11.3300,-11.0500){\makebox(0,0){$1$}}%
% STR 2 0 3 0
% 3 1028 1624 1028 1694 5 0
% $1$
\put(10.2800,-12.9400){\makebox(0,0){$1$}}%
% STR 2 0 3 0
% 3 853 1715 853 1785 5 0
% $1$
\put(8.5300,-13.8500){\makebox(0,0){$1$}}%
% STR 2 0 3 0
% 3 755 1904 755 1974 5 0
% $1$
\put(7.5500,-15.7400){\makebox(0,0){$1$}}%
% STR 2 0 3 0
% 3 580 2009 580 2079 5 0
% $1$
\put(5.8000,-16.7900){\makebox(0,0){$1$}}%
% STR 2 0 3 0
% 3 1133 1715 1133 1785 5 0
% $1$
\put(11.3300,-13.8500){\makebox(0,0){$1$}}%
% STR 2 0 3 0
% 3 1028 1904 1028 1974 5 0
% $1$
\put(10.2800,-15.7400){\makebox(0,0){$1$}}%
% STR 2 0 3 0
% 3 853 2002 853 2072 5 0
% $1$
\put(8.5300,-16.7200){\makebox(0,0){$1$}}%
% STR 2 0 3 0
% 3 755 2184 755 2254 5 0
% $1$
\put(7.5500,-18.5400){\makebox(0,0){$1$}}%
% STR 2 0 3 0
% 3 1133 2002 1133 2072 5 0
% $1$
\put(11.3300,-16.7200){\makebox(0,0){$1$}}%
% STR 2 0 3 0
% 3 1028 2184 1028 2254 5 0
% $1$
\put(10.2800,-18.5400){\makebox(0,0){$1$}}%
% STR 2 0 3 0
% 3 853 2282 853 2352 5 0
% $1$
\put(8.5300,-19.5200){\makebox(0,0){$1$}}%
% STR 2 0 3 0
% 3 1133 2275 1133 2345 5 0
% $1$
\put(11.3300,-19.4500){\makebox(0,0){$1$}}%
% STR 2 0 3 0
% 3 1028 2464 1028 2534 5 0
% $1$
\put(10.2800,-21.3400){\makebox(0,0){$1$}}%
% STR 2 0 3 0
% 3 1133 2562 1133 2632 5 0
% $1$
\put(11.3300,-22.3200){\makebox(0,0){$1$}}%
% STR 2 0 3 0
% 3 951 1029 951 1099 2 0
% $1$
\put(9.5100,-6.9900){\makebox(0,0)[lb]{$1$}}%
% STR 2 0 3 0
% 3 811 1169 811 1239 2 0
% $1$
\put(8.1100,-8.3900){\makebox(0,0)[lb]{$1$}}%
% STR 2 0 3 0
% 3 664 1323 664 1393 2 0
% $1$
\put(6.6400,-9.9300){\makebox(0,0)[lb]{$1$}}%
% STR 2 0 3 0
% 3 538 1456 538 1526 2 0
% $1$
\put(5.3800,-11.2600){\makebox(0,0)[lb]{$1$}}%
% STR 2 0 3 0
% 3 391 1610 391 1680 2 0
% $1$
\put(3.9100,-12.8000){\makebox(0,0)[lb]{$1$}}%
% STR 2 0 3 0
% 3 244 1736 244 1806 2 0
% $1$
\put(2.4400,-14.0600){\makebox(0,0)[lb]{$1$}}%
% STR 2 0 3 0
% 3 230 2072 230 2142 2 0
% $1$
\put(2.3000,-17.4200){\makebox(0,0)[lb]{$1$}}%
% STR 2 0 3 0
% 3 384 2268 384 2338 2 0
% $1$
\put(3.8400,-19.3800){\makebox(0,0)[lb]{$1$}}%
% STR 2 0 3 0
% 3 517 2380 517 2450 2 0
% $1$
\put(5.1700,-20.5000){\makebox(0,0)[lb]{$1$}}%
% STR 2 0 3 0
% 3 657 2548 657 2618 2 0
% $1$
\put(6.5700,-22.1800){\makebox(0,0)[lb]{$1$}}%
% STR 2 0 3 0
% 3 783 2681 783 2751 2 0
% $1$
\put(7.8300,-23.5100){\makebox(0,0)[lb]{$1$}}%
% STR 2 0 3 0
% 3 937 2821 937 2891 2 0
% $1$
\put(9.3700,-24.9100){\makebox(0,0)[lb]{$1$}}%
% STR 2 0 3 0
% 3 1273 980 1273 1050 5 0
% $0$
\put(12.7300,-6.5000){\makebox(0,0){$0$}}%
% STR 2 0 3 0
% 3 1273 2821 1273 2891 5 0
% $0$
\put(12.7300,-24.9100){\makebox(0,0){$0$}}%
\end{picture}%

\vspace{12pt}

%WinTpicVersion2.16
\unitlength 0.1in
\begin{picture}(27.70,20.17)(-4.82,-25.46)
% LINE 0 0 3 0
% 2 1273 1141 1273 2814
% 
\special{pn 20}%
\special{pa 1273 741}%
\special{pa 1273 2414}%
\special{fp}%
% VECTOR 1 0 3 0
% 2 1273 2793 1273 2821
% 
\special{pn 13}%
\special{pa 1273 2393}%
\special{pa 1273 2421}%
\special{fp}%
\special{sh 1}%
\special{pa 1273 2421}%
\special{pa 1293 2354}%
\special{pa 1273 2368}%
\special{pa 1253 2354}%
\special{pa 1273 2421}%
\special{fp}%
% VECTOR 2 0 3 0
% 6 1553 1134 1553 2807 1840 1414 1840 2534 2113 1694 2113 2254
% 
\special{pn 8}%
\special{pa 1553 734}%
\special{pa 1553 2407}%
\special{fp}%
\special{sh 1}%
\special{pa 1553 2407}%
\special{pa 1573 2340}%
\special{pa 1553 2354}%
\special{pa 1533 2340}%
\special{pa 1553 2407}%
\special{fp}%
\special{pa 1840 1014}%
\special{pa 1840 2134}%
\special{fp}%
\special{sh 1}%
\special{pa 1840 2134}%
\special{pa 1860 2067}%
\special{pa 1840 2081}%
\special{pa 1820 2067}%
\special{pa 1840 2134}%
\special{fp}%
\special{pa 2113 1294}%
\special{pa 2113 1854}%
\special{fp}%
\special{sh 1}%
\special{pa 2113 1854}%
\special{pa 2133 1787}%
\special{pa 2113 1801}%
\special{pa 2093 1787}%
\special{pa 2113 1854}%
\special{fp}%
% VECTOR 2 0 3 0
% 6 993 1134 993 2814 713 1414 713 2527 433 1694 433 2247
% 
\special{pn 8}%
\special{pa 993 734}%
\special{pa 993 2414}%
\special{fp}%
\special{sh 1}%
\special{pa 993 2414}%
\special{pa 1013 2347}%
\special{pa 993 2361}%
\special{pa 973 2347}%
\special{pa 993 2414}%
\special{fp}%
\special{pa 713 1014}%
\special{pa 713 2127}%
\special{fp}%
\special{sh 1}%
\special{pa 713 2127}%
\special{pa 733 2060}%
\special{pa 713 2074}%
\special{pa 693 2060}%
\special{pa 713 2127}%
\special{fp}%
\special{pa 433 1294}%
\special{pa 433 1847}%
\special{fp}%
\special{sh 1}%
\special{pa 433 1847}%
\special{pa 453 1780}%
\special{pa 433 1794}%
\special{pa 413 1780}%
\special{pa 433 1847}%
\special{fp}%
% VECTOR 2 0 3 0
% 6 2260 1834 293 1834 1973 1547 580 1547 1693 1274 853 1274
% 
\special{pn 8}%
\special{pa 2260 1434}%
\special{pa 293 1434}%
\special{fp}%
\special{sh 1}%
\special{pa 293 1434}%
\special{pa 360 1454}%
\special{pa 346 1434}%
\special{pa 360 1414}%
\special{pa 293 1434}%
\special{fp}%
\special{pa 1973 1147}%
\special{pa 580 1147}%
\special{fp}%
\special{sh 1}%
\special{pa 580 1147}%
\special{pa 647 1167}%
\special{pa 633 1147}%
\special{pa 647 1127}%
\special{pa 580 1147}%
\special{fp}%
\special{pa 1693 874}%
\special{pa 853 874}%
\special{fp}%
\special{sh 1}%
\special{pa 853 874}%
\special{pa 920 894}%
\special{pa 906 874}%
\special{pa 920 854}%
\special{pa 853 874}%
\special{fp}%
% VECTOR 2 0 3 0
% 6 2253 2114 293 2114 1973 2394 580 2394 1693 2674 853 2674
% 
\special{pn 8}%
\special{pa 2253 1714}%
\special{pa 293 1714}%
\special{fp}%
\special{sh 1}%
\special{pa 293 1714}%
\special{pa 360 1734}%
\special{pa 346 1714}%
\special{pa 360 1694}%
\special{pa 293 1714}%
\special{fp}%
\special{pa 1973 1994}%
\special{pa 580 1994}%
\special{fp}%
\special{sh 1}%
\special{pa 580 1994}%
\special{pa 647 2014}%
\special{pa 633 1994}%
\special{pa 647 1974}%
\special{pa 580 1994}%
\special{fp}%
\special{pa 1693 2274}%
\special{pa 853 2274}%
\special{fp}%
\special{sh 1}%
\special{pa 853 2274}%
\special{pa 920 2294}%
\special{pa 906 2274}%
\special{pa 920 2254}%
\special{pa 853 2274}%
\special{fp}%
% STR 2 0 3 0
% 3 1273 2961 1273 3031 5 0
% $(2\Lambda_0,\Lambda_0+\Lambda_1)$-Case
\put(12.7300,-26.3100){\makebox(0,0){$(2\Lambda_0,\Lambda_0+\Lambda_1)$-Case}}%
% STR 2 0 3 0
% 3 1574 1043 1574 1113 2 0
% $1$
\put(15.7400,-7.1300){\makebox(0,0)[lb]{$1$}}%
% STR 2 0 3 0
% 3 1721 1176 1721 1246 2 0
% $1$
\put(17.2100,-8.4600){\makebox(0,0)[lb]{$1$}}%
% STR 2 0 3 0
% 3 1868 1323 1868 1393 2 0
% $1$
\put(18.6800,-9.9300){\makebox(0,0)[lb]{$1$}}%
% STR 2 0 3 0
% 3 1994 1456 1994 1526 2 0
% $1$
\put(19.9400,-11.2600){\makebox(0,0)[lb]{$1$}}%
% STR 2 0 3 0
% 3 2134 1603 2134 1673 2 0
% $1$
\put(21.3400,-12.7300){\makebox(0,0)[lb]{$1$}}%
% STR 2 0 3 0
% 3 2288 1736 2288 1806 2 0
% $1$
\put(22.8800,-14.0600){\makebox(0,0)[lb]{$1$}}%
% STR 2 0 3 0
% 3 1413 1169 1413 1239 5 0
% $1$
\put(14.1300,-8.3900){\makebox(0,0){$1$}}%
% STR 2 0 3 0
% 3 1693 1442 1693 1512 5 0
% $1$
\put(16.9300,-11.1200){\makebox(0,0){$1$}}%
% STR 2 0 3 0
% 3 1980 1722 1980 1792 5 0
% $1$
\put(19.8000,-13.9200){\makebox(0,0){$1$}}%
% STR 2 0 3 0
% 3 1308 1344 1308 1414 5 0
% $1$
\put(13.0800,-10.1400){\makebox(0,0){$1$}}%
% STR 2 0 3 0
% 3 1595 1344 1595 1414 5 0
% $1$
\put(15.9500,-10.1400){\makebox(0,0){$1$}}%
% STR 2 0 3 0
% 3 1413 1435 1413 1505 5 0
% $1$
\put(14.1300,-11.0500){\makebox(0,0){$1$}}%
% STR 2 0 3 0
% 3 1308 1624 1308 1694 5 0
% $0$
\put(13.0800,-12.9400){\makebox(0,0){$0$}}%
% STR 2 0 3 0
% 3 1308 1904 1308 1974 5 0
% $1$
\put(13.0800,-15.7400){\makebox(0,0){$1$}}%
% STR 2 0 3 0
% 3 1308 2184 1308 2254 5 0
% $0$
\put(13.0800,-18.5400){\makebox(0,0){$0$}}%
% STR 2 0 3 0
% 3 1308 2464 1308 2534 5 0
% $1$
\put(13.0800,-21.3400){\makebox(0,0){$1$}}%
% STR 2 0 3 0
% 3 1413 1715 1413 1785 5 0
% $1$
\put(14.1300,-13.8500){\makebox(0,0){$1$}}%
% STR 2 0 3 0
% 3 1595 1624 1595 1694 5 0
% $1$
\put(15.9500,-12.9400){\makebox(0,0){$1$}}%
% STR 2 0 3 0
% 3 1882 1624 1882 1694 5 0
% $1$
\put(18.8200,-12.9400){\makebox(0,0){$1$}}%
% STR 2 0 3 0
% 3 1693 1715 1693 1785 5 0
% $1$
\put(16.9300,-13.8500){\makebox(0,0){$1$}}%
% STR 2 0 3 0
% 3 1595 1904 1595 1974 5 0
% $1$
\put(15.9500,-15.7400){\makebox(0,0){$1$}}%
% STR 2 0 3 0
% 3 1413 2009 1413 2079 5 0
% $1$
\put(14.1300,-16.7900){\makebox(0,0){$1$}}%
% STR 2 0 3 0
% 3 1875 1904 1875 1974 5 0
% $1$
\put(18.7500,-15.7400){\makebox(0,0){$1$}}%
% STR 2 0 3 0
% 3 1588 2184 1588 2254 5 0
% $1$
\put(15.8800,-18.5400){\makebox(0,0){$1$}}%
% STR 2 0 3 0
% 3 1413 2282 1413 2352 5 0
% $1$
\put(14.1300,-19.5200){\makebox(0,0){$1$}}%
% STR 2 0 3 0
% 3 2148 1904 2148 1974 5 0
% $1$
\put(21.4800,-15.7400){\makebox(0,0){$1$}}%
% STR 2 0 3 0
% 3 1973 2002 1973 2072 5 0
% $1$
\put(19.7300,-16.7200){\makebox(0,0){$1$}}%
% STR 2 0 3 0
% 3 1875 2184 1875 2254 5 0
% $1$
\put(18.7500,-18.5400){\makebox(0,0){$1$}}%
% STR 2 0 3 0
% 3 1693 2275 1693 2345 5 0
% $1$
\put(16.9300,-19.4500){\makebox(0,0){$1$}}%
% STR 2 0 3 0
% 3 1693 2009 1693 2079 5 0
% $1$
\put(16.9300,-16.7900){\makebox(0,0){$1$}}%
% STR 2 0 3 0
% 3 1588 2464 1588 2534 5 0
% $1$
\put(15.8800,-21.3400){\makebox(0,0){$1$}}%
% STR 2 0 3 0
% 3 1413 2562 1413 2632 5 0
% $1$
\put(14.1300,-22.3200){\makebox(0,0){$1$}}%
% STR 2 0 3 0
% 3 2288 2072 2288 2142 2 0
% $1$
\put(22.8800,-17.4200){\makebox(0,0)[lb]{$1$}}%
% STR 2 0 3 0
% 3 2141 2240 2141 2310 2 0
% $1$
\put(21.4100,-19.1000){\makebox(0,0)[lb]{$1$}}%
% STR 2 0 3 0
% 3 2015 2373 2015 2443 2 0
% $1$
\put(20.1500,-20.4300){\makebox(0,0)[lb]{$1$}}%
% STR 2 0 3 0
% 3 1868 2548 1868 2618 2 0
% $1$
\put(18.6800,-22.1800){\makebox(0,0)[lb]{$1$}}%
% STR 2 0 3 0
% 3 1721 2646 1721 2716 2 0
% $1$
\put(17.2100,-23.1600){\makebox(0,0)[lb]{$1$}}%
% STR 2 0 3 0
% 3 1588 2807 1588 2877 2 0
% $1$
\put(15.8800,-24.7700){\makebox(0,0)[lb]{$1$}}%
% STR 2 0 3 0
% 3 1133 1162 1133 1232 5 0
% $0$
\put(11.3300,-8.3200){\makebox(0,0){$0$}}%
% STR 2 0 3 0
% 3 1028 1344 1028 1414 5 0
% $0$
\put(10.2800,-10.1400){\makebox(0,0){$0$}}%
% STR 2 0 3 0
% 3 755 1624 755 1694 5 0
% $0$
\put(7.5500,-12.9400){\makebox(0,0){$0$}}%
% STR 2 0 3 0
% 3 853 1442 853 1512 5 0
% $0$
\put(8.5300,-11.1200){\makebox(0,0){$0$}}%
% STR 2 0 3 0
% 3 580 1722 580 1792 5 0
% $0$
\put(5.8000,-13.9200){\makebox(0,0){$0$}}%
% STR 2 0 3 0
% 3 475 1904 475 1974 5 0
% $0$
\put(4.7500,-15.7400){\makebox(0,0){$0$}}%
% STR 2 0 3 0
% 3 1133 1435 1133 1505 5 0
% $2$
\put(11.3300,-11.0500){\makebox(0,0){$2$}}%
% STR 2 0 3 0
% 3 1028 1624 1028 1694 5 0
% $2$
\put(10.2800,-12.9400){\makebox(0,0){$2$}}%
% STR 2 0 3 0
% 3 853 1715 853 1785 5 0
% $2$
\put(8.5300,-13.8500){\makebox(0,0){$2$}}%
% STR 2 0 3 0
% 3 755 1904 755 1974 5 0
% $2$
\put(7.5500,-15.7400){\makebox(0,0){$2$}}%
% STR 2 0 3 0
% 3 580 2009 580 2079 5 0
% $2$
\put(5.8000,-16.7900){\makebox(0,0){$2$}}%
% STR 2 0 3 0
% 3 1133 1715 1133 1785 5 0
% $0$
\put(11.3300,-13.8500){\makebox(0,0){$0$}}%
% STR 2 0 3 0
% 3 1028 1904 1028 1974 5 0
% $0$
\put(10.2800,-15.7400){\makebox(0,0){$0$}}%
% STR 2 0 3 0
% 3 853 2002 853 2072 5 0
% $0$
\put(8.5300,-16.7200){\makebox(0,0){$0$}}%
% STR 2 0 3 0
% 3 755 2184 755 2254 5 0
% $0$
\put(7.5500,-18.5400){\makebox(0,0){$0$}}%
% STR 2 0 3 0
% 3 1133 2002 1133 2072 5 0
% $2$
\put(11.3300,-16.7200){\makebox(0,0){$2$}}%
% STR 2 0 3 0
% 3 1028 2184 1028 2254 5 0
% $2$
\put(10.2800,-18.5400){\makebox(0,0){$2$}}%
% STR 2 0 3 0
% 3 853 2282 853 2352 5 0
% $2$
\put(8.5300,-19.5200){\makebox(0,0){$2$}}%
% STR 2 0 3 0
% 3 1133 2275 1133 2345 5 0
% $0$
\put(11.3300,-19.4500){\makebox(0,0){$0$}}%
% STR 2 0 3 0
% 3 1028 2464 1028 2534 5 0
% $0$
\put(10.2800,-21.3400){\makebox(0,0){$0$}}%
% STR 2 0 3 0
% 3 1133 2562 1133 2632 5 0
% $2$
\put(11.3300,-22.3200){\makebox(0,0){$2$}}%
% STR 2 0 3 0
% 3 951 1029 951 1099 2 0
% $2$
\put(9.5100,-6.9900){\makebox(0,0)[lb]{$2$}}%
% STR 2 0 3 0
% 3 811 1169 811 1239 2 0
% $2$
\put(8.1100,-8.3900){\makebox(0,0)[lb]{$2$}}%
% STR 2 0 3 0
% 3 664 1323 664 1393 2 0
% $2$
\put(6.6400,-9.9300){\makebox(0,0)[lb]{$2$}}%
% STR 2 0 3 0
% 3 538 1456 538 1526 2 0
% $2$
\put(5.3800,-11.2600){\makebox(0,0)[lb]{$2$}}%
% STR 2 0 3 0
% 3 391 1610 391 1680 2 0
% $2$
\put(3.9100,-12.8000){\makebox(0,0)[lb]{$2$}}%
% STR 2 0 3 0
% 3 244 1736 244 1806 2 0
% $2$
\put(2.4400,-14.0600){\makebox(0,0)[lb]{$2$}}%
% STR 2 0 3 0
% 3 230 2072 230 2142 2 0
% $0$
\put(2.3000,-17.4200){\makebox(0,0)[lb]{$0$}}%
% STR 2 0 3 0
% 3 384 2268 384 2338 2 0
% $2$
\put(3.8400,-19.3800){\makebox(0,0)[lb]{$2$}}%
% STR 2 0 3 0
% 3 517 2380 517 2450 2 0
% $0$
\put(5.1700,-20.5000){\makebox(0,0)[lb]{$0$}}%
% STR 2 0 3 0
% 3 657 2548 657 2618 2 0
% $2$
\put(6.5700,-22.1800){\makebox(0,0)[lb]{$2$}}%
% STR 2 0 3 0
% 3 783 2681 783 2751 2 0
% $0$
\put(7.8300,-23.5100){\makebox(0,0)[lb]{$0$}}%
% STR 2 0 3 0
% 3 937 2821 937 2891 2 0
% $2$
\put(9.3700,-24.9100){\makebox(0,0)[lb]{$2$}}%
% STR 2 0 3 0
% 3 1273 1008 1273 1078 5 0
% $0$
\put(12.7300,-6.7800){\makebox(0,0){$0$}}%
% STR 2 0 3 0
% 3 1273 2814 1273 2884 5 0
% $0$
\put(12.7300,-24.8400){\makebox(0,0){$0$}}%
\end{picture}%
%WinTpicVersion2.16
\unitlength 0.1in
\begin{picture}(27.70,20.17)(-4.82,-25.46)
% LINE 0 0 3 0
% 2 1273 1141 1273 2814
% 
\special{pn 20}%
\special{pa 1273 741}%
\special{pa 1273 2414}%
\special{fp}%
% VECTOR 1 0 3 0
% 2 1273 2793 1273 2821
% 
\special{pn 13}%
\special{pa 1273 2393}%
\special{pa 1273 2421}%
\special{fp}%
\special{sh 1}%
\special{pa 1273 2421}%
\special{pa 1293 2354}%
\special{pa 1273 2368}%
\special{pa 1253 2354}%
\special{pa 1273 2421}%
\special{fp}%
% VECTOR 2 0 3 0
% 6 1553 1134 1553 2807 1840 1414 1840 2534 2113 1694 2113 2254
% 
\special{pn 8}%
\special{pa 1553 734}%
\special{pa 1553 2407}%
\special{fp}%
\special{sh 1}%
\special{pa 1553 2407}%
\special{pa 1573 2340}%
\special{pa 1553 2354}%
\special{pa 1533 2340}%
\special{pa 1553 2407}%
\special{fp}%
\special{pa 1840 1014}%
\special{pa 1840 2134}%
\special{fp}%
\special{sh 1}%
\special{pa 1840 2134}%
\special{pa 1860 2067}%
\special{pa 1840 2081}%
\special{pa 1820 2067}%
\special{pa 1840 2134}%
\special{fp}%
\special{pa 2113 1294}%
\special{pa 2113 1854}%
\special{fp}%
\special{sh 1}%
\special{pa 2113 1854}%
\special{pa 2133 1787}%
\special{pa 2113 1801}%
\special{pa 2093 1787}%
\special{pa 2113 1854}%
\special{fp}%
% VECTOR 2 0 3 0
% 6 993 1134 993 2814 713 1414 713 2527 433 1694 433 2247
% 
\special{pn 8}%
\special{pa 993 734}%
\special{pa 993 2414}%
\special{fp}%
\special{sh 1}%
\special{pa 993 2414}%
\special{pa 1013 2347}%
\special{pa 993 2361}%
\special{pa 973 2347}%
\special{pa 993 2414}%
\special{fp}%
\special{pa 713 1014}%
\special{pa 713 2127}%
\special{fp}%
\special{sh 1}%
\special{pa 713 2127}%
\special{pa 733 2060}%
\special{pa 713 2074}%
\special{pa 693 2060}%
\special{pa 713 2127}%
\special{fp}%
\special{pa 433 1294}%
\special{pa 433 1847}%
\special{fp}%
\special{sh 1}%
\special{pa 433 1847}%
\special{pa 453 1780}%
\special{pa 433 1794}%
\special{pa 413 1780}%
\special{pa 433 1847}%
\special{fp}%
% VECTOR 2 0 3 0
% 6 2260 1834 293 1834 1973 1547 580 1547 1693 1274 853 1274
% 
\special{pn 8}%
\special{pa 2260 1434}%
\special{pa 293 1434}%
\special{fp}%
\special{sh 1}%
\special{pa 293 1434}%
\special{pa 360 1454}%
\special{pa 346 1434}%
\special{pa 360 1414}%
\special{pa 293 1434}%
\special{fp}%
\special{pa 1973 1147}%
\special{pa 580 1147}%
\special{fp}%
\special{sh 1}%
\special{pa 580 1147}%
\special{pa 647 1167}%
\special{pa 633 1147}%
\special{pa 647 1127}%
\special{pa 580 1147}%
\special{fp}%
\special{pa 1693 874}%
\special{pa 853 874}%
\special{fp}%
\special{sh 1}%
\special{pa 853 874}%
\special{pa 920 894}%
\special{pa 906 874}%
\special{pa 920 854}%
\special{pa 853 874}%
\special{fp}%
% VECTOR 2 0 3 0
% 6 2253 2114 293 2114 1973 2394 580 2394 1693 2674 853 2674
% 
\special{pn 8}%
\special{pa 2253 1714}%
\special{pa 293 1714}%
\special{fp}%
\special{sh 1}%
\special{pa 293 1714}%
\special{pa 360 1734}%
\special{pa 346 1714}%
\special{pa 360 1694}%
\special{pa 293 1714}%
\special{fp}%
\special{pa 1973 1994}%
\special{pa 580 1994}%
\special{fp}%
\special{sh 1}%
\special{pa 580 1994}%
\special{pa 647 2014}%
\special{pa 633 1994}%
\special{pa 647 1974}%
\special{pa 580 1994}%
\special{fp}%
\special{pa 1693 2274}%
\special{pa 853 2274}%
\special{fp}%
\special{sh 1}%
\special{pa 853 2274}%
\special{pa 920 2294}%
\special{pa 906 2274}%
\special{pa 920 2254}%
\special{pa 853 2274}%
\special{fp}%
% STR 2 0 3 0
% 3 1273 2961 1273 3031 5 0
% $(2\Lambda_1,\Lambda_0+\Lambda_1)$-Case
\put(12.7300,-26.3100){\makebox(0,0){$(2\Lambda_1,\Lambda_0+\Lambda_1)$-Case}}%
% STR 2 0 3 0
% 3 1574 1043 1574 1113 2 0
% $1$
\put(15.7400,-7.1300){\makebox(0,0)[lb]{$1$}}%
% STR 2 0 3 0
% 3 1721 1176 1721 1246 2 0
% $1$
\put(17.2100,-8.4600){\makebox(0,0)[lb]{$1$}}%
% STR 2 0 3 0
% 3 1868 1323 1868 1393 2 0
% $1$
\put(18.6800,-9.9300){\makebox(0,0)[lb]{$1$}}%
% STR 2 0 3 0
% 3 1994 1456 1994 1526 2 0
% $1$
\put(19.9400,-11.2600){\makebox(0,0)[lb]{$1$}}%
% STR 2 0 3 0
% 3 2134 1603 2134 1673 2 0
% $1$
\put(21.3400,-12.7300){\makebox(0,0)[lb]{$1$}}%
% STR 2 0 3 0
% 3 2288 1736 2288 1806 2 0
% $1$
\put(22.8800,-14.0600){\makebox(0,0)[lb]{$1$}}%
% STR 2 0 3 0
% 3 1413 1169 1413 1239 5 0
% $1$
\put(14.1300,-8.3900){\makebox(0,0){$1$}}%
% STR 2 0 3 0
% 3 1693 1442 1693 1512 5 0
% $1$
\put(16.9300,-11.1200){\makebox(0,0){$1$}}%
% STR 2 0 3 0
% 3 1980 1722 1980 1792 5 0
% $1$
\put(19.8000,-13.9200){\makebox(0,0){$1$}}%
% STR 2 0 3 0
% 3 1308 1344 1308 1414 5 0
% $0$
\put(13.0800,-10.1400){\makebox(0,0){$0$}}%
% STR 2 0 3 0
% 3 1595 1344 1595 1414 5 0
% $1$
\put(15.9500,-10.1400){\makebox(0,0){$1$}}%
% STR 2 0 3 0
% 3 1413 1435 1413 1505 5 0
% $1$
\put(14.1300,-11.0500){\makebox(0,0){$1$}}%
% STR 2 0 3 0
% 3 1308 1624 1308 1694 5 0
% $1$
\put(13.0800,-12.9400){\makebox(0,0){$1$}}%
% STR 2 0 3 0
% 3 1308 1904 1308 1974 5 0
% $0$
\put(13.0800,-15.7400){\makebox(0,0){$0$}}%
% STR 2 0 3 0
% 3 1308 2184 1308 2254 5 0
% $1$
\put(13.0800,-18.5400){\makebox(0,0){$1$}}%
% STR 2 0 3 0
% 3 1308 2464 1308 2534 5 0
% $0$
\put(13.0800,-21.3400){\makebox(0,0){$0$}}%
% STR 2 0 3 0
% 3 1413 1715 1413 1785 5 0
% $1$
\put(14.1300,-13.8500){\makebox(0,0){$1$}}%
% STR 2 0 3 0
% 3 1595 1624 1595 1694 5 0
% $1$
\put(15.9500,-12.9400){\makebox(0,0){$1$}}%
% STR 2 0 3 0
% 3 1882 1624 1882 1694 5 0
% $1$
\put(18.8200,-12.9400){\makebox(0,0){$1$}}%
% STR 2 0 3 0
% 3 1693 1715 1693 1785 5 0
% $1$
\put(16.9300,-13.8500){\makebox(0,0){$1$}}%
% STR 2 0 3 0
% 3 1595 1904 1595 1974 5 0
% $1$
\put(15.9500,-15.7400){\makebox(0,0){$1$}}%
% STR 2 0 3 0
% 3 1413 2009 1413 2079 5 0
% $1$
\put(14.1300,-16.7900){\makebox(0,0){$1$}}%
% STR 2 0 3 0
% 3 1875 1904 1875 1974 5 0
% $1$
\put(18.7500,-15.7400){\makebox(0,0){$1$}}%
% STR 2 0 3 0
% 3 1588 2184 1588 2254 5 0
% $1$
\put(15.8800,-18.5400){\makebox(0,0){$1$}}%
% STR 2 0 3 0
% 3 1413 2282 1413 2352 5 0
% $1$
\put(14.1300,-19.5200){\makebox(0,0){$1$}}%
% STR 2 0 3 0
% 3 2148 1904 2148 1974 5 0
% $1$
\put(21.4800,-15.7400){\makebox(0,0){$1$}}%
% STR 2 0 3 0
% 3 1973 2002 1973 2072 5 0
% $1$
\put(19.7300,-16.7200){\makebox(0,0){$1$}}%
% STR 2 0 3 0
% 3 1875 2184 1875 2254 5 0
% $1$
\put(18.7500,-18.5400){\makebox(0,0){$1$}}%
% STR 2 0 3 0
% 3 1693 2275 1693 2345 5 0
% $1$
\put(16.9300,-19.4500){\makebox(0,0){$1$}}%
% STR 2 0 3 0
% 3 1693 2009 1693 2079 5 0
% $1$
\put(16.9300,-16.7900){\makebox(0,0){$1$}}%
% STR 2 0 3 0
% 3 1588 2464 1588 2534 5 0
% $1$
\put(15.8800,-21.3400){\makebox(0,0){$1$}}%
% STR 2 0 3 0
% 3 1413 2562 1413 2632 5 0
% $1$
\put(14.1300,-22.3200){\makebox(0,0){$1$}}%
% STR 2 0 3 0
% 3 2288 2072 2288 2142 2 0
% $1$
\put(22.8800,-17.4200){\makebox(0,0)[lb]{$1$}}%
% STR 2 0 3 0
% 3 2141 2240 2141 2310 2 0
% $1$
\put(21.4100,-19.1000){\makebox(0,0)[lb]{$1$}}%
% STR 2 0 3 0
% 3 2015 2373 2015 2443 2 0
% $1$
\put(20.1500,-20.4300){\makebox(0,0)[lb]{$1$}}%
% STR 2 0 3 0
% 3 1868 2548 1868 2618 2 0
% $1$
\put(18.6800,-22.1800){\makebox(0,0)[lb]{$1$}}%
% STR 2 0 3 0
% 3 1721 2646 1721 2716 2 0
% $1$
\put(17.2100,-23.1600){\makebox(0,0)[lb]{$1$}}%
% STR 2 0 3 0
% 3 1588 2807 1588 2877 2 0
% $1$
\put(15.8800,-24.7700){\makebox(0,0)[lb]{$1$}}%
% STR 2 0 3 0
% 3 1133 1162 1133 1232 5 0
% $2$
\put(11.3300,-8.3200){\makebox(0,0){$2$}}%
% STR 2 0 3 0
% 3 1028 1344 1028 1414 5 0
% $2$
\put(10.2800,-10.1400){\makebox(0,0){$2$}}%
% STR 2 0 3 0
% 3 755 1624 755 1694 5 0
% $2$
\put(7.5500,-12.9400){\makebox(0,0){$2$}}%
% STR 2 0 3 0
% 3 853 1442 853 1512 5 0
% $2$
\put(8.5300,-11.1200){\makebox(0,0){$2$}}%
% STR 2 0 3 0
% 3 580 1722 580 1792 5 0
% $2$
\put(5.8000,-13.9200){\makebox(0,0){$2$}}%
% STR 2 0 3 0
% 3 475 1904 475 1974 5 0
% $2$
\put(4.7500,-15.7400){\makebox(0,0){$2$}}%
% STR 2 0 3 0
% 3 1133 1435 1133 1505 5 0
% $0$
\put(11.3300,-11.0500){\makebox(0,0){$0$}}%
% STR 2 0 3 0
% 3 1028 1624 1028 1694 5 0
% $0$
\put(10.2800,-12.9400){\makebox(0,0){$0$}}%
% STR 2 0 3 0
% 3 853 1715 853 1785 5 0
% $0$
\put(8.5300,-13.8500){\makebox(0,0){$0$}}%
% STR 2 0 3 0
% 3 755 1904 755 1974 5 0
% $0$
\put(7.5500,-15.7400){\makebox(0,0){$0$}}%
% STR 2 0 3 0
% 3 580 2009 580 2079 5 0
% $0$
\put(5.8000,-16.7900){\makebox(0,0){$0$}}%
% STR 2 0 3 0
% 3 1133 1715 1133 1785 5 0
% $2$
\put(11.3300,-13.8500){\makebox(0,0){$2$}}%
% STR 2 0 3 0
% 3 1028 1904 1028 1974 5 0
% $2$
\put(10.2800,-15.7400){\makebox(0,0){$2$}}%
% STR 2 0 3 0
% 3 853 2002 853 2072 5 0
% $2$
\put(8.5300,-16.7200){\makebox(0,0){$2$}}%
% STR 2 0 3 0
% 3 755 2184 755 2254 5 0
% $2$
\put(7.5500,-18.5400){\makebox(0,0){$2$}}%
% STR 2 0 3 0
% 3 1133 2002 1133 2072 5 0
% $0$
\put(11.3300,-16.7200){\makebox(0,0){$0$}}%
% STR 2 0 3 0
% 3 1028 2184 1028 2254 5 0
% $0$
\put(10.2800,-18.5400){\makebox(0,0){$0$}}%
% STR 2 0 3 0
% 3 853 2282 853 2352 5 0
% $0$
\put(8.5300,-19.5200){\makebox(0,0){$0$}}%
% STR 2 0 3 0
% 3 1133 2275 1133 2345 5 0
% $2$
\put(11.3300,-19.4500){\makebox(0,0){$2$}}%
% STR 2 0 3 0
% 3 1028 2464 1028 2534 5 0
% $2$
\put(10.2800,-21.3400){\makebox(0,0){$2$}}%
% STR 2 0 3 0
% 3 1133 2562 1133 2632 5 0
% $0$
\put(11.3300,-22.3200){\makebox(0,0){$0$}}%
% STR 2 0 3 0
% 3 951 1029 951 1099 2 0
% $0$
\put(9.5100,-6.9900){\makebox(0,0)[lb]{$0$}}%
% STR 2 0 3 0
% 3 811 1169 811 1239 2 0
% $0$
\put(8.1100,-8.3900){\makebox(0,0)[lb]{$0$}}%
% STR 2 0 3 0
% 3 664 1323 664 1393 2 0
% $0$
\put(6.6400,-9.9300){\makebox(0,0)[lb]{$0$}}%
% STR 2 0 3 0
% 3 538 1456 538 1526 2 0
% $0$
\put(5.3800,-11.2600){\makebox(0,0)[lb]{$0$}}%
% STR 2 0 3 0
% 3 391 1610 391 1680 2 0
% $0$
\put(3.9100,-12.8000){\makebox(0,0)[lb]{$0$}}%
% STR 2 0 3 0
% 3 244 1736 244 1806 2 0
% $0$
\put(2.4400,-14.0600){\makebox(0,0)[lb]{$0$}}%
% STR 2 0 3 0
% 3 230 2072 230 2142 2 0
% $2$
\put(2.3000,-17.4200){\makebox(0,0)[lb]{$2$}}%
% STR 2 0 3 0
% 3 384 2268 384 2338 2 0
% $0$
\put(3.8400,-19.3800){\makebox(0,0)[lb]{$0$}}%
% STR 2 0 3 0
% 3 517 2380 517 2450 2 0
% $2$
\put(5.1700,-20.5000){\makebox(0,0)[lb]{$2$}}%
% STR 2 0 3 0
% 3 657 2548 657 2618 2 0
% $0$
\put(6.5700,-22.1800){\makebox(0,0)[lb]{$0$}}%
% STR 2 0 3 0
% 3 783 2681 783 2751 2 0
% $2$
\put(7.8300,-23.5100){\makebox(0,0)[lb]{$2$}}%
% STR 2 0 3 0
% 3 937 2821 937 2891 2 0
% $0$
\put(9.3700,-24.9100){\makebox(0,0)[lb]{$0$}}%
% STR 2 0 3 0
% 3 1273 1008 1273 1078 5 0
% $1$
\put(12.7300,-6.7800){\makebox(0,0){$1$}}%
% STR 2 0 3 0
% 3 1273 2814 1273 2884 5 0
% $1$
\put(12.7300,-24.8400){\makebox(0,0){$1$}}%
\end{picture}%
\begin{center}
 {\phantom{************************}\bf Figure 4}
\end{center}

Let us set
a configuration $C$ be an assignment of spins.
Hence there are $4N^2+4N+1$ configurations 
for each boundary condition $(\lambda,\mu)$.
We introduce a probability measure in the set of
all configurations, assigning a statistical weight
$W_N^{(\lambda,\mu)}(C)$ to each configuration $C$
attached to the boundary condition
$(\lambda,\mu)$.
The weight $W_N^{(\lambda,\mu)}(C)$ is given as 
the product over all vertices
\begin{eqnarray}
W_N^{(\lambda,\mu)}(C)
=\prod_{vertex} R^{(1,1)}(\zeta)_{i'j'}^{ij}
\prod_{vertex}
R^{\left(\frac{1}{2},1\right)}(\zeta)_{k'l'}^{kl}.
\nonumber
\end{eqnarray}
Here we muliply $R$-matrices under the 
boundary condition $(\lambda,\mu)$. 
The probability for the configuration $C$ to take place
is 
$\frac{1}{Z_N^{(\lambda,\mu)}}
W_N^{(\lambda,\mu)}(C)$, where
\begin{eqnarray}
Z^{(\lambda, \mu)}_N=\sum_{C}W_N^{(\lambda,\mu)}(C).\nonumber
\end{eqnarray}
This normalization factor 
$Z_N^{(\lambda,\mu)}$ is called the Partition function.
The probability that
the vertex of the center
of our lattice takes value $\epsilon$
is given as follows
\begin{eqnarray}
P_{\epsilon}^{(\lambda,\mu)}(N)
=\displaystyle
\frac{\displaystyle \sum_{C~ ({\rm s.t.}
\epsilon(C)=\epsilon)}W_N^{(\lambda,\mu)}(C)}
{\displaystyle Z^{(\lambda, \mu)}_N}.
\end{eqnarray}
Here the suffix $(\lambda,\mu)$ represents
the boundary conditions.
In this article we are interested in 
the probability functions in the infinite
volume limit
defined by
\begin{eqnarray}
P_{\epsilon}^{(\lambda,\mu)}=
\lim_{N \to \infty}P_{\epsilon}^{(\lambda,\mu)}(N).
\end{eqnarray}
We consider the infinite volume limit
in the region given by
\begin{eqnarray}
-1<q<0,~~1<\zeta<(-q)^{-1}.\nonumber
\end{eqnarray}
From the symmetry arguments, we have the relations
between the probability functions:
\begin{eqnarray}
P_{\epsilon}^{(\Lambda_0+\Lambda_1,2\Lambda_0)}
=P_{1-\epsilon}^{(\Lambda_0+\Lambda_1,2\Lambda_1)}
=P_{\epsilon}^{(2\Lambda_1,\Lambda_0+\Lambda_1)}
=P_{1-\epsilon}^{(2\Lambda_0,\Lambda_0+\Lambda_1)},~
(\epsilon=0,1).\nonumber
\end{eqnarray}
We will show that
the probability functions have the following formulae.
\begin{eqnarray}
&&P_{0}^{(\Lambda_0+\Lambda_1,2\Lambda_0)}\label{eqn:P0}
\\
&=&
\frac{1}{2}\frac{(q^4;q^2)_\infty}
{(q^4;-q^2)_\infty}
-\frac{1}{1-q^2}\frac{(q^{16};q^{16})_\infty}
{(q^4;q^4)_\infty}\left\{
\frac{1}{2}
\left(-1-q^2+\frac{4q^4}{1-q^4}\right)
\frac{(-q^8;q^{16})_\infty^2}{(-q^2;q^4)_\infty^2}\right.
\nonumber
\\
&+& \left(-q^2-q^4+
\frac{4q^6}{1-q^4}\right)
\frac{(-q^{16};q^{16})_\infty^2}
{(-q^2;q^4)_\infty^2}
\left.+
\left(1+q^2-\frac{2q^2}{1-q^4}\right)
\frac{(-q^4;q^8)_\infty}{(-q^4;q^4)_\infty^2}\right\},
\nonumber
\end{eqnarray}
and
\begin{eqnarray}
&&P_{1}^{(\Lambda_0+\Lambda_1,2\Lambda_0)}\label{eqn:P1}
\\
&=&
\frac{1}{2}\frac{(q^4;q^2)_\infty}
{(q^4;-q^2)_\infty}
-\frac{1}{1-q^2}\frac{(q^{16};q^{16})_\infty}
{(q^4;q^4)_\infty}\left\{
\frac{1}{2}
\left(-1-q^2+\frac{4q^2}{1-q^4}\right)
\frac{(-q^8;q^{16})_\infty^2}{(-q^2;q^4)_\infty^2}\right.
\nonumber
\\
&+& \left(-q^2-q^4+
\frac{4q^4}{1-q^4}\right)
\frac{(-q^{16};q^{16})_\infty^2}
{(-q^2;q^4)_\infty^2}
\left.-\frac{2q^4}{1-q^4}
\frac{(-q^4;q^8)_\infty}{(-q^4;q^4)_\infty^2}\right\}.
\nonumber
\end{eqnarray}
The following is the main result of our paper,
which is a direct consequence of relations
(\ref{eqn:P0}) and (\ref{eqn:P1}).

\vspace{12pt}
{\bf Main Result}~~~
\begin{em}
The Spontaneous Staggered Polarization
of our model has the following infinite 
product formula
\begin{eqnarray}
&&P_{0}^{(\Lambda_0+\Lambda_1,2\Lambda_0)}
-
P_{1}^{(\Lambda_0+\Lambda_1,2\Lambda_0)}
\label{eqn:main}\\
&=&
\displaystyle
-\frac{1}{1-q^4}
\frac{(q^{16};q^{16})_\infty}
{(q^{4};q^{4})_\infty}
\left\{(1+q^{4})\frac{(-q^4;q^8)_\infty}
{(-q^4;q^4)_\infty^2}
-2q^{2}\frac{(-q^8;q^{16})_\infty^2}
{(-q^2;q^4)_\infty^2}
-4q^4\frac{(-q^{16};q^{16})_\infty^2}
{(-q^2;q^4)_\infty^2}\right\}.\nonumber
\end{eqnarray}
\end{em}
\vspace{12pt}

In fact, the Spontaneous Staggered Polarization
is independent of the spectral parameter $\zeta$.

~\\
{\it Remark.~
From the relation
(\ref{eqn:id}) and
the trace formulae
(\ref{eqn:trace}),
we get}
$$P_0^{(\Lambda_0+\Lambda_1,2\Lambda_0)}
+P_1^{(\Lambda_0+\Lambda_1,2\Lambda_0)}=1.$$

~\\
In the sequel
we explain how to derive this formula.

\section{Derivation}
The purpose of this section is to show
Main Result.
\subsection{Infinite Volume Limit}

We consider the infinite volume limit $N \to
\infty$.
For simplicity,
we concentrate ourselves
to the boundary condition
$(\Lambda_0+\Lambda_1,2\Lambda_0)$.

A path is defined as a sequence
of $0,1,2$, denoted by
$|p\rangle=\{p(j)\}_{j\geq 1}$.
For weights 
$\lambda=2\Lambda_0, \Lambda_0+\Lambda_1, 2\Lambda_1$,
consider the set of paths
$P_{2\Lambda_0}, P_{\Lambda_0+\Lambda_1},
P_{2\Lambda_1}$ by
\begin{eqnarray}
P_{2\Lambda_0}&=&
\left\{|p\rangle|p(j)=1+(-1)^{j}, {\rm for}~~ j>>0
\right\},\nonumber \\
P_{\Lambda_0+\Lambda_1}&=&
\left\{|p\rangle |p(j)=1, {\rm for}~~ j>>0
\right\},\nonumber \\
P_{2\Lambda_1}&=&
\left\{|p\rangle|p(j)=1+(-1)^{j+1}, {\rm for}~~ j>>0
\right\}.\nonumber
\end{eqnarray}

The infinite lattice so defined may be split into 
six pieces, consisting of four corners and two half
columns(see figure 5).
The associated Corner Transfer Matrix are labelled
$A(\zeta),B(\zeta),C(\zeta)$ and $D(\zeta)$. 
Two lines are 
labelled $\Phi_{UP,\epsilon}(\zeta)$ 
and $\Phi_{LOW,\epsilon}(\zeta)$.

\vspace{12pt}

%WinTpicVersion2.13
\unitlength 0.1in
\begin{picture}(45.25,36.65)(4.65,-38.10)
% VECTOR 2 0 3 0
% 16 3000 1000 2600 1000 3000 1400 2600 1400 3000 1800 2600 1800 3000 2200 2600 2200 3000 2600 2600 2600 3000 3000 2600 3000 3000 3400 2600 3400 3000 3800 2600 3800
% 
\special{pn 8}%
\special{pa 3000 600}%
\special{pa 2600 600}%
\special{fp}%
\special{sh 1}%
\special{pa 2600 600}%
\special{pa 2667 620}%
\special{pa 2653 600}%
\special{pa 2667 580}%
\special{pa 2600 600}%
\special{fp}%
\special{pa 3000 1000}%
\special{pa 2600 1000}%
\special{fp}%
\special{sh 1}%
\special{pa 2600 1000}%
\special{pa 2667 1020}%
\special{pa 2653 1000}%
\special{pa 2667 980}%
\special{pa 2600 1000}%
\special{fp}%
\special{pa 3000 1400}%
\special{pa 2600 1400}%
\special{fp}%
\special{sh 1}%
\special{pa 2600 1400}%
\special{pa 2667 1420}%
\special{pa 2653 1400}%
\special{pa 2667 1380}%
\special{pa 2600 1400}%
\special{fp}%
\special{pa 3000 1800}%
\special{pa 2600 1800}%
\special{fp}%
\special{sh 1}%
\special{pa 2600 1800}%
\special{pa 2667 1820}%
\special{pa 2653 1800}%
\special{pa 2667 1780}%
\special{pa 2600 1800}%
\special{fp}%
\special{pa 3000 2200}%
\special{pa 2600 2200}%
\special{fp}%
\special{sh 1}%
\special{pa 2600 2200}%
\special{pa 2667 2220}%
\special{pa 2653 2200}%
\special{pa 2667 2180}%
\special{pa 2600 2200}%
\special{fp}%
\special{pa 3000 2600}%
\special{pa 2600 2600}%
\special{fp}%
\special{sh 1}%
\special{pa 2600 2600}%
\special{pa 2667 2620}%
\special{pa 2653 2600}%
\special{pa 2667 2580}%
\special{pa 2600 2600}%
\special{fp}%
\special{pa 3000 3000}%
\special{pa 2600 3000}%
\special{fp}%
\special{sh 1}%
\special{pa 2600 3000}%
\special{pa 2667 3020}%
\special{pa 2653 3000}%
\special{pa 2667 2980}%
\special{pa 2600 3000}%
\special{fp}%
\special{pa 3000 3400}%
\special{pa 2600 3400}%
\special{fp}%
\special{sh 1}%
\special{pa 2600 3400}%
\special{pa 2667 3420}%
\special{pa 2653 3400}%
\special{pa 2667 3380}%
\special{pa 2600 3400}%
\special{fp}%
% VECTOR 2 0 3 0
% 16 4800 2200 3400 2200 4400 1800 3400 1800 4000 1400 3400 1400 3600 1000 3400 1000 4800 2600 3400 2600 4400 3000 3400 3000 4000 3400 3400 3400 3600 3800 3400 3800
% 
\special{pn 8}%
\special{pa 4800 1800}%
\special{pa 3400 1800}%
\special{fp}%
\special{sh 1}%
\special{pa 3400 1800}%
\special{pa 3467 1820}%
\special{pa 3453 1800}%
\special{pa 3467 1780}%
\special{pa 3400 1800}%
\special{fp}%
\special{pa 4400 1400}%
\special{pa 3400 1400}%
\special{fp}%
\special{sh 1}%
\special{pa 3400 1400}%
\special{pa 3467 1420}%
\special{pa 3453 1400}%
\special{pa 3467 1380}%
\special{pa 3400 1400}%
\special{fp}%
\special{pa 4000 1000}%
\special{pa 3400 1000}%
\special{fp}%
\special{sh 1}%
\special{pa 3400 1000}%
\special{pa 3467 1020}%
\special{pa 3453 1000}%
\special{pa 3467 980}%
\special{pa 3400 1000}%
\special{fp}%
\special{pa 3600 600}%
\special{pa 3400 600}%
\special{fp}%
\special{sh 1}%
\special{pa 3400 600}%
\special{pa 3467 620}%
\special{pa 3453 600}%
\special{pa 3467 580}%
\special{pa 3400 600}%
\special{fp}%
\special{pa 4800 2200}%
\special{pa 3400 2200}%
\special{fp}%
\special{sh 1}%
\special{pa 3400 2200}%
\special{pa 3467 2220}%
\special{pa 3453 2200}%
\special{pa 3467 2180}%
\special{pa 3400 2200}%
\special{fp}%
\special{pa 4400 2600}%
\special{pa 3400 2600}%
\special{fp}%
\special{sh 1}%
\special{pa 3400 2600}%
\special{pa 3467 2620}%
\special{pa 3453 2600}%
\special{pa 3467 2580}%
\special{pa 3400 2600}%
\special{fp}%
\special{pa 4000 3000}%
\special{pa 3400 3000}%
\special{fp}%
\special{sh 1}%
\special{pa 3400 3000}%
\special{pa 3467 3020}%
\special{pa 3453 3000}%
\special{pa 3467 2980}%
\special{pa 3400 3000}%
\special{fp}%
\special{pa 3600 3400}%
\special{pa 3400 3400}%
\special{fp}%
\special{sh 1}%
\special{pa 3400 3400}%
\special{pa 3467 3420}%
\special{pa 3453 3400}%
\special{pa 3467 3380}%
\special{pa 3400 3400}%
\special{fp}%
% VECTOR 0 0 3 0
% 4 2800 800 2800 2330 2810 2490 2810 4090
% 
\special{pn 20}%
\special{pa 2800 400}%
\special{pa 2800 1930}%
\special{fp}%
\special{sh 1}%
\special{pa 2800 1930}%
\special{pa 2820 1863}%
\special{pa 2800 1877}%
\special{pa 2780 1863}%
\special{pa 2800 1930}%
\special{fp}%
\special{pa 2810 2090}%
\special{pa 2810 3690}%
\special{fp}%
\special{sh 1}%
\special{pa 2810 3690}%
\special{pa 2830 3623}%
\special{pa 2810 3637}%
\special{pa 2790 3623}%
\special{pa 2810 3690}%
\special{fp}%
% VECTOR 2 0 3 0
% 2 2190 2600 790 2600
% 
\special{pn 8}%
\special{pa 2190 2200}%
\special{pa 790 2200}%
\special{fp}%
\special{sh 1}%
\special{pa 790 2200}%
\special{pa 857 2220}%
\special{pa 843 2200}%
\special{pa 857 2180}%
\special{pa 790 2200}%
\special{fp}%
% VECTOR 2 0 3 0
% 14 2190 3000 1190 3000 2190 3400 1590 3400 2190 3800 1990 3800 2190 2200 790 2200 2190 1800 1190 1800 2190 1400 1590 1400 2190 1000 1990 1000
% 
\special{pn 8}%
\special{pa 2190 2600}%
\special{pa 1190 2600}%
\special{fp}%
\special{sh 1}%
\special{pa 1190 2600}%
\special{pa 1257 2620}%
\special{pa 1243 2600}%
\special{pa 1257 2580}%
\special{pa 1190 2600}%
\special{fp}%
\special{pa 2190 3000}%
\special{pa 1590 3000}%
\special{fp}%
\special{sh 1}%
\special{pa 1590 3000}%
\special{pa 1657 3020}%
\special{pa 1643 3000}%
\special{pa 1657 2980}%
\special{pa 1590 3000}%
\special{fp}%
\special{pa 2190 3400}%
\special{pa 1990 3400}%
\special{fp}%
\special{sh 1}%
\special{pa 1990 3400}%
\special{pa 2057 3420}%
\special{pa 2043 3400}%
\special{pa 2057 3380}%
\special{pa 1990 3400}%
\special{fp}%
\special{pa 2190 1800}%
\special{pa 790 1800}%
\special{fp}%
\special{sh 1}%
\special{pa 790 1800}%
\special{pa 857 1820}%
\special{pa 843 1800}%
\special{pa 857 1780}%
\special{pa 790 1800}%
\special{fp}%
\special{pa 2190 1400}%
\special{pa 1190 1400}%
\special{fp}%
\special{sh 1}%
\special{pa 1190 1400}%
\special{pa 1257 1420}%
\special{pa 1243 1400}%
\special{pa 1257 1380}%
\special{pa 1190 1400}%
\special{fp}%
\special{pa 2190 1000}%
\special{pa 1590 1000}%
\special{fp}%
\special{sh 1}%
\special{pa 1590 1000}%
\special{pa 1657 1020}%
\special{pa 1643 1000}%
\special{pa 1657 980}%
\special{pa 1590 1000}%
\special{fp}%
\special{pa 2190 600}%
\special{pa 1990 600}%
\special{fp}%
\special{sh 1}%
\special{pa 1990 600}%
\special{pa 2057 620}%
\special{pa 2043 600}%
\special{pa 2057 580}%
\special{pa 1990 600}%
\special{fp}%
% LINE 2 1 3 0
% 6 3190 610 3190 4210 2390 610 2390 4210 4990 2410 590 2410
% 
\special{pn 8}%
\special{pa 3190 210}%
\special{pa 3190 3810}%
\special{da 0.070}%
\special{pa 2390 210}%
\special{pa 2390 3810}%
\special{da 0.070}%
\special{pa 4990 2010}%
\special{pa 590 2010}%
\special{da 0.070}%
% VECTOR 2 0 3 0
% 4 2100 910 2100 2310 2100 2510 2100 3910
% 
\special{pn 8}%
\special{pa 2100 510}%
\special{pa 2100 1910}%
\special{fp}%
\special{sh 1}%
\special{pa 2100 1910}%
\special{pa 2120 1843}%
\special{pa 2100 1857}%
\special{pa 2080 1843}%
\special{pa 2100 1910}%
\special{fp}%
\special{pa 2100 2110}%
\special{pa 2100 3510}%
\special{fp}%
\special{sh 1}%
\special{pa 2100 3510}%
\special{pa 2120 3443}%
\special{pa 2100 3457}%
\special{pa 2080 3443}%
\special{pa 2100 3510}%
\special{fp}%
% VECTOR 2 0 3 0
% 4 3530 890 3530 2290 3530 2490 3530 3890
% 
\special{pn 8}%
\special{pa 3530 490}%
\special{pa 3530 1890}%
\special{fp}%
\special{sh 1}%
\special{pa 3530 1890}%
\special{pa 3550 1823}%
\special{pa 3530 1837}%
\special{pa 3510 1823}%
\special{pa 3530 1890}%
\special{fp}%
\special{pa 3530 2090}%
\special{pa 3530 3490}%
\special{fp}%
\special{sh 1}%
\special{pa 3530 3490}%
\special{pa 3550 3423}%
\special{pa 3530 3437}%
\special{pa 3510 3423}%
\special{pa 3530 3490}%
\special{fp}%
% VECTOR 2 0 3 0
% 24 3930 1290 3930 2290 3930 2490 3930 3490 4330 1690 4330 2290 4330 2490 4330 3090 4730 2090 4730 2290 4730 2490 4730 2690 1730 1290 1730 2290 1730 2490 1730 3490 1330 1690 1330 2290 1330 2490 1330 3090 930 2090 930 2290 930 2490 930 2690
% 
\special{pn 8}%
\special{pa 3930 890}%
\special{pa 3930 1890}%
\special{fp}%
\special{sh 1}%
\special{pa 3930 1890}%
\special{pa 3950 1823}%
\special{pa 3930 1837}%
\special{pa 3910 1823}%
\special{pa 3930 1890}%
\special{fp}%
\special{pa 3930 2090}%
\special{pa 3930 3090}%
\special{fp}%
\special{sh 1}%
\special{pa 3930 3090}%
\special{pa 3950 3023}%
\special{pa 3930 3037}%
\special{pa 3910 3023}%
\special{pa 3930 3090}%
\special{fp}%
\special{pa 4330 1290}%
\special{pa 4330 1890}%
\special{fp}%
\special{sh 1}%
\special{pa 4330 1890}%
\special{pa 4350 1823}%
\special{pa 4330 1837}%
\special{pa 4310 1823}%
\special{pa 4330 1890}%
\special{fp}%
\special{pa 4330 2090}%
\special{pa 4330 2690}%
\special{fp}%
\special{sh 1}%
\special{pa 4330 2690}%
\special{pa 4350 2623}%
\special{pa 4330 2637}%
\special{pa 4310 2623}%
\special{pa 4330 2690}%
\special{fp}%
\special{pa 4730 1690}%
\special{pa 4730 1890}%
\special{fp}%
\special{sh 1}%
\special{pa 4730 1890}%
\special{pa 4750 1823}%
\special{pa 4730 1837}%
\special{pa 4710 1823}%
\special{pa 4730 1890}%
\special{fp}%
\special{pa 4730 2090}%
\special{pa 4730 2290}%
\special{fp}%
\special{sh 1}%
\special{pa 4730 2290}%
\special{pa 4750 2223}%
\special{pa 4730 2237}%
\special{pa 4710 2223}%
\special{pa 4730 2290}%
\special{fp}%
\special{pa 1730 890}%
\special{pa 1730 1890}%
\special{fp}%
\special{sh 1}%
\special{pa 1730 1890}%
\special{pa 1750 1823}%
\special{pa 1730 1837}%
\special{pa 1710 1823}%
\special{pa 1730 1890}%
\special{fp}%
\special{pa 1730 2090}%
\special{pa 1730 3090}%
\special{fp}%
\special{sh 1}%
\special{pa 1730 3090}%
\special{pa 1750 3023}%
\special{pa 1730 3037}%
\special{pa 1710 3023}%
\special{pa 1730 3090}%
\special{fp}%
\special{pa 1330 1290}%
\special{pa 1330 1890}%
\special{fp}%
\special{sh 1}%
\special{pa 1330 1890}%
\special{pa 1350 1823}%
\special{pa 1330 1837}%
\special{pa 1310 1823}%
\special{pa 1330 1890}%
\special{fp}%
\special{pa 1330 2090}%
\special{pa 1330 2690}%
\special{fp}%
\special{sh 1}%
\special{pa 1330 2690}%
\special{pa 1350 2623}%
\special{pa 1330 2637}%
\special{pa 1310 2623}%
\special{pa 1330 2690}%
\special{fp}%
\special{pa 930 1690}%
\special{pa 930 1890}%
\special{fp}%
\special{sh 1}%
\special{pa 930 1890}%
\special{pa 950 1823}%
\special{pa 930 1837}%
\special{pa 910 1823}%
\special{pa 930 1890}%
\special{fp}%
\special{pa 930 2090}%
\special{pa 930 2290}%
\special{fp}%
\special{sh 1}%
\special{pa 930 2290}%
\special{pa 950 2223}%
\special{pa 930 2237}%
\special{pa 910 2223}%
\special{pa 930 2290}%
\special{fp}%
% STR 2 0 3 0
% 3 2900 2210 2900 2310 5 0
% $\epsilon$
\put(29.0000,-19.1000){\makebox(0,0){$\epsilon$}}%
% STR 2 0 3 0
% 3 2900 2410 2900 2510 5 0
% $\epsilon$
\put(29.0000,-21.1000){\makebox(0,0){$\epsilon$}}%
% STR 2 0 3 0
% 3 4390 3910 4390 4010 5 0
% \bf Figure 5
\put(43.9000,-36.1000){\makebox(0,0){\bf Figure 5}}%
% STR 2 0 3 0
% 3 2800 530 2800 630 5 0
% $\Phi_{UP,\epsilon}$
\put(28.0000,-2.3000){\makebox(0,0){$\Phi_{UP,\epsilon}$}}%
% STR 2 0 3 0
% 3 2800 4130 2800 4230 5 0
% $\Phi_{LOW,\epsilon}$
\put(28.0000,-38.3000){\makebox(0,0){$\Phi_{LOW,\epsilon}$}}%
% STR 2 0 3 0
% 3 1010 980 1010 1080 5 0
% \huge \bf B
\put(10.1000,-6.8000){\makebox(0,0){\huge \bf B}}%
% STR 2 0 3 0
% 3 4610 980 4610 1080 5 0
% \huge \bf A
\put(46.1000,-6.8000){\makebox(0,0){\huge \bf A}}%
% STR 2 0 3 0
% 3 4610 3580 4610 3680 5 0
% \huge \bf D
\put(46.1000,-32.8000){\makebox(0,0){\huge \bf D}}%
% STR 2 0 3 0
% 3 960 3600 960 3700 5 0
% \huge \bf C
\put(9.6000,-33.0000){\makebox(0,0){\huge \bf C}}%
\end{picture}%

\vspace{36pt}

Following Baxter we define the 
Corner Transfer Matrix $O^{(1)}(\zeta),
O^{(2)}(\zeta)$ 
in the infinite volume limit $N\to \infty$, by 
the sum over the spin configurations in the 
interior,
$$
\left(O^{(b)}(\zeta)\right)^{|p'\rangle}
_{|p \rangle}=
\sum_{\rm interior \  edges}
\prod R^{\epsilon'_1 \epsilon'_2}
_{\epsilon_1 \epsilon_2},
$$
where we take summations with the following boundary
conditions related to the suffix $b=1,2$.
For $b=1$, the paths $|p\rangle, |p'\rangle$
belong to the set of paths $P_{\Lambda_0+\Lambda_1}$,
and North-West boundary is
fixed by $b=1$(see figure 6).
For $b=2$, the paths $|p\rangle, |p'\rangle$
belong to the set of paths $P_{2\Lambda_0}$,
and North-West boundary is
fixed by $b=2$(see figure 7).
The Corner Transfer Matrix $O^{(1)}(\zeta)$
and $O^{(2)}(\zeta)$ act on the path spaces
$P_{\Lambda_0+\Lambda_1}$ and
$P_{2\Lambda_0}$, respectively.

%WinTpicVersion2.13
\unitlength 0.1in
\begin{picture}(17.73,19.66)(18.24,-27.05)
% VECTOR 2 0 3 0
% 10 3359 1189 3359 2589 3079 1469 3079 2589 2799 1749 2799 2589 2519 2029 2519 2589 2239 2309 2239 2589
% 
\special{pn 8}%
\special{pa 3359 789}%
\special{pa 3359 2189}%
\special{fp}%
\special{sh 1}%
\special{pa 3359 2189}%
\special{pa 3379 2122}%
\special{pa 3359 2136}%
\special{pa 3339 2122}%
\special{pa 3359 2189}%
\special{fp}%
\special{pa 3079 1069}%
\special{pa 3079 2189}%
\special{fp}%
\special{sh 1}%
\special{pa 3079 2189}%
\special{pa 3099 2122}%
\special{pa 3079 2136}%
\special{pa 3059 2122}%
\special{pa 3079 2189}%
\special{fp}%
\special{pa 2799 1349}%
\special{pa 2799 2189}%
\special{fp}%
\special{sh 1}%
\special{pa 2799 2189}%
\special{pa 2819 2122}%
\special{pa 2799 2136}%
\special{pa 2779 2122}%
\special{pa 2799 2189}%
\special{fp}%
\special{pa 2519 1629}%
\special{pa 2519 2189}%
\special{fp}%
\special{sh 1}%
\special{pa 2519 2189}%
\special{pa 2539 2122}%
\special{pa 2519 2136}%
\special{pa 2499 2122}%
\special{pa 2519 2189}%
\special{fp}%
\special{pa 2239 1909}%
\special{pa 2239 2189}%
\special{fp}%
\special{sh 1}%
\special{pa 2239 2189}%
\special{pa 2259 2122}%
\special{pa 2239 2136}%
\special{pa 2219 2122}%
\special{pa 2239 2189}%
\special{fp}%
% VECTOR 2 0 3 0
% 10 3499 2449 2099 2449 3499 2169 2379 2169 3499 1889 2659 1889 3499 1609 2939 1609 3499 1329 3219 1329
% 
\special{pn 8}%
\special{pa 3499 2049}%
\special{pa 2099 2049}%
\special{fp}%
\special{sh 1}%
\special{pa 2099 2049}%
\special{pa 2166 2069}%
\special{pa 2152 2049}%
\special{pa 2166 2029}%
\special{pa 2099 2049}%
\special{fp}%
\special{pa 3499 1769}%
\special{pa 2379 1769}%
\special{fp}%
\special{sh 1}%
\special{pa 2379 1769}%
\special{pa 2446 1789}%
\special{pa 2432 1769}%
\special{pa 2446 1749}%
\special{pa 2379 1769}%
\special{fp}%
\special{pa 3499 1489}%
\special{pa 2659 1489}%
\special{fp}%
\special{sh 1}%
\special{pa 2659 1489}%
\special{pa 2726 1509}%
\special{pa 2712 1489}%
\special{pa 2726 1469}%
\special{pa 2659 1489}%
\special{fp}%
\special{pa 3499 1209}%
\special{pa 2939 1209}%
\special{fp}%
\special{sh 1}%
\special{pa 2939 1209}%
\special{pa 3006 1229}%
\special{pa 2992 1209}%
\special{pa 3006 1189}%
\special{pa 2939 1209}%
\special{fp}%
\special{pa 3499 929}%
\special{pa 3219 929}%
\special{fp}%
\special{sh 1}%
\special{pa 3219 929}%
\special{pa 3286 949}%
\special{pa 3272 929}%
\special{pa 3286 909}%
\special{pa 3219 929}%
\special{fp}%
% STR 2 0 3 0
% 3 1959 2379 1959 2449 5 0
% $1$
\put(19.5900,-20.4900){\makebox(0,0){$1$}}%
% STR 2 0 3 0
% 3 2099 2239 2099 2309 5 0
% $1$
\put(20.9900,-19.0900){\makebox(0,0){$1$}}%
% STR 2 0 3 0
% 3 2239 2099 2239 2169 5 0
% $1$
\put(22.3900,-17.6900){\makebox(0,0){$1$}}%
% STR 2 0 3 0
% 3 2386 1973 2386 2043 5 0
% $1$
\put(23.8600,-16.4300){\makebox(0,0){$1$}}%
% STR 2 0 3 0
% 3 2526 1833 2526 1903 5 0
% $1$
\put(25.2600,-15.0300){\makebox(0,0){$1$}}%
% STR 2 0 3 0
% 3 2659 1686 2659 1756 5 0
% $1$
\put(26.5900,-13.5600){\makebox(0,0){$1$}}%
% STR 2 0 3 0
% 3 2799 1546 2799 1616 5 0
% $1$
\put(27.9900,-12.1600){\makebox(0,0){$1$}}%
% STR 2 0 3 0
% 3 2939 1406 2939 1476 5 0
% $1$
\put(29.3900,-10.7600){\makebox(0,0){$1$}}%
% STR 2 0 3 0
% 3 3079 1266 3079 1336 5 0
% $1$
\put(30.7900,-9.3600){\makebox(0,0){$1$}}%
% STR 2 0 3 0
% 3 3247 1155 3247 1224 5 0
% $1$
\put(32.4700,-8.2400){\makebox(0,0){$1$}}%
% STR 2 0 3 0
% 3 3548 2414 3548 2484 2 0
% $p'(1)$
\put(35.4800,-20.8400){\makebox(0,0)[lb]{$p'(1)$}}%
% STR 2 0 3 0
% 3 3548 2134 3548 2204 2 0
% $p'(2)$
\put(35.4800,-18.0400){\makebox(0,0)[lb]{$p'(2)$}}%
% STR 2 0 3 0
% 3 3541 1854 3541 1924 2 0
% $p'(3)$
\put(35.4100,-15.2400){\makebox(0,0)[lb]{$p'(3)$}}%
% STR 2 0 3 0
% 3 3359 2582 3359 2652 5 0
% $p(1)$
\put(33.5900,-22.5200){\makebox(0,0){$p(1)$}}%
% STR 2 0 3 0
% 3 3079 2575 3079 2645 5 0
% $p(2)$
\put(30.7900,-22.4500){\makebox(0,0){$p(2)$}}%
% STR 2 0 3 0
% 3 2799 2575 2799 2645 5 0
% $p(3)$
\put(27.9900,-22.4500){\makebox(0,0){$p(3)$}}%
% LINE 1 2 3 0
% 2 2449 2652 2267 2652
% 
\special{pn 13}%
\special{pa 2449 2252}%
\special{pa 2267 2252}%
\special{dt 0.045}%
\special{pa 2267 2252}%
\special{pa 2268 2252}%
\special{dt 0.045}%
% LINE 1 2 3 0
% 2 3597 1581 3597 1406
% 
\special{pn 13}%
\special{pa 3597 1181}%
\special{pa 3597 1006}%
\special{dt 0.045}%
\special{pa 3597 1006}%
\special{pa 3597 1007}%
\special{dt 0.045}%
% STR 2 0 3 0
% 3 2940 3090 2940 3190 5 0
% {\bf Figure 6}
\put(29.4000,-27.9000){\makebox(0,0){{\bf Figure 6}}}%
% STR 2 0 3 0
% 3 2960 2810 2960 2910 5 0
% $|p \rangle$
\put(29.6000,-25.1000){\makebox(0,0){$|p \rangle$}}%
% STR 2 0 3 0
% 3 3980 1940 3980 2040 5 0
% $|p' \rangle$
\put(39.8000,-16.4000){\makebox(0,0){$|p' \rangle$}}%
\end{picture}%
\hspace{36pt}
%WinTpicVersion2.13
\unitlength 0.1in
\begin{picture}(17.73,19.56)(18.24,-26.95)
% VECTOR 2 0 3 0
% 10 3359 1189 3359 2589 3079 1469 3079 2589 2799 1749 2799 2589 2519 2029 2519 2589 2239 2309 2239 2589
% 
\special{pn 8}%
\special{pa 3359 789}%
\special{pa 3359 2189}%
\special{fp}%
\special{sh 1}%
\special{pa 3359 2189}%
\special{pa 3379 2122}%
\special{pa 3359 2136}%
\special{pa 3339 2122}%
\special{pa 3359 2189}%
\special{fp}%
\special{pa 3079 1069}%
\special{pa 3079 2189}%
\special{fp}%
\special{sh 1}%
\special{pa 3079 2189}%
\special{pa 3099 2122}%
\special{pa 3079 2136}%
\special{pa 3059 2122}%
\special{pa 3079 2189}%
\special{fp}%
\special{pa 2799 1349}%
\special{pa 2799 2189}%
\special{fp}%
\special{sh 1}%
\special{pa 2799 2189}%
\special{pa 2819 2122}%
\special{pa 2799 2136}%
\special{pa 2779 2122}%
\special{pa 2799 2189}%
\special{fp}%
\special{pa 2519 1629}%
\special{pa 2519 2189}%
\special{fp}%
\special{sh 1}%
\special{pa 2519 2189}%
\special{pa 2539 2122}%
\special{pa 2519 2136}%
\special{pa 2499 2122}%
\special{pa 2519 2189}%
\special{fp}%
\special{pa 2239 1909}%
\special{pa 2239 2189}%
\special{fp}%
\special{sh 1}%
\special{pa 2239 2189}%
\special{pa 2259 2122}%
\special{pa 2239 2136}%
\special{pa 2219 2122}%
\special{pa 2239 2189}%
\special{fp}%
% VECTOR 2 0 3 0
% 10 3499 2449 2099 2449 3499 2169 2379 2169 3499 1889 2659 1889 3499 1609 2939 1609 3499 1329 3219 1329
% 
\special{pn 8}%
\special{pa 3499 2049}%
\special{pa 2099 2049}%
\special{fp}%
\special{sh 1}%
\special{pa 2099 2049}%
\special{pa 2166 2069}%
\special{pa 2152 2049}%
\special{pa 2166 2029}%
\special{pa 2099 2049}%
\special{fp}%
\special{pa 3499 1769}%
\special{pa 2379 1769}%
\special{fp}%
\special{sh 1}%
\special{pa 2379 1769}%
\special{pa 2446 1789}%
\special{pa 2432 1769}%
\special{pa 2446 1749}%
\special{pa 2379 1769}%
\special{fp}%
\special{pa 3499 1489}%
\special{pa 2659 1489}%
\special{fp}%
\special{sh 1}%
\special{pa 2659 1489}%
\special{pa 2726 1509}%
\special{pa 2712 1489}%
\special{pa 2726 1469}%
\special{pa 2659 1489}%
\special{fp}%
\special{pa 3499 1209}%
\special{pa 2939 1209}%
\special{fp}%
\special{sh 1}%
\special{pa 2939 1209}%
\special{pa 3006 1229}%
\special{pa 2992 1209}%
\special{pa 3006 1189}%
\special{pa 2939 1209}%
\special{fp}%
\special{pa 3499 929}%
\special{pa 3219 929}%
\special{fp}%
\special{sh 1}%
\special{pa 3219 929}%
\special{pa 3286 949}%
\special{pa 3272 929}%
\special{pa 3286 909}%
\special{pa 3219 929}%
\special{fp}%
% STR 2 0 3 0
% 3 1959 2379 1959 2449 5 0
% $2$
\put(19.5900,-20.4900){\makebox(0,0){$2$}}%
% STR 2 0 3 0
% 3 2099 2239 2099 2309 5 0
% $2$
\put(20.9900,-19.0900){\makebox(0,0){$2$}}%
% STR 2 0 3 0
% 3 2239 2099 2239 2169 5 0
% $2$
\put(22.3900,-17.6900){\makebox(0,0){$2$}}%
% STR 2 0 3 0
% 3 2386 1973 2386 2043 5 0
% $2$
\put(23.8600,-16.4300){\makebox(0,0){$2$}}%
% STR 2 0 3 0
% 3 2526 1833 2526 1903 5 0
% $2$
\put(25.2600,-15.0300){\makebox(0,0){$2$}}%
% STR 2 0 3 0
% 3 2659 1686 2659 1756 5 0
% $2$
\put(26.5900,-13.5600){\makebox(0,0){$2$}}%
% STR 2 0 3 0
% 3 2799 1546 2799 1616 5 0
% $2$
\put(27.9900,-12.1600){\makebox(0,0){$2$}}%
% STR 2 0 3 0
% 3 2939 1406 2939 1476 5 0
% $2$
\put(29.3900,-10.7600){\makebox(0,0){$2$}}%
% STR 2 0 3 0
% 3 3079 1266 3079 1336 5 0
% $2$
\put(30.7900,-9.3600){\makebox(0,0){$2$}}%
% STR 2 0 3 0
% 3 3247 1155 3247 1224 5 0
% $2$
\put(32.4700,-8.2400){\makebox(0,0){$2$}}%
% STR 2 0 3 0
% 3 3548 2414 3548 2484 2 0
% $p'(1)$
\put(35.4800,-20.8400){\makebox(0,0)[lb]{$p'(1)$}}%
% STR 2 0 3 0
% 3 3548 2134 3548 2204 2 0
% $p'(2)$
\put(35.4800,-18.0400){\makebox(0,0)[lb]{$p'(2)$}}%
% STR 2 0 3 0
% 3 3541 1854 3541 1924 2 0
% $p'(3)$
\put(35.4100,-15.2400){\makebox(0,0)[lb]{$p'(3)$}}%
% STR 2 0 3 0
% 3 3359 2582 3359 2652 5 0
% $p(1)$
\put(33.5900,-22.5200){\makebox(0,0){$p(1)$}}%
% STR 2 0 3 0
% 3 3079 2575 3079 2645 5 0
% $p(2)$
\put(30.7900,-22.4500){\makebox(0,0){$p(2)$}}%
% STR 2 0 3 0
% 3 2799 2575 2799 2645 5 0
% $p(3)$
\put(27.9900,-22.4500){\makebox(0,0){$p(3)$}}%
% LINE 1 2 3 0
% 2 2449 2652 2267 2652
% 
\special{pn 13}%
\special{pa 2449 2252}%
\special{pa 2267 2252}%
\special{dt 0.045}%
\special{pa 2267 2252}%
\special{pa 2268 2252}%
\special{dt 0.045}%
% LINE 1 2 3 0
% 2 3597 1581 3597 1406
% 
\special{pn 13}%
\special{pa 3597 1181}%
\special{pa 3597 1006}%
\special{dt 0.045}%
\special{pa 3597 1006}%
\special{pa 3597 1007}%
\special{dt 0.045}%
% STR 2 0 3 0
% 3 2920 3080 2920 3180 5 0
% {\bf Figure 7}
\put(29.2000,-27.8000){\makebox(0,0){{\bf Figure 7}}}%
% STR 2 0 3 0
% 3 2940 2800 2940 2900 5 0
% $|p \rangle$
\put(29.4000,-25.0000){\makebox(0,0){$|p \rangle$}}%
% STR 2 0 3 0
% 3 3990 1950 3990 2050 5 0
% $|p' \rangle$
\put(39.9000,-16.5000){\makebox(0,0){$|p' \rangle$}}%
\end{picture}%

~\\

Define the operators $S :P_{2\Lambda_0}
\cup P_{\Lambda_0+\Lambda_1}\cup P_{2\Lambda_1} 
\to
P_{2\Lambda_0}
\cup P_{\Lambda_0+\Lambda_1}\cup P_{2\Lambda_1} 
$
by $p(j) \to 2-p(j)$.
The Corner Transfer Matrix $A(\zeta), D(\zeta)$
act on the path space 
$P_{2\Lambda_0}$.
The Corner Transfer Matrix $B(\zeta), C(\zeta)$
act on the path space 
$P_{\Lambda_0+\Lambda_1}$.
Using the crossing symmerty of R-matrix,
we can write
\begin{eqnarray}
A(\zeta)=O^{(2)}(-q^{-1}\zeta^{-1})
\cdot S|_{P_{2\Lambda_0}},~~
B(\zeta)=O^{(1)}(\zeta)|_{P_{\Lambda_0
+\Lambda_1}},\nonumber \\
C(\zeta)=S\cdot O^{(1)}
(-q^{-1}\zeta^{-1})|_{P_{\Lambda_0
+\Lambda_1}},~~
D(\zeta)=S\cdot O^{(2)}(\zeta)\cdot S|_{P_{2\Lambda_0}}.
\nonumber 
\end{eqnarray}

Baxter's argument \cite{book:Bax} implies that 
$O^{(1)}(\zeta)=const. 
\zeta^{-H_{CTM}|_{P_{\Lambda_0+\Lambda_1}}}$,
and
$O^{(2)}(\zeta)=const. 
\zeta^{-H_{CTM}|_{P_{2\Lambda_0}}}$,
where 
$H_{CTM}|_{P_{\lambda}}$ does not depend on
the spectral parameter $\zeta$.
Kyoto-school's conjecture
is to identify the path spaces 
$P_{2\Lambda_0},
P_{\Lambda_0+\Lambda_1}$ and
$P_{2\Lambda_1}$
with
the highest weight modules of $U_q(\widehat{sl_2})$, 
$V(2\Lambda_0),V(\Lambda_0+\Lambda_1)$ and
$V(2\Lambda_1)$
, which has been proved
at $q=0$ by crystal base argument.
Under this identification
the degree operator $H_{CTM}|_{P_{\lambda}}$ is realized
as $H_{CTM}|_{P_{\lambda}}=
D|_{V(\lambda)}=-\rho +(\rho, \lambda)$, where $\lambda=
2\Lambda_0,\Lambda_0+\Lambda_1$ and $2\Lambda_1$.
The semi-infinite chain
$\Phi_{UP,\epsilon}(\zeta)$
is identified with the Type-I Vertex operator 
$\Phi_{2\Lambda_0,\epsilon}^{\Lambda_0+\Lambda_1}(\zeta)$ 
defined by
\begin{eqnarray}
\Phi_{2\Lambda_0}^{\Lambda_0+\Lambda_1}
(\zeta)=\sum_{\epsilon=0,1}
\Phi_{2\Lambda_0,\epsilon}^{\Lambda_0+\Lambda_1}
(\zeta)
\otimes v_\epsilon,\nonumber
\end{eqnarray}
where 
$U_q(\widehat{sl_2})$-intertwiner
$\Phi_{2\Lambda_0}^{\Lambda_0+\Lambda_1}
(\zeta)$ is defined by
\begin{eqnarray}
\Phi_{2\Lambda_0}^{\Lambda_0+\Lambda_1}
(\zeta): V(2\Lambda_0) \longrightarrow
V(\Lambda_0+\Lambda_1)
\otimes V^{\left(\frac{1}{2}\right)}_{\zeta}.
\nonumber
\end{eqnarray} 
The semi-infinite chain
$\Phi_{LOW,\epsilon}(\zeta)$
is identified with the Type-I Vertex operator 
\begin{eqnarray}
\Phi_{LOW,\epsilon}(\zeta)=
S\cdot \Phi_{\Lambda_0+\Lambda_1,1-\epsilon}^{2\Lambda_0}
(\zeta)\cdot S.\nonumber
\end{eqnarray}
The Type-I
Vertex operator 
$\Phi_{\Lambda_0+\Lambda_1,\epsilon}^{2\Lambda_0}
(\zeta)$ is defined as the same manner
\begin{eqnarray}
\Phi_{\Lambda_0+\Lambda_1}^{2\Lambda_0}
(\zeta)=\sum_{\epsilon=0,1}
\Phi_{\Lambda_0+\Lambda_1,\epsilon}^{2\Lambda_0}
(\zeta)
\otimes v_\epsilon,\nonumber
\end{eqnarray}
where 
$U_q(\widehat{sl_2})$-intertwiner
$\Phi_{\Lambda_0+\Lambda_1}^{2\Lambda_0}
(\zeta)$ is defined by
\begin{eqnarray}
\Phi_{\Lambda_0+\Lambda_1}^{2\Lambda_0}
(\zeta): V(\Lambda_0+\Lambda_1) \longrightarrow
V(2\Lambda_0)
\otimes V^{\left(\frac{1}{2}\right)}_{\zeta}.
\nonumber
\end{eqnarray} 
We assume, along the line of XXZ-chain \cite{JM},
that the Vertex operators satisfy the Homogeneity condition, 
\begin{eqnarray}
\xi^{-D}\cdot \Phi_{\Lambda_0+\Lambda_1,\epsilon}
^{2\Lambda_0}(\zeta) \cdot \xi^{D}
=\Phi_{\Lambda_0+\Lambda_1,\epsilon}
^{2\Lambda_0}(\zeta/\xi).\nonumber
\end{eqnarray}
From this condition we have
\begin{eqnarray}
P_\epsilon^{(\Lambda_0+\Lambda_1,2\Lambda_0)}=
\frac{
{\rm tr}_{V(2\Lambda_0)}\left(q^{2D}
\Phi^{2\Lambda_0}
_{\Lambda_0+\Lambda_1,1-\epsilon}(-q^{-1}\zeta)
\Phi^{\Lambda_0+\Lambda_1}
_{2\Lambda_0,\epsilon}(\zeta)\right)}
{\sum_{\epsilon=0,1}
{\rm tr}_{V(2\Lambda_0)}\left(q^{2D}
\Phi^{2\Lambda_0}
_{\Lambda_0+\Lambda_1,1-\epsilon}(-q^{-1}\zeta)
\Phi^{\Lambda_0+\Lambda_1}
_{2\Lambda_0,\epsilon}(\zeta)\right)
}.\nonumber
\end{eqnarray}
We adopt the normalizations
\begin{eqnarray}
\langle \Lambda_0+\Lambda_1 |
\Phi^{\Lambda_0+\Lambda_1}
_{2\Lambda_0,1}(\zeta)
|2\Lambda_0 \rangle=1,~~\langle 2\Lambda_0 |
\Phi^{2\Lambda_0}
_{\Lambda_0+\Lambda_1,0}(\zeta)
|\Lambda_0+\Lambda_1 \rangle=1.\nonumber
\end{eqnarray}

The Vertex operator $\Phi^{2\Lambda_0}
_{\Lambda_0+\Lambda_1,\epsilon}(-q^{-1}\zeta)$
is identified with the dual Vertex operator
\begin{eqnarray}
\Phi^{2\Lambda_0}
_{\Lambda_0+\Lambda_1,\epsilon}(-q^{-1}\zeta)
=\Phi^{2\Lambda_0~*}
_{\Lambda_0+\Lambda_1,1-\epsilon}(\zeta).
\nonumber
\end{eqnarray}
The dual vertex operator $\Phi^{2\Lambda_0~*}
_{\Lambda_0+\Lambda_1,\epsilon}(\zeta)$
is defined by 
\begin{eqnarray}
\Phi_{\Lambda_0+\Lambda_1,\epsilon}
^{2\Lambda_0~*}
(\zeta)|v\rangle=
\Phi_{\Lambda_0+\Lambda_1}^{2\Lambda_0~*}
(\zeta)(|v\rangle \otimes v_\epsilon).
\nonumber
\end{eqnarray}
where $U_q(\widehat{sl_2})$-intertwiner
$\Phi_{\Lambda_0+\Lambda_1}^{2\Lambda_0~*}
(\zeta)$ is defined by
\begin{eqnarray}
\Phi_{\Lambda_0+\Lambda_1}^{2\Lambda_0~*}
(\zeta): V(\Lambda_0+\Lambda_1) 
\otimes V^{\left(\frac{1}{2}\right)}_{\zeta}
\longrightarrow
V(2\Lambda_0).\nonumber
\end{eqnarray} 
We adopt the normalization.
\begin{eqnarray}
\langle 2\Lambda_0 |
\Phi^{2\Lambda_0~*}
_{\Lambda_0+\Lambda_1,1}(\zeta)
|\Lambda_0+\Lambda_1 \rangle=1.\nonumber
\end{eqnarray}
Because the operator $\sum_{\epsilon=0,1}
\Phi^{2\Lambda_0~*}
_{\Lambda_0+\Lambda_1,\epsilon}(\zeta)
\Phi^{\Lambda_0+\Lambda_1}
_{2\Lambda_0,\epsilon}(\zeta)$
commutes with $U_q(\widehat{sl_2})$ on the irreducible
module $V(2\Lambda_0)$, it 
becomes a constant $g_{2\Lambda_0}^{-1}$
on $V(2\Lambda_0)$.
The constant $g_{2\Lambda_0}^{-1}$ can be determined by
solving the q-KZ equation for variables $\zeta_1/\zeta_2$,
which is satisfied by
the vacuum expectation
value 
$\langle 2\Lambda_0 |
\Phi^{2\Lambda_0~*}
_{\Lambda_0+\Lambda_1,\epsilon}(\zeta_1)
\Phi^{\Lambda_0+\Lambda_1}
_{2\Lambda_0,\epsilon}(\zeta_2) |2\Lambda_0\rangle$
\cite{IIJMNT}. 
\begin{eqnarray}
g_{2\Lambda_0}
\sum_{\epsilon=0,1}\Phi^{2\Lambda_0~*}
_{\Lambda_0+\Lambda_1,\epsilon}(\zeta)
\Phi^{\Lambda_0+\Lambda_1}
_{2\Lambda_0,\epsilon}(\zeta)=id.\label{eqn:id}
\end{eqnarray}
We get the following formulae
\begin{eqnarray}
P_\epsilon^{(\Lambda_0+\Lambda_1, 2\Lambda_0)}=
\frac{
{\rm tr}_{V(2\Lambda_0)}\left(q^{2D}
\Phi^{2\Lambda_0~*}
_{\Lambda_0+\Lambda_1,\epsilon}(\zeta)
\Phi^{\Lambda_0+\Lambda_1}
_{2\Lambda_0,\epsilon}(\zeta)\right)}
{g_{2\Lambda_0}^{-1}
{\rm tr}_{V(2\Lambda_0)}\left(q^{2D}\right)
}.\label{eqn:trace}
\end{eqnarray}
Here we have used 
\begin{eqnarray}
g_{2\Lambda_0}^{-1}
=(1+q^2)\frac{(q^{12};q^8)_\infty (q^{10};q^4,q^4)_\infty}
{(q^8;q^8)_\infty (q^{12};q^4,q^8)_\infty^2},
\nonumber
\end{eqnarray}
and
\begin{eqnarray}
{\rm tr}_{V(2\Lambda_0)}\left(q^{2D}\right)
=(-q^2;q^2)_\infty (-q^4;q^4)_\infty.
\nonumber
\end{eqnarray}
Here we have used the notation
\begin{eqnarray}
(z;p_1,p_2,\cdots,p_k)_\infty=
\prod_{m_1,m_2,\cdots,m_k=0}^\infty
(1-p_1^{m_1} p_2^{m_2} \cdots p_k^{m_k}z).\nonumber
\end{eqnarray}
By the same arguments we have the followings
formulae for the boundary conditions
$(\mu, \lambda)=(\Lambda_0+\Lambda_1, 2\Lambda_1),
(2\Lambda_0, \Lambda_0+\Lambda_1)$ or
$(2\Lambda_1, \Lambda_0+\Lambda_1)$.
\begin{eqnarray}
P_\epsilon^{(\mu, \lambda)}=
\frac{
{\rm tr}_{V(\lambda)}\left(q^{2D}
\Phi^{\lambda~*}
_{\mu, \epsilon}(\zeta)
\Phi^{\mu}
_{\lambda, \epsilon}(\zeta)\right)}
{g_{\lambda}^{-1}
{\rm tr}_{V(\lambda)}\left(q^{2D}\right)
}.
\end{eqnarray}
Here we have used
$$g_{2\Lambda_1}^{-1}
=(1+q^2)\frac{(q^{12};q^8)_\infty (q^{10};q^4,q^4)_\infty}
{(q^8;q^8)_\infty (q^{12};q^4,q^8)_\infty^2},$$
$$g_{\Lambda_0+\Lambda_1}^{-1}
=\frac{(q^{6};q^4)_\infty (q^{10};q^4,q^4)_\infty}
{(q^4;q^4)_\infty (q^{12};q^4,q^8)_\infty^2},$$
and
$$
{\rm tr}_{V(2\Lambda_1)}\left(q^{2D}\right)
=(-q^2;q^2)_\infty (-q^4;q^4)_\infty,
$$
$$
{\rm tr}_{V(\Lambda_0+\Lambda_1)}\left(q^{2D}\right)
=(-q^2;q^2)_\infty (-q^2;q^4)_\infty.
$$
The vertex operators are defined as the same manner.
From the cyclic property of trace, we obtain
the following relations
$$P_\epsilon^{(2\Lambda_0,\Lambda_0+\Lambda_1)}=
P_{-\epsilon}^{(\Lambda_0+\Lambda_1,2\Lambda_0)},~~~~
P_\epsilon^{(2\Lambda_1,\Lambda_0+\Lambda_1)}=
P_{-\epsilon}^{(\Lambda_0+\Lambda_1,2\Lambda_1)}.
$$
From symmetries we easily know the following relations
$$P_\epsilon^{(2\Lambda_0,\Lambda_0+\Lambda_1)}=
P_{-\epsilon}^{(2\Lambda_1,\Lambda_0+\Lambda_1)},~~~~
P_\epsilon^{(\Lambda_0+\Lambda_1,2\Lambda_0)}=
P_{-\epsilon}^{(\Lambda_0+\Lambda_1,2\Lambda_1)}.
$$
From the commutation relation of Vertex operators
\cite{IIJMNT} and the cyclic property of trace,
we can write down q-diffence equation for
parameter $\zeta_1/\zeta_2$, which
the trace ${\rm tr}_{V(2\Lambda_0)}
(q^{2D}
\Phi_{\Lambda_0+\Lambda_1,\epsilon}
^{2\Lambda_0~*}(\zeta_1)\Phi_{2\Lambda_0,\epsilon}
^{\Lambda_0+\Lambda_1}(\zeta_2) )
$ satisfies.
However we cannot solve this q-difference equation, now.
In order to get exact formulae of the probability
functions, we will use another 
method - free field realizations.

\subsection{Free field realization}
In order to calculate the trace of vertex operators
$$
{\rm tr}_{V(2\Lambda_0)}\left(q^{2D}
\Phi^{2\Lambda_0~*}
_{\Lambda_0+\Lambda_1,\epsilon}(\zeta)
\Phi^{\Lambda_0+\Lambda_1}
_{2\Lambda_0,\epsilon}(\zeta)\right),
$$
we use the free field realization
obtained by Y.Hara \cite{Hara}.
For readers' convenience, we summarize his result.
The formulae in this paper are slightly different
from Hara's paper, because his paper includes 
a few mistakes, which is serious for our purpose.
We use current type generators of $U_q'(\widehat{sl_2})$
introduced by 
Drinfeld.
Let $A$ be an algebra generated by $x_m^\pm (m\in 
{\mathbb Z}), a_m (m \in {\mathbb Z}_{\neq 0}),
\gamma$ and $K$ with relations
$$\gamma : {\rm central},$$
$$[a_m,a_n]=\delta_{m+n,0}\frac{[2m]}{m}
\frac{\gamma^m-\gamma^{-m}}{q-q^{-1}},$$
$$[a_m,K]=0,$$
$$K x_m^\pm K^{-1}=q^{\pm 2} x_m^\pm ,$$
$$[a_m,x_n^\pm]=\pm 
\frac{[2m]}{[m]}\gamma^{\mp\frac{|m|}{2}}
x_{m+n}^\pm,$$
$$x_{m+1}^\pm x_n^\pm-q^{\pm 2}x_n^\pm x_{m+1}^\pm
=q^{\pm 2}x_m^\pm x_{n+1}^\pm
-x_{n+1}^\pm x_m^\pm,$$
$$[x_m^+,x_n^-]=\frac{1}{q-q^{-1}}
(\gamma^{\frac{1}{2}(m-n)}\psi_{m+n}-
\gamma^{-\frac{1}{2}(m-n)}\varphi_{m+n}),
$$
where
$$
\sum_{m=0}^\infty \psi_m z^{-m}=
K \exp\left[ (q-q^{-1})\sum_{m=1}^\infty
a_m z^{-m}\right],$$
$$
\sum_{m=0}^\infty \varphi_{-m} z^{m}=
K^{-1} \exp\left[-(q-q^{-1})\sum_{m=1}^\infty
a_{-m} z^{m}\right],$$
and $\psi_{-m}=\varphi_m=0$ for $m>0$.
Drinfeld showed that the algebra $A$
is isomorphic to $U_q'(\widehat{sl_2})$.
The Chevalley generators are given by the identification
$$t_0=\gamma K^{-1},~
t_1=K,~e_1=x_0^+,~f_1=x_0^-,~e_0=x_1^-K^{-1},
~f_0=Kx_{-1}^+.$$
We will give explicit constructions of level 2 
irreducible highest weight modules.
Let us put $\gamma=q^2$ since we want to construct 
level $2$ modules. 
In the sequel we use a parameter $x=-q$ 
for our convenience.
Commutation and Anti-commutation relations of
bosons and fermions are given by
$$[a_m,a_n]=\delta_{m+n,0}\frac{[2m]^2}{m},$$
$$\{\phi_m,\phi_n\}=\delta_{m+n,0}\eta_m,$$
$$\eta_m=x^{2m}+x^{-2m},$$
where $m,n \in {\mathbb Z}+\frac{1}{2}$
or $ \in {\mathbb Z}$ for Neuveu-Schwartz-sector
or Ramond-sector respectively.
Fock spaces and vacuum vectors are denoted as
$F^a,F^{\phi^{NS}}, F^{\phi^{R}}$ and
$|vac\rangle, |NS\rangle, |R\rangle$ for
boson and Neuveu-Schwartz and Ramond fermion 
respectively.
Fermion currents are defined as
$$\phi^{NS}(z)=\sum_{n \in {\mathbb Z}+\frac{1}{2}}
\phi_n^{NS}z^{-n},
~~\phi^{R}(z)=\sum_{n \in {\mathbb Z}}
\phi_n^{R}z^{-n},
$$
Let us set the degree of the monomial of fermions,
$\phi_{n_1}^{NS}
\phi_{n_2}^{NS}
\cdots \phi_{n_s}^{NS}$ as
$n_1+n_2+\cdots+n_s$,
and
$\phi_{n_1}^{R}
\phi_{n_2}^{R}
\cdots \phi_{n_r}^{R}$ as
$n_1+n_2+\cdots+n_r$.
$Q={\mathbb Z}\alpha$ is the root lattice of $sl_2$
and $F[Q]$ is the group algebra.
We use $\partial$ as
$$[\partial,\alpha]=2.$$
The irreducible highest weight module $V(2\Lambda_0)$
is identified with the Fock space 
$$F^{(0)}_+=
F^a \otimes \left\{ \left(F_{+}^{\phi^{NS}}
\otimes F[2Q]\right) \oplus 
\left(F_{-}^{\phi^{NS}}
\otimes e^{\alpha}
F[2Q]\right)\right\},$$
where $F_{+}^{\phi^{NS}}$
represents the subspace of Fermion Fock space
which is spanned by $even$ degree fermions, and
$F_{-}^{\phi^{NS}}$ that by $odd$ ones.
The highest weight vector is
$|vac\rangle \otimes |NS\rangle \otimes 1$.
The irreducible highest weight module $V(2\Lambda_1)$
is identified with the Fock space 
$$F^{(0)}_-=
F^a \otimes \left\{ \left(F_{+}^{\phi^{NS}}
\otimes e^{\alpha}F[2Q]\right) \oplus 
\left(F_{-}^{\phi^{NS}}
\otimes 
F[2Q]\right)\right\}.$$
The highest weight vector is
$|vac\rangle \otimes |NS\rangle \otimes e^{\alpha}$.
We define the actions of Drinfeld generators as
$$\gamma=q^2,~~~K=q^{\partial},$$
\begin{eqnarray}
x^{\pm}(z)=\sum_{m \in {\mathbb Z}}x_m^\pm z^{-m}
=E^\pm_{<}(z)E^\pm_{>}(z)\phi^{NS}(z)
e^{\pm \alpha}z^{\frac{1}{2}\pm \frac{1}{2}\partial},
\label{eqn:action}
\end{eqnarray}
where
$$E^\pm_{<}(z)
=\exp\left(\pm \sum_{m>0}
\frac{a_{-m}}{[2m]}q^{\mp m}z^m\right),~
E^\pm_{>}(z)
=\exp\left(\mp \sum_{m>0}
\frac{a_{m}}{[2m]}q^{\mp m}z^{-m}\right).$$
The irreducible highest weight module 
$V(\Lambda_0+\Lambda_1)$
is identified with the Fock space 
$$F^{(1)}=
F^a \otimes F^{\phi^{R}}
\otimes e^{\frac{\alpha}{2}}F[Q],$$
where 
$$ \phi_0^R|R\rangle =|R\rangle.$$
The highest weight vector is
$|vac\rangle \otimes |R\rangle \otimes 
e^{\frac{\alpha}{2}}$.
For the actions of Drinfeld generators,
we just replace $\phi^{NS}(z)$ with $\phi^R(z)$
in (\ref{eqn:action}). 
The free field realizations of vertex operators 
\cite{Hara} are constructed by
\begin{eqnarray}
\Phi_{2\Lambda_0,\epsilon}^{\Lambda_0+\Lambda_1}(\zeta)
=\zeta ^{1-\epsilon} \Phi_\epsilon(\zeta),\nonumber
\\
\Phi_{\Lambda_0+\Lambda_1,\epsilon}
^{2\Lambda_0}(\zeta)
=-x \zeta^{\frac{1}{2}-\epsilon}\Phi_\epsilon(\zeta).
\nonumber
\end{eqnarray}
Here we have set
\begin{eqnarray}
\Phi_1(\zeta)&=&B_{I,<}(\zeta)B_{I,>}(\zeta)
\Omega_{NS}^R(\zeta)e^{\frac{\alpha}{2}}x^\partial 
\zeta^{\frac{\partial}{2}},
\label{def:Phi_1}\\
\Phi_0(\zeta)&=&\oint \frac{dw}{2\pi i}
B_{I,<}(\zeta)E_{<}^-(w)
B_{I,>}(\zeta)E_{>}^-(w)
\Omega_{NS}^R(\zeta)
\phi^{NS}(w)e^{-\frac{\alpha}{2}}x^\partial 
\zeta^{\frac{\partial}{2}}w^{-\frac{\partial}{2}}
\nonumber \\
&\times&x^{-2}\zeta^{-1}w^{-\frac{3}{2}}
\frac{\displaystyle
\left(-\frac{w}{x^3\zeta^2}; x^4\right)_\infty}
{\displaystyle
\left(-\frac{w}{x \zeta^2}; x^4\right)_\infty}
\left\{\frac{w}{\displaystyle 1+\frac{w}{x^3\zeta^2}}
+\frac{x^5\zeta^2}{
\displaystyle 1+\frac{x^5\zeta^2}{w}}\right\},
\label{def:Phi_0}
\end{eqnarray}
where
$$B_{I,<}(\zeta)=\exp\left(\sum_{n=1}^\infty
\frac{[n]a_{-n}}{[2n]^2}(-x^5\zeta^2)^n\right),$$
$$B_{I,>}(\zeta)=\exp\left(-\sum_{n=1}^\infty
\frac{[n]a_{n}}{[2n]^2}(-x^3\zeta^2)^{-n}\right).
$$
Note that the sign above differs from 
the one in \cite{Hara}. 
The integrand of $\Phi_0(\zeta)$ has poles only at
$w=-x^5\zeta^2,-x^3\zeta^2$ except for $w=0,\infty$
and the contor of integration encloses $w=0,-x^5\zeta^2$ 
as in figure 8.

\vspace{6pt}

%WinTpicVersion2.13
\unitlength 0.1in
\begin{picture}(23.93,15.84)(16.21,-26.14)
% DOT 0 0 3 0
% 2 3214 2214 3214 2214
% 
\special{pn 20}%
\special{sh 1}%
\special{ar 3214 1814 10 10 0  6.28318530717959E+0000}%
\special{sh 1}%
\special{ar 3214 1814 10 10 0  6.28318530717959E+0000}%
% CIRCLE 2 0 3 0
% 4 3214 2222 2422 2222 2422 2222 2422 2222
% 
\special{pn 8}%
\special{ar 3214 1822 792 792  0.0000000 6.2831853}%
% DOT 0 0 3 0
% 3 2670 2214 2670 2214 2670 2214
% 
\special{pn 20}%
\special{sh 1}%
\special{ar 2670 1814 10 10 0  6.28318530717959E+0000}%
\special{sh 1}%
\special{ar 2670 1814 10 10 0  6.28318530717959E+0000}%
\special{sh 1}%
\special{ar 2670 1814 10 10 0  6.28318530717959E+0000}%
% DOT 0 0 3 0
% 2 2206 2214 2206 2214
% 
\special{pn 20}%
\special{sh 1}%
\special{ar 2206 1814 10 10 0  6.28318530717959E+0000}%
\special{sh 1}%
\special{ar 2206 1814 10 10 0  6.28318530717959E+0000}%
% LINE 2 2 3 0
% 2 2830 2214 3102 2214
% 
\special{pn 8}%
\special{pa 2830 1814}%
\special{pa 3102 1814}%
\special{dt 0.045}%
\special{pa 3102 1814}%
\special{pa 3101 1814}%
\special{dt 0.045}%
% LINE 2 2 3 0
% 2 1710 2214 2054 2214
% 
\special{pn 8}%
\special{pa 1710 1814}%
\special{pa 2054 1814}%
\special{dt 0.045}%
\special{pa 2054 1814}%
\special{pa 2053 1814}%
\special{dt 0.045}%
% STR 2 0 3 0
% 3 3206 2254 3206 2334 5 0
% $0$
\put(32.0600,-19.3400){\makebox(0,0){$0$}}%
% STR 2 0 3 0
% 3 2670 2254 2670 2334 5 0
% $-x^5\zeta^2$
\put(26.7000,-19.3400){\makebox(0,0){$-x^5\zeta^2$}}%
% STR 2 0 3 0
% 3 2206 2254 2206 2334 5 0
% $-x^3\zeta^2$
\put(22.0600,-19.3400){\makebox(0,0){$-x^3\zeta^2$}}%
% VECTOR 1 0 3 0
% 2 2510 1854 2478 1918
% 
\special{pn 13}%
\special{pa 2510 1454}%
\special{pa 2478 1518}%
\special{fp}%
\special{sh 1}%
\special{pa 2478 1518}%
\special{pa 2526 1467}%
\special{pa 2502 1470}%
\special{pa 2490 1449}%
\special{pa 2478 1518}%
\special{fp}%
% STR 2 0 3 0
% 3 4014 2838 4014 2918 2 0
% Figure 8
\put(40.1400,-25.1800){\makebox(0,0)[lb]{\bf Figure 8}}%
\end{picture}%

\vspace{6pt}

For those of $\Phi_{\Lambda_0+\Lambda_1,\epsilon}
^{2\Lambda_0}(\zeta)$ we just replace 
$\Omega_{NS}^R(\zeta), \phi^{NS}(w)$
with
$\Omega_{R}^{NS}(\zeta), \phi^{R}(w)$
in (\ref{def:Phi_1}), (\ref{def:Phi_0}).
The fermion part $\Omega(\zeta)$'s are intertwiners
between different fermion sectors and satisfy
\begin{eqnarray}
\phi^{NS}(w)\Omega_{R}^{NS}(\zeta)
=x^2\zeta w^{-\frac{1}{2}}
\frac{\displaystyle
\left(-\frac{w}{x^3 \zeta^2};x^4\right)_\infty
\left(-\frac{x^7\zeta^2}{w};x^4\right)_\infty}
{\displaystyle
\left(-\frac{w}{x\zeta^2};x^4\right)_\infty
\left(-\frac{x^5\zeta^2}{w};x^4\right)_\infty}
\Omega_R^{NS}(\zeta)\phi^R(w),
\nonumber
\end{eqnarray}
and exactly the same equation except subscripts
for fermion sectors are exchanged. 
The homogeneity condition of the fermion parts is 
given in \cite{Hara}
\begin{equation}
\xi^{d^R}\cdot \Omega_{NS}^R(\zeta) 
\cdot \xi^{-d^{NS}}
=\Omega_{NS}^R(\zeta/\xi).\label{eqn:ferhom}
\end{equation}
The fermion parts $\Omega_{NS}^R(\zeta),
\Omega_{R}^{NS}(\zeta)$
are identified with Type-I Vertex operators of
the two-dimensional Ising model
$\Phi_{NS}^R(\zeta),
\Phi_R^{NS}(\zeta)$ 
which were investigated in details \cite{FJMMN}:

\begin{eqnarray}
\Omega_{NS}^R(\zeta)=\Phi_{NS}^R \left(
-\frac{i}{ x^{ \frac{3}{2}} \zeta} \right),~~
\Omega_{R}^{NS}(\zeta)=\Phi_{R}^{NS} \left(
-\frac{i}{ x^{ \frac{3}{2}} \zeta} \right).\nonumber
\end{eqnarray}
For readers' convenience we summarize the definitions
and properties of vertex operators
$\Phi_{NS}^R(\zeta)$
and $\Phi_{R}^{NS}(\zeta)$, which will be used later.
The Type-I Vertex operators of the two-dimentional
Ising model are operators on Fock spaces
$$\Phi_{NS}^{R}(\zeta):F^{\phi^{NS}} \rightarrow 
F^{\phi^R},$$
$$\Phi_{R}^{NS}(\zeta):F^{\phi^{R}} \rightarrow 
F^{\phi^{NS}}.$$
Define the subsectors as
$$\Phi_{NS,\sigma}^{R}(\zeta)=\Phi_{NS}^{R}(\zeta)
|_{V^{\phi^{NS}}_{\sigma}},~~~
\Phi_{R,\sigma}^{NS}(\zeta)=
P^{\sigma}\Phi_{R}^{NS}(\zeta),~~{\rm for}~~\sigma=\pm,
$$
where $P^\sigma$ denots the projection onto subspace
$F^{\phi^{NS}}_{\sigma}$.
The intertwing relations are given by
$$
\phi^{NS}(w)\Phi_{R}^{NS,\sigma}(\zeta)
=f(w\zeta^2)\Phi_{R}^{NS,-\sigma}(\zeta)\phi^R(w),
$$
$$
\phi^{R}(w)\Phi_{NS,\sigma}^{R}(\zeta)
=f(w\zeta^2)\Phi_{NS,-\sigma}^{R}(\zeta)\phi^{NS}(w).
$$
Here we set
$$
f(z)=-\sqrt{\frac{2{\pi}x}{I}}
(x^4;x^4)_{\infty}(-x^4;x^4)^2_{\infty}{\rm sn}(v),$$
where $z=\exp(\pi i v/I)$ and ${\rm sn}(v)$ 
is Jacobi elliptic
function with half periods $I, iI'$.
Because of the intertwining relations, the following
relations hold.
\begin{eqnarray}
\sum_{\sigma}\Phi_{NS,\sigma}^{R}(x\zeta)
\Phi_{R}^{NS,\sigma}(\zeta)=g^R \times 
id_{F^{\phi^{R}}},\\
\Phi_{R}^{NS,\sigma}(x\zeta)
\Phi_{NS,\sigma}^{R}(\zeta)=g^{NS} \times 
id_{F^{\phi^{NS}}_{\sigma}},\label{eqn:inter}
\end{eqnarray}
where the constants are
$$
g^R=\frac{(x^4;x^4,x^8)^2_\infty}
{(x^2;x^4,x^4)_\infty},~~
g^{NS}=\frac{(x^8;x^4,x^8)_\infty^2}
{(x^6;x^4,x^4)_\infty}.
$$
We will use the following intertwing relations
in the next section.
\begin{eqnarray}
\sigma 
\Phi_{R}^{NS,\sigma}(\zeta)
=\Phi_{R}^{NS,\sigma}(\zeta) \psi^R_1(\zeta),\\
\sigma \Phi_{NS,\sigma}^{R}(\zeta)
=-i \Phi_{NS,-\sigma}^{R}(\zeta)
\psi_1^{NS}(\zeta),\label{eqn:inter'}
\end{eqnarray}
where we have set
\begin{eqnarray}
\psi_1^{R}(\zeta)=\oint \frac{dz}{2\pi i z}
f_0^{R}(z)\phi^{R}(z/\zeta^2),\nonumber
\\
\psi_1^{NS}(\zeta)=\oint \frac{dz}{2\pi i z}
f_0^{NS}(z)\phi^{NS}(z/\zeta^2).\nonumber
\end{eqnarray}
Here we have set
$$f_0^{NS}(z)=
2\sqrt{x}(x^4;x^4)_{\infty}(-x^4;x^4)^2_{\infty}
{\rm cn}(v),~~
f_0^R(z)=\sqrt{\frac{2I}{\pi}}{\rm dn}(v),$$
where
$z=\exp(\pi i v/I)$ and ${\rm cn}(v),~{\rm dn}(v)$ 
are Jacobi elliptic
functions with half periods $I, iI'$.

\subsection{Integral Formulae}
In this subsection we will calculate trace of 
a product of 
two vertex operators and derive an integral formulae
of the probability function.
The free field realizations of the degree operators
are given by
\begin{equation}
D|_{V(\lambda)}=-\rho=
-2\bar{d}^{a}-2\bar{d}^\phi+\frac{1}{4}\partial_\alpha^2
-\frac{1}{2}\partial_\alpha
-\frac{(\lambda,\lambda)}{2},
(\lambda=2\Lambda_0, 2\Lambda_1, \Lambda_0+\Lambda_1).
\label{eqn:degree}
\end{equation}
Here we set
$$
\bar{d}^{a}=\sum_{m=0}^\infty mN_m^a 
,~~\bar{d}^\phi=\sum_{k>0}k N_k^\phi,
$$
where
$$
N_m^a=\frac{m}{[2m]^2}a_{-m} a_m,~~~
N_k^\phi=\frac{1}{x^{2k}+x^{-2k}}\phi_{-k}\phi_k.
$$
We divide trace on $V(2\Lambda_0)
\simeq F^{(0)}_+$ into three parts;
\begin{eqnarray}
{\rm tr}_{V(2\Lambda_0)}(x^{2D}\cdots)
&=&{\rm tr}_{F^a}(x^{-4\bar{d}^a}\cdot \cdot)\cdot
{\rm tr}_{F^{\phi^{NS}}_+}(x^{-4d^{\phi^{NS}}}\cdot
\cdot)
\cdot
{\rm tr}_{F[2Q]}(
x^{\frac{1}{2}\partial^2_\alpha-\partial_\alpha}
\cdot \cdot)\nonumber \\
&+&
{\rm tr}_{F^a}(x^{-4\bar{d}^a}\cdot \cdot)\cdot
{\rm tr}_{F^{\phi^{NS}}_-}(x^{-4d^{\phi^{NS}}}\cdot
\cdot)
\cdot
{\rm tr}_{e^\alpha F[2Q]}(
x^{\frac{1}{2}\partial^2_\alpha-\partial_\alpha}
\cdot \cdot)\nonumber 
\end{eqnarray}
The fermion parts can be written as
$$
{\rm tr}_{F^{\phi^{NS}}_\pm}(x^{-4d^{\phi^{NS}}}
\cdots)=\frac{1}{2}
\left({\rm tr}_{F^{\phi^{NS}}}(x^{-4d^{\phi^{NS}}}\cdot
\cdot)\pm{\rm tr}_{F^{\phi^{NS}}}((ix)^{-4d^{\phi^{NS}}}\cdot
\cdot)
\right).$$
Now we consider the trace of 
a product of two vertex operators
$${\rm tr}_{V(2\Lambda_0)}
\left(x^{2D}
\Phi_{\epsilon}(x^{-1}\zeta)\Phi_{1-\epsilon}(\zeta)
\right),~~~{\rm for}~~\epsilon=0,1.
$$
The trace taken over bosonic space is direct consequence of
the following formulae.
\begin{eqnarray}
&&{\rm tr}_{F^a}\left(y^{-2\bar{d}^a}
\exp\left(\sum_{n=1}^\infty A_n a_{-n}\right)
\exp\left(\sum_{n=1}^\infty B_n a_{n} \right) \right)
\nonumber \\
&=&\exp\left(\sum_{n=1}^\infty
\sum_{m=1}^\infty y^{2mn}A_nB_n\frac{[2n]^2}{n}\right)
\prod_{m=1}^\infty \frac{1}{1-y^{2m}},\nonumber
\end{eqnarray}
and
\begin{eqnarray}
&&\displaystyle
\left(x^6\left(\frac{\zeta_1}{\zeta_2}\right)^2
;x^4,x^4\right)_\infty^2
B_{I,>}(\zeta_2)B_{I,<}(\zeta_1)\nonumber \\
&=&
\displaystyle
\left(x^4\left(\frac{\zeta_1}{\zeta_2}\right)^2
;x^4,x^4\right)_\infty
\left(x^8\left(\frac{\zeta_1}{\zeta_2}\right)^2
;x^4,x^4\right)_\infty
B_{I,<}(\zeta_1)B_{I,>}(\zeta_2)
.\label{eqn:change}
\end{eqnarray}
The trace taken over bosonic space $F^a$ can be written 
as infinite products.

\begin{eqnarray}
&&{\rm tr}_{F^a}\left(y^{-2\bar{d}}
B_{I,<}(\zeta_2)B_{I,<}(\zeta_1)E_<^-(w)
B_{I,>}(\zeta_2)B_{I,>}(\zeta_1)E_>^-(w)
\right)\nonumber\\
&=&
\frac{(x^4y^2;x^4,x^4,y^2)_\infty^2
 (x^8y^2;x^4,x^4,y^2)_\infty^2}
{(x^6y^2;x^4,x^4,y^2)_\infty^4}\cdot
\frac{(x^2y^2;y^2)_\infty}{(y^2;y^2)_\infty}\nonumber \\
&\times&\left[\frac{
\displaystyle \left(x^4y^2\left(\frac{\zeta_2}
{\zeta_1}\right)^2; x^4, x^4, y^2\right)_\infty
\left(x^8y^2\left(\frac{\zeta_2}
{\zeta_1}\right)^2; x^4, x^4, y^2\right)_\infty}
{\displaystyle 
\left(x^6y^2\left(\frac{\zeta_2}
{\zeta_1}\right)^2; x^4, x^4, y^2\right)_\infty^2}
\right. \nonumber \\
&\times& \left.
\frac{\displaystyle
\left(-x^9y^2\frac{\zeta_2^2}{w}
;x^4,y^2\right)_\infty
\left(-x y^2\frac{w}{\zeta_2^2}
;x^4,y^2\right)_\infty
}
{\displaystyle
\left(-x^7y^2\frac{\zeta_2^2}{w}
;x^4,y^2\right)_\infty
\left(-x^{-1} y^2\frac{w}{\zeta_2^2}
;x^4,y^2\right)_\infty}\times 
(\zeta_2 \leftrightarrow \zeta_1)\right].
\label{essential1}
\end{eqnarray}

The trace taken over
lattice space have the following 
theta function formulae.
\begin{eqnarray}
{\rm tr}_{F[2Q]}
\left(x^{\frac{1}{2}\partial_\alpha^2
-\partial_\alpha} f^{\partial_\alpha}\right)=
\sum_{l \in {\mathbb Z}}x^{8l^2-4l}f^{4l}=
\Theta_{x^{16}}(-x^4f^4),
\label{essential2}\\
{\rm tr}_{e^{\alpha}F[2Q]}
\left(x^{\frac{1}{2}\partial_\alpha^2
-\partial_\alpha} f^{\partial_\alpha}\right)=
\sum_{l \in {\mathbb Z}}x^{8l^2+4l}f^{4l+2}=
f^2\Theta_{x^{16}}(-x^{12}f^4).
\label{essential2'}
\end{eqnarray}
Here we have used the standard notation of
the theta function defined as 
$$
\Theta_p(z)=(p;p)_\infty
(z;p)_\infty (pz^{-1};p)_\infty.
$$
Now we concentrate ourselves to trace 
taken over the fermionic space:
\begin{eqnarray}
&&{\rm tr}_{F^{\phi^{NS}}_\pm}\left(x^{-4\bar{d}^\phi}
\Omega_{R}^{NS}(\zeta/x)
\Omega_{NS}^R(\zeta)\phi^{NS}(w)\right)
\nonumber\\
&=&
{\rm tr}_{F^{\phi^{NS}}_\pm}\left(x^{-4\bar{d}^\phi}
\Phi_{R}^{NS,\pm}\left(-\frac{i}{x^{\frac{1}{2}}\zeta}
\right)
\Phi_{NS,\mp}^R\left(-\frac{i}{x^\frac{3}{2}\zeta}
\right)\phi^{NS}(w)\right),
\label{eqn:f1}
\end{eqnarray}
where the symbols $\Phi_{NS,\pm}^R(\zeta)$
and $\Phi_R^{NS,\pm}(\zeta)$ are Type-I Vertex operators
of two dimensional Ising model.
From the relations (\ref{eqn:inter}) and 
(\ref{eqn:inter'}), we 
deform
the vertex operators in (\ref{eqn:f1}) 
to the fermion currents.
Only the fermion currents and the degree operator
appear in trace.
\begin{eqnarray}
&\pm \frac{i}{2} g^{NS}
&\left\{
{\rm tr}_{F^{\phi^{NS}}}
\left(x^{-4\bar{d}^\phi}
\psi_1^{NS}
\left(-\frac{i}{x^\frac{3}{2}\zeta}\right)
\phi^{NS}(w)\right)\right.\nonumber \\
&&\pm\left.
{\rm tr}_{F^{\phi^{NS}}}
\left((ix)^{-4\bar{d}^\phi}\psi_1^{NS}
\left(-\frac{i}{x^\frac{3}{2}\zeta}\right)\phi^{NS}(w)
\right)
\right\}.
\label{eqn:f2}
\end{eqnarray}
Here we have used
$$\psi_1^{NS}(\zeta)=\oint \frac{dz}{2\pi i z}
f_0^{NS}(z)\phi^{NS}(z/\zeta^2),~~~
f_0^{NS}(z)=
2\sqrt{x}(x^4;x^4)_{\infty}(-x^4;x^4)^2_{\infty}
{\rm cn}(v).$$
To take the trace of (\ref{eqn:f2})
we invoke the following simple relation
$$\frac{{\rm tr}_{F^{\phi^{NS}}}
\left(\xi^{-2\bar{d}^\phi}\phi^{NS}(w_1)
\phi^{NS}(w_2)\right)}
{{\rm tr}_{F^{\phi^{NS}}}
\left(\xi^{-2\bar{d}^\phi}\right)}=
\sum_{m \in {\mathbb Z}+\frac{1}{2}}
\frac{x^{2m}+x^{-2m}}{1+\xi^{2m}}
\left(\frac{w_2}{w_1}\right)^m.
$$
We calculate the trace taken over fermionic space and
calculate integrals in (\ref{eqn:f2}), 
using the
Fourier expansion of coefficient function
$f_0^{NS}(z)$ given by
$$
f_0^{NS}(z)
=\frac{1}{\sqrt{x}(x^4;x^4)_{\infty}(-x^4;x^4)^2_{\infty}}
\sum_{r \in {\mathbb Z}+\frac{1}{2}}
\frac{1}{x^{2r}+x^{-2r}}z^r.
$$
We have the following infinite sum formulae
\begin{eqnarray}
&&\pm \frac{i}{2} g^{NS}
\frac{1}{\sqrt{x}(x^4;x^4)_{\infty}(-x^4;x^4)^2_{\infty}}
\nonumber \\
&\times&\left\{
{\rm tr}_{F^{\phi^{NS}}}\left(x^{-4\bar{d}^\phi}\right)
\sum_{m \in {\mathbb Z}+\frac{1}{2}}
\frac{1}{x^{2m}+x^{-2m}}\left(-\frac{w}{x^5\zeta^2}
\right)^m \right.
\nonumber \\
&& \mp \left.
{\rm tr}_{F^{\phi^{NS}}}\left((ix)^{-4\bar{d}^\phi}\right)
\sum_{m \in {\mathbb Z}+\frac{1}{2}}
\frac{1}{x^{2m}-x^{-2m}}\left(-\frac{w}{x^5\zeta^2}\right)^m
\right\}.\nonumber
\end{eqnarray}
Using the following theta function's identities
$$
\sum_{m \in {\mathbb Z}+\frac{1}{2}}
\frac{1}{x^{2m} \pm x^{-2m}}z^m=
\pm x\sqrt{z}
\frac{(x^4;x^4)_\infty^2}{(\mp x^2;x^4)_\infty^2}
\frac{\Theta_{x^4}(\mp x^4z)}
{\Theta_{x^4}(x^2z)},
$$
we get the infinite product formulae of trace taken over
fermionic space
\begin{eqnarray}
&&{\rm tr}_{F^{\phi^{NS}}_\pm}\left(x^{-4\bar{d}^\phi}
\Omega_{R}^{NS}(\zeta/x)
\Omega_{NS}^R(\zeta)\phi^{NS}(w)\right)=
\mp \frac{1}{2}\frac{w^{\frac{1}{2}}}{\zeta x^2}
\frac{(x^4;x^4)_\infty^2}
{(-x^4;x^4)_\infty^2}
\frac{(x^8;x^4,x^8)_\infty^2}{(x^6;x^4,x^4)_\infty} 
\nonumber
\\
&\times&
\frac{1}
{
\displaystyle
\Theta_{x^4}\left(-\frac{w}{x^3\zeta^2}\right)}
\left\{
(-x^2;x^4)_\infty
\frac{\Theta_{x^4} \left(\frac{w}{x\zeta^2}\right)}{
\displaystyle \Theta_{x^4}(-x^2)}\pm 
(x^2;x^4)_\infty
\frac{\displaystyle \Theta_{x^4}
\left(-\frac{w}{x\zeta^2}\right)}{
\displaystyle \Theta_{x^4}(x^2)}
\right\}.\label{essential3}
\end{eqnarray}
Here we have
used the character formulae 
of fermion Fock space;
$$
{\rm tr}_{F^{\phi^{NS}}}
(x^{-4\bar{d}^\phi})=(-x^2;x^4)_\infty.
$$
Combining the relations 
(\ref{eqn:change}), (\ref{essential1}),
(\ref{essential2}), (\ref{essential2'}) and 
(\ref{essential3}),
we get an integral formulae of 
the spontaneous staggered polarizations. 
We can summarize the conclusion just obtained 
as follows: 

\vspace{12pt}

\begin{em}
The trace of a product of two
vertex operators has following integral formulae.

\begin{eqnarray}
&&{\rm tr}_{V(2{\Lambda}_0)}
\left(x^{2D}{\Phi}_{\Lambda_0+\Lambda_1,\epsilon}
^{2\Lambda_0~*}(\zeta)
{\Phi}_{2\Lambda_0,\epsilon}^{\Lambda_0+\Lambda_1}
(\zeta)\right)\nonumber\\
&=&\frac{1}{2x^3 \zeta^2} 
\frac{(x^8;x^4,x^8)^2_{\infty}
(x^{10};x^4,x^4)_{\infty}
}{(x^8;x^4,x^4)^2_{\infty}}
\frac{(x^6;x^4)^2_{\infty}
(x^4;x^4)^2_{\infty}}{
(-x^4;x^4)^2_\infty}\nonumber\\
&\times&{\oint}_{C_{1+\epsilon}}\frac{dw}{2{\pi}i w}
\left\{
(1+x^2)w+2x^{5-2\epsilon}{\zeta}^2\right\}
\frac{1}{\Theta_{x^4}\left(-\displaystyle{
\frac{w}{x{\zeta}^2}}\right)
\Theta_{x^4}\left(-\displaystyle{
\frac{w}{x^3{\zeta}^2}}\right)}\nonumber\\
&\times&\left[\left\{\frac{\Theta_{x^4}
\left(\displaystyle{\frac{w}{x{\zeta}^2}}\right)}
{\Theta_{x^4}(-x^2)}(-x^2;x^4)_{\infty}
+\frac{\Theta_{x^4}\left(-\displaystyle{
\frac{w}{x{\zeta}^2}}\right)}{\Theta_{x^4}(x^2)}
(x^2;x^4)_{\infty}\right\}\Theta_{x^{16}}
\left(-x^6\left(
\frac{w}{{\zeta}^2}\right)^2\right)\right.
\label{eqn:trace1}\\
&-&\left.\left\{\frac{\Theta_{x^4}\left(
\displaystyle{\frac{w}{x{\zeta}^2}}\right)}
{\Theta_{x^4}(-x^2)}(-x^2;x^4)_{\infty}-
\frac{\Theta_{x^4}\left(-\displaystyle{
\frac{w}{x{\zeta}^2}}\right)}{\Theta_{x^4}(x^2)}
(x^2;x^4)_{\infty}\right\}x^3\frac{\zeta^2}{w}
\Theta_{x^{16}}\left(-\frac{1}{x^2}
\left(\frac{w}{{\zeta}^2}\right)^2
\right)\right],\nonumber
\end{eqnarray}
Here the contours encircle $w=0$ in such a way that
\begin{flushleft}
\hspace{3cm}$C_1:$ $-x^5\zeta^2$ 
is inside and $-x^3\zeta^2$ is outside,\par
\hspace{3cm}$C_2:$ $-x^3\zeta^2$ is inside 
and $-x\zeta^2$ is outside,
\end{flushleft}
as in figure 9.
\end{em}

\vspace{6pt}

%WinTpicVersion2.13
\unitlength 0.1in
\begin{picture}(43.75,14.39)(-0.62,-19.68)
% DOT 0 0 3 0
% 2 1432 1615 1432 1615
% 
\special{pn 20}%
\special{sh 1}%
\special{ar 1432 1215 10 10 0  6.28318530717959E+0000}%
\special{sh 1}%
\special{ar 1432 1215 10 10 0  6.28318530717959E+0000}%
% CIRCLE 2 0 3 0
% 4 1435 1618 771 1626 771 1626 787 1626
% 
\special{pn 8}%
\special{ar 1435 1218 664 664  3.1292476 3.1295450}%
% CIRCLE 2 0 3 0
% 4 3621 1621 2929 1615 2929 1615 2929 1615
% 
\special{pn 8}%
\special{ar 3621 1221 692 692  0.0000000 6.2831853}%
% STR 2 0 3 0
% 3 2507 2389 2507 2453 5 0
% Figure 9
\put(25.0700,-20.5300){\makebox(0,0){\bf Figure 9}}%
% DOT 0 0 3 0
% 2 971 1626 971 1626
% 
\special{pn 20}%
\special{sh 1}%
\special{ar 971 1226 10 10 0  6.28318530717959E+0000}%
\special{sh 1}%
\special{ar 971 1226 10 10 0  6.28318530717959E+0000}%
% DOT 0 0 3 0
% 2 523 1626 523 1626
% 
\special{pn 20}%
\special{sh 1}%
\special{ar 523 1226 10 10 0  6.28318530717959E+0000}%
\special{sh 1}%
\special{ar 523 1226 10 10 0  6.28318530717959E+0000}%
% DOT 0 0 3 0
% 2 3171 1626 3171 1626
% 
\special{pn 20}%
\special{sh 1}%
\special{ar 3171 1226 10 10 0  6.28318530717959E+0000}%
\special{sh 1}%
\special{ar 3171 1226 10 10 0  6.28318530717959E+0000}%
% DOT 0 0 3 0
% 2 2707 1626 2707 1626
% 
\special{pn 20}%
\special{sh 1}%
\special{ar 2707 1226 10 10 0  6.28318530717959E+0000}%
\special{sh 1}%
\special{ar 2707 1226 10 10 0  6.28318530717959E+0000}%
% LINE 2 2 3 0
% 2 3315 1626 3531 1626
% 
\special{pn 8}%
\special{pa 3315 1226}%
\special{pa 3531 1226}%
\special{dt 0.045}%
\special{pa 3531 1226}%
\special{pa 3530 1226}%
\special{dt 0.045}%
% LINE 2 2 3 0
% 2 2427 1626 2611 1626
% 
\special{pn 8}%
\special{pa 2427 1226}%
\special{pa 2611 1226}%
\special{dt 0.045}%
\special{pa 2611 1226}%
\special{pa 2610 1226}%
\special{dt 0.045}%
% LINE 2 2 3 0
% 2 1139 1626 1307 1626
% 
\special{pn 8}%
\special{pa 1139 1226}%
\special{pa 1307 1226}%
\special{dt 0.045}%
\special{pa 1307 1226}%
\special{pa 1306 1226}%
\special{dt 0.045}%
% LINE 2 2 3 0
% 2 171 1626 403 1626
% 
\special{pn 8}%
\special{pa 171 1226}%
\special{pa 403 1226}%
\special{dt 0.045}%
\special{pa 403 1226}%
\special{pa 402 1226}%
\special{dt 0.045}%
% VECTOR 1 0 3 0
% 2 859 1250 835 1282
% 
\special{pn 13}%
\special{pa 859 850}%
\special{pa 835 882}%
\special{fp}%
\special{sh 1}%
\special{pa 835 882}%
\special{pa 891 841}%
\special{pa 867 839}%
\special{pa 859 817}%
\special{pa 835 882}%
\special{fp}%
% VECTOR 1 0 3 0
% 2 3083 1186 3035 1242
% 
\special{pn 13}%
\special{pa 3083 786}%
\special{pa 3035 842}%
\special{fp}%
\special{sh 1}%
\special{pa 3035 842}%
\special{pa 3094 804}%
\special{pa 3070 802}%
\special{pa 3063 778}%
\special{pa 3035 842}%
\special{fp}%
% STR 2 0 3 0
% 3 3619 1650 3619 1730 5 0
% $0$
\put(36.1900,-13.3000){\makebox(0,0){$0$}}%
% STR 2 0 3 0
% 3 3171 1658 3171 1738 5 0
% $-x^3\zeta^2$
\put(31.7100,-13.3800){\makebox(0,0){$-x^3\zeta^2$}}%
% STR 2 0 3 0
% 3 2707 1650 2707 1730 5 0
% $-x\zeta^2$
\put(27.0700,-13.3000){\makebox(0,0){$-x\zeta^2$}}%
% STR 2 0 3 0
% 3 1435 1642 1435 1722 5 0
% $0$
\put(14.3500,-13.2200){\makebox(0,0){$0$}}%
% STR 2 0 3 0
% 3 523 1634 523 1714 5 0
% $-x^3\zeta^2$
\put(5.2300,-13.1400){\makebox(0,0){$-x^3\zeta^2$}}%
% STR 2 0 3 0
% 3 1435 2370 1435 2450 5 0
% $C_1$
\put(14.3500,-20.5000){\makebox(0,0){$C_1$}}%
% CIRCLE 2 0 3 0
% 4 1435 1626 739 1626 739 1626 739 1626
% 
\special{pn 8}%
\special{ar 1435 1226 696 696  0.0000000 6.2831853}%
% STR 2 0 3 0
% 3 971 1634 971 1714 5 0
% $-x^5\zeta^2$
\put(9.7100,-13.1400){\makebox(0,0){$-x^5\zeta^2$}}%
% STR 2 0 3 0
% 3 3610 2350 3610 2450 5 0
% $C_2$
\put(36.1000,-20.5000){\makebox(0,0){$C_2$}}%
% DOT 0 0 3 0
% 3 3620 1630 3620 1630 3620 1630
% 
\special{pn 20}%
\special{sh 1}%
\special{ar 3620 1230 10 10 0  6.28318530717959E+0000}%
\special{sh 1}%
\special{ar 3620 1230 10 10 0  6.28318530717959E+0000}%
\special{sh 1}%
\special{ar 3620 1230 10 10 0  6.28318530717959E+0000}%
\end{picture}%

\vspace{6pt}

\subsection{Infinite Product Formulae}
The purpose of this subsection is
to calculate integral in (\ref{eqn:trace1}) 
and derive an infinite product
formula of spontaneous staggered polarization
in Main Result.\\
Let us use the following abbreviations:
\begin{eqnarray}
p_1(x)&=&
\frac{(x^{16};x^{16})_\infty}
{(x^4;x^4)_\infty^3}(-x^4;x^8)_\infty,\nonumber \\
p_2(x)&=&
\frac{(x^{16};x^{16})_\infty}
{(x^2;x^2)_\infty^2 (x^4;x^4)_\infty}(-x^4;x^4)_\infty^2
(-x^8;x^{16})^2_\infty,\nonumber \\
p_3(x)&=&
\frac{(x^{16};x^{16})_\infty}
{(x^2;x^2)_\infty^2 (x^4;x^4)_\infty}(-x^4;x^4)_\infty^2
(-x^{16};x^{16})^2_\infty,\nonumber \\
p_4(x)&=&
\frac{(x^{16};x^{16})_\infty}
{(x^2;x^2)_\infty^2 (x^4;x^4)_\infty}(-x^2;x^4)_\infty^2
(-x^4;x^{8})_\infty.\nonumber 
\end{eqnarray}
Now we consider the following integral
$$
\oint \frac{dw}{2\pi i w}w
\frac{\Theta_{x^4}\left(\frac{w}{x \zeta^2}\right)
\Theta_{x^{16}}\left(-x^6\left(\frac{w}{\zeta^2}\right)^2
\right)}{
\Theta_{x^4}\left(-\frac{w}{x \zeta^2}\right)
\Theta_{x^{4}}\left(-\frac{w}{x^3\zeta^2}
\right)}.
$$
The integrand function $I(z)=z
\frac{\displaystyle \Theta_{x^4}\left(z\right)
\Theta_{x^{16}}\left(-x^8z^2
\right)}{\displaystyle
\Theta_{x^4}\left(-z\right)
\Theta_{x^{4}}\left(-z/x^2\right)}~(z=w/x\zeta^2)$
is an elliptic function and has odd invariance;
$$
I(x^8z)=I(z),~~I(z)=-I(z^{-1}).
$$
Therefore, taking the residue of Cauchy's
principal value at $z=-1$, we have
$$
\oint_{|z|=1{-0}}\frac{dz}{2 \pi iz}
I(z)=-\frac{1}{2}{\rm Res}_{z=-1}I(z).
$$
Taking the residues near $w=-x\zeta^2$, we get
the following formulae
\begin{eqnarray}
\oint_C \frac{dw}{2\pi i w}w
\frac{\Theta_{x^4}\left(\frac{w}{x \zeta^2}\right)
\Theta_{x^{16}}\left(-x^6\left(\frac{w}{\zeta^2}\right)^2
\right)}{
\Theta_{x^4}\left(-\frac{w}{x \zeta^2}\right)
\Theta_{x^{4}}\left(-\frac{w}{x^3\zeta^2}
\right)}=x^3 \zeta^2 \times
\left\{
\begin{array}{cc}
p_2(x)-p_4(x),&~~C=C_1,\\
p_2(x),&~~C=C_2.
\end{array}\right. \label{PD1}
\end{eqnarray}
As the same arguments above, we get
\begin{eqnarray}
\oint_C \frac{dw}{2\pi i w}
\frac{\Theta_{x^4}\left(\frac{w}{x \zeta^2}\right)
\Theta_{x^{16}}\left(-\frac{1}{x^2}
\left(\frac{w}{\zeta^2}\right)^2
\right)}{
\Theta_{x^4}\left(-\frac{w}{x \zeta^2}\right)
\Theta_{x^{4}}\left(-\frac{w}{x^3\zeta^2}
\right)}=
\left\{
\begin{array}{cc}
-2x^2 p_3(x)+p_4(x),&~~~~C=C_1,\\
-2x^2 p_3(x),&~~~~C=C_2.
\end{array}\right.  \label{PD2}
\end{eqnarray}
We consider the following integral
$$
\oint \frac{dw}{2\pi i w}
w \frac{\Theta_{x^{16}}\left(-x^6\left(
\frac{w}{\zeta^2}\right)^2\right)}
{\Theta_{x^4}\left(-\frac{w}{x^3 \zeta^2}\right)}.
$$
The integrand function $J(z)=
\frac{1}{\displaystyle z}
\frac{\displaystyle \Theta_{x^{16}}(-z^2)}
{\displaystyle \Theta_{x^4}(-x^2 z)}~(z=wx^3/\zeta^2)$ 
is an elliptic function
and is composed of a product of two odd invariant
functions $J_1(z)=\frac{\displaystyle \Theta_{x^{16}}(-z^2)}
{\displaystyle \Theta_{x^{16}}(z^2)}$ and $J_2(z)
=\frac{1}{\displaystyle z}
\frac{\displaystyle \Theta_{x^{16}}(z^2)}
{\displaystyle \Theta_{x^4}(-x^2 z)}
$:
$$
J(x^8z)=J(z),~~J(z)=J_1(z)J_2(z),
$$
$$
J_k(x^8z)=-J_k(z),~~J_k(z)=-J_k(z^{-1})
~~{\rm for}~~k=1,2.
$$
From the odd invariance property,
the constant term of Fourier expansion for variable $u$
such that
$z=e^{iu}$
becomes zero.
Therefore we have
$$
\oint_{|z|=1}\frac{dz}{2\pi iz}J_1(z)
J_2(z)=0.
$$ 
Taking the residue near $w=-x^3/\zeta^2$,
we get the following formulae
\begin{eqnarray}
\oint_C \frac{dw}{2\pi i w}w
\frac{\Theta_{x^{16}}\left(-x^6
\left(\frac{w}{\zeta^2}\right)^2
\right)}{
\Theta_{x^{4}}\left(-\frac{w}{x^3\zeta^2}
\right)}=
\left\{
\begin{array}{cc}
0,&~~~~C=C_1,\\
x^3\zeta^2 p_1(x),&~~~~C=C_2.
\end{array}\right.  \label{PD3}
\end{eqnarray}
As the same arguments above, we get 
the following formulae
\begin{eqnarray}
\oint_C \frac{dw}{2\pi i w}
\frac{\Theta_{x^{16}}\left(-\frac{1}{x^2}
\left(\frac{w}{\zeta^2}\right)^2
\right)}{
\Theta_{x^{4}}\left(-\frac{w}{x^3\zeta^2}
\right)}=
\left\{
\begin{array}{cc}
p_1(x),&~~~~C=C_1,\\
0,&~~~~C=C_2.
\end{array}\right.  \label{PD4}
\end{eqnarray}
We consider the following integral
$$
\oint \frac{dw}{2\pi i w}
\frac{\Theta_{x^{16}}\left(-x^6
\left(\frac{w}{\zeta^2}\right)^2
\right)}{
\Theta_{x^{4}}\left(-\frac{w}{x^3\zeta^2}
\right)}.
$$
The integrand function $K(z)=
\frac{\displaystyle \Theta_{x^{16}}(-z^2)}
{\displaystyle 
\Theta_{x^4}(-z/x^6)},~(z=x^3w/\zeta^2)$ satisfies
the quasi-periodicity.
$$
K(z)=x^8K(x^8z).
$$
Therefore we have
\begin{eqnarray}
&&\oint_{|z|=1}\frac{dz}{2\pi iz}K(z)=
\frac{-1}{1-x^{8}}\left\{
\oint_{z=-x^{2}}+\oint_{z=-x^{6}}\right\}
\frac{dz}{2\pi iz}
K(z),\nonumber
\end{eqnarray}
where we take the residues near $\infty$.
Now we get the following formulae.
\begin{eqnarray}
\oint_C \frac{dw}{2\pi i w}
\frac{\Theta_{x^{16}}\left(-x^6
\left(\frac{w}{\zeta^2}\right)^2
\right)}{
\Theta_{x^{4}}\left(-\frac{w}{x^3\zeta^2}
\right)}
=\left\{
\begin{array}{cc}
\frac{1}{1+x^4} p_1(x),&~~~C=C_1,\\
-\frac{x^4}{1+x^4} p_1(x),&~~~C=C_2.
\end{array}\right. \label{PD5}
\end{eqnarray}
As the same arguments above we get the following
formulae:

\begin{eqnarray}
\oint_C \frac{dw}{2\pi i w}\frac{1}{w}
\frac{\Theta_{x^{16}}\left(-\frac{1}{x^2}
\left(\frac{w}{\zeta^2}\right)^2
\right)}{
\Theta_{x^{4}}\left(-\frac{w}{x^3\zeta^2}
\right)}
=\frac{1}{x^3 \zeta^2}\left\{
\begin{array}{cc}
-\frac{1}{1+x^4} p_1(x),&~~~C=C_1,\\
\frac{x^4}{1+x^4} p_1(x),&~~~C=C_2,
\end{array}\right. \label{PD6}
\end{eqnarray}
and
\begin{eqnarray}
&&\oint_C \frac{dw}{2\pi i w}\frac{1}{w}
\frac{\Theta_{x^4}\left(\frac{w}{x \zeta^2}\right)
\Theta_{x^{16}}\left(-\frac{1}{x^2}
\left(\frac{w}{\zeta^2}\right)^2
\right)}{
\Theta_{x^4}\left(-\frac{w}{x \zeta^2}\right)
\Theta_{x^{4}}\left(-\frac{w}{x^3\zeta^2}
\right)}
=\frac{-1}{x\zeta^2(1-x^8)}\nonumber\\
&\times&\left\{
\begin{array}{cc}
-2x^4 p_2(x)-4x^2 p_3(x) +(x^2+x^{-2})p_4(x),&~~~~C=C_1,\\
-2x^4 p_2(x)-4x^2 p_3(x)+(x^2+x^6)p_4(x),&~~~~C=C_2,
\end{array}\right. \label{PD7}
\end{eqnarray}
and
\begin{eqnarray}
&&\oint_C \frac{dw}{2\pi i w}
\frac{\Theta_{x^4}\left(\frac{w}{x \zeta^2}\right)
\Theta_{x^{16}}\left(-x^6
\left(\frac{w}{\zeta^2}\right)^2
\right)}{
\Theta_{x^4}\left(-\frac{w}{x \zeta^2}\right)
\Theta_{x^{4}}\left(-\frac{w}{x^3\zeta^2}
\right)}
=\frac{-1}{1-x^8}\nonumber\\
&\times&\left\{
\begin{array}{cc}
2x^2 p_2(x)+4x^8 p_3(x) -(1+x^4)p_4(x),&~~~~C=C_1,\\
2x^2 p_2(x)+4x^8 p_3(x)-(x^4+x^8)p_4(x),&~~~~C=C_2.
\end{array}\right. \label{PD8}
\end{eqnarray}
Inserting the relations (\ref{PD1}),
(\ref{PD2}), (\ref{PD3}), (\ref{PD4}),
(\ref{PD5}), (\ref{PD6}), (\ref{PD7}) 
and (\ref{PD8}) into
integral formulae (\ref{eqn:trace1}), we arrive at the
formulae (\ref{eqn:P0}) and (\ref{eqn:P1}).
Now we have proved Main Result in page 9.

~
\\
{\sl Acknowledgements}~~~~~~~
We
want to thank to Professor Tetsuji Miwa 
and Professor Kimio Ueno for their encouragements.
This work is partly supported by Waseda University 
Grant for Special Research Projects, 
the Grant from
Research Institute of Science and Technology,
Nihon University, and the Grant from the Ministry
of Education (11740099).

\end{document}